\documentclass[aps,rmp,reprint]{revtex4-1} %

\usepackage[colorlinks=true, allcolors=blue]{hyperref}
\usepackage[dvipsnames]{xcolor}
\usepackage[T1]{fontenc}
\usepackage{booktabs}
\usepackage{amsmath}
\usepackage{comment}
\usepackage{amssymb}
\usepackage{graphicx}
\usepackage{commath}
\usepackage{xspace}
\usepackage{comment}

\renewcommand{\selectlanguage}[1]{} %

\newcommand{\owner}[1]{}%
\newcommand{\MS}[1]{}%
\newcommand{\FB}[1]{}%
\newcommand{\fb}[1]{}%
\newcommand{\MK}[1]{}%
\newcommand{\DG}[1]{}%
\newcommand{\GI}[1]{}%
\newcommand{\JL}[1]{}%
\newcommand{\KSW}[1]{}%
\newcommand{\ltg}[1]{}%
\newcommand{\rtg}[1]{}%
\newcommand{\lrtg}[1]{}%
\newcommand{\tg}[1]{}%

\newcommand{\eref}[1]{Eq.~(\ref{#1})}

\newcommand{\sref}[1]{Sec.~\ref{#1}}
\newcommand{\fref}[1]{Fig.~\ref{#1}}

\newcommand{\tref}[1]{Table~\ref{#1}}

\newcommand{\msvm}{MSVm\xspace} %
\newcommand{\avm}{AVm\xspace} %
\newcommand{\dw}{DWm\xspace} %
\newcommand{\abcm}{ABCm\xspace} %
\newcommand{\hk}{HKm\xspace} %

\newcommand{\me}{ME\xspace} %
\newcommand{\mes}{MEs\xspace} %
\newcommand{\fpe}{Fokker--Planck equation\xspace} %
\newcommand{\ame}{AME\xspace} %
\newcommand{\ames}{AMEs\xspace} %

\begin{document}
\title{Opinion dynamics: Statistical physics and beyond}

\author{Michele Starnini}
\affiliation{Department of Engineering,~ Universitat Pompeu Fabra, E-08018 Barcelona, Spain}

\author{Fabian Baumann}
\affiliation{Department of Biology, University of Pennsylvania, Philadelphia, PA 19104, USA}

\author{Tobias Galla}
\affiliation{Institute for Cross-Disciplinary Physics and Complex Systems (IFISC UIB-CSIC), Edifici Instituts Universitaris de Recerca, 
Campus Universitat de les Illes Balears,
E-07122 Palma, Spain}
 
\author{David Garcia}
\affiliation{Department of Politics and Public Administration, University of Konstanz, DE-78457 Konstanz, Germany, \\~\\
Complexity Science Hub, Vienna, Austria\\~\\
Barcelona Supercomputing Center, Barcelona, Spain}

\author{Gerardo I\~niguez}
\affiliation{Tampere Complexity Lab, Data Science Research Centre, Tampere University, FI-33720 Tampere, Finland \\~\\Centro de Ciencias de la Complejidad, Universidad Nacional Autonóma de México, 04510 Ciudad de México, Mexico}

\author{M\'arton Karsai}
\affiliation{Department of Network and Data Science, Central European University, Vienna, Austria \\~\\
HUN-REN Rényi Institute of Mathematics, Budapest, Hungary}

\author{Jan Lorenz}
\affiliation{School of Business, Social and Decision Sciences, Constructor University, Bremen, Germany}

\author{Katarzyna Sznajd-Weron}
\affiliation{Wroclaw University of Science and Technology, Faculty of Management, Wyb. Wyspiańskiego 27, Wroclaw, 50-370, Poland}
\date{\today{}}

\begin{abstract}
Opinion dynamics, the study of how individual beliefs and collective public opinion evolve, is a fertile domain for applying the framework of statistical physics to complex social phenomena. 
Like physical systems, societies exhibit macroscopic regularities arising from numerous localized interactions, leading to outcomes such as consensus or fragmentation. 
Opinion dynamics has grown from a fringe interest to an active field of physics, attracting interdisciplinary methods from computer science, economics, and sociology. 
This review covers the ongoing rapid progress of the field, driven by an unprecedented surge in available large-scale behavioral data, bridging the inherently interdisciplinary literature. 
We begin with essential concepts and definitions, encompassing the nature of opinions and their microscopic and macroscopic dynamics. 
This foundation leads to an overview of empirical research, from lab experiments to large-scale data analysis, which informs and validates models of opinion dynamics. 
We then categorize models by the macroscopic phenomena they describe (e.g., consensus, polarization, echo chambers) and the microscopic mechanisms of opinion change they encode (e.g., homophily, assimilation). 
The review covers common analytical and computational tools to study these models, such as stochastic processes, exact and approximate treatments, simulation techniques, and optimization methods. 
Finally, we explore emerging frontiers, including the ongoing effort to establish a closer connection from empirical data to models, and the recent developments in building artificial intelligence agents as testbeds for novel social phenomena. 
By systematizing terminology across disciplines and emphasizing analogies with traditional physics, we highlight connections between theoretical models, concepts in social psychology, and observable societal phenomena. 
This review aims to consolidate current knowledge, provide a robust theoretical foundation, and shape future research directions in opinion dynamics.
\end{abstract}

\maketitle
\tableofcontents{}

\section{Introduction}

The study of collective phenomena in complex systems extends far beyond the traditional boundaries of physics. 
It finds applications in social systems, particularly those formed by millions of individuals. 
Similar to condensed matter, fluids, electronic materials, and many other archetypal systems in statistical mechanics, societal dynamics show stunning macroscopic regularities. 
There are transitions from disorder to order, including the emergence of common languages and cultures, the rise of consensus and polarization in opinion landscapes, and universal patterns across various social systems, such as in the distribution of votes in political elections. 
These collective phenomena are compatible with a statistical physics perspective of social behavior: the attempt to understand macroscopic regularities in terms of general principles and tracing them back to the interactions among many individuals, modeled as relatively simple but intertwined entities. 

Among the many aspects of interest in social systems, the dynamics of opinions have captured the long-term interest of physicists. 
Since pioneering works framing social interactions {via statistical physics} \cite{weidlich_statistical_1971,holley_ergodic_1975,galam_sociophysics_1982}, the study of opinion dynamics has progressed substantially. 
This has enabled a more formal and quantitative modeling of concepts originating from the social and political sciences, such as opinion depolarization or issue alignment, enriching the field with interdisciplinary insights.
After the pioneering statistical physics review of social dynamics by \textcite{castellano_statistical_2009}, 
the subfield of opinion dynamics has become significantly more active, 
spawning multiple comprehensive reviews:
\textcite{xia_opinion_2011} have further highlighted that opinion dynamics stems from combining social processes with the analytical tools of mathematics and physics, while \textcite{sen_sociophysics_2014} have stressed the challenges 
of applying statistical physics models to social dynamics.
Existing reviews focus on different formal representations of opinions \cite{jedrzejewski_statistical_2019,olsson_analogies_2024},
the expression of public vs private opinions \cite{kaminska_impact_2025}, and applications of opinion dynamics within finance and business \cite{zha_opinion_2021}.
Several works have reviewed specific models of opinion dynamics, such as consensus-reaching processes \cite{hassani_classical_2022}, 
voter dynamics \cite{redner_reality-inspired_2019,zschaler_adaptive-network_2012}, the Sznajd model \cite{sznajd-weron_review_2021}, kinetic exchange models \cite{biswas_social_2023}, majority vote models  \cite{vieira_phase_2016}, $q$-voter models \cite{jedrzejewski_statistical_2019}, and coevolutionary or adaptive opinion dynamics \cite{berner_adaptive_2023}.
This review provides an updated synthesis of opinion dynamics, structured around three objectives to help the reader navigate the modern literature.

First, over the last two decades, the field has advanced considerably, and several new perspectives, concepts, and methods have been introduced.
Crucially, the increased availability of extensive and detailed datasets, collected across numerous social settings, has driven a transformation in computational social science \cite{lazer_computational_2020}. Even when providing indirect or biased information on opinions, such data has enabled researchers to pursue a better empirical validation of theoretical models that describe social phenomena. To emphasize the role of data, this review dedicates \sref{sec:empirical_data} to studies specifically aimed at quantifying opinion dynamics phenomena in empirical data. 
This close integration of empirical works provides robust avenues for the validation and refinement of theoretical models, while expanding researchers' awareness of and access to relevant empirical datasets.

Second, opinion dynamics has consistently been an interdisciplinary field, with relevant literature scattered across physics, mathematics, computer science, and social science journals. 
Such interdisciplinary work has become even more pronounced in recent years.
This dispersion presents a challenge for researchers across disciplines, as it makes it harder to keep abreast of the growing body of relevant literature. 
This review %
organizes material and references, connecting different research questions, main findings, and open problems. 
Our goal is twofold.
On one hand, we %
provide statistical physicists with a comprehensive resource on opinion dynamics, introducing them to relevant themes from other fields and highlighting connections to broader societal and political issues.
On the other hand, we hope our review will act as a reference for researchers beyond physics, enabling them to better understand and relate to concepts and insights derived from statistical physics. 
To achieve both objectives, we systematize terminology across fields, mainly from a modeling perspective. 
While statistical physics guides our navigation of the literature, we also cover works well outside physics.
In doing so, we highlight how concepts like stochastic dynamics, phase transitions, scaling, and metastability are relevant to social phenomena, drawing analogies with traditional physics topics which may be evident for the physicist, but not to other scientists.

Third, and centrally, we employ a novel organizational structure that moves beyond a taxonomy of model classes (e.g., binary opinion models, voter models, majority-rule models, and so). 
Instead, we systematically categorize models based on: (i) the macroscopic phenomena they describe (e.g., consensus, fragmentation, polarization); and (ii) the microscopic mechanisms of opinion change they incorporate (e.g., assimilation, distancing, directional updating, similarity bias, noise, adaptivity). 
This division is not always clear-cut. %
Some overlap is inevitable, and the same model class may feature in several sections of the review.
Nonetheless, this structure provides a clearer entry point for the interdisciplinary scientist. 
It highlights fundamental connections between social psychology concepts and macroscopic observations in social and political science, fostering a deeper understanding of the emergent phenomena in contemporary society.

The review is organized as follows. \sref{sec:building_blocks_od} establishes essential concepts, defining what we mean by an ``opinion'', detailing microscopic mechanisms of change, and describing macroscopic phenomena such as consensus and polarization. 
\sref{sec:empirical_data} provides an overview of empirical research, including controlled experiments and large-scale data analysis. 
In \sref{sec:models} we present individual-based opinion dynamics models, categorized by the macroscopic phenomena they describe: consensus, fragmentation, polarization, echo chambers, and coevolution of networks and opinions. 
\sref{sec:methods} reviews common analytical and computational tools used to study the models presented earlier, covering stochastic processes, exact and approximate analytical treatments, simulation techniques, and optimization approaches. 
Finally, \sref{sec:outlook} is devoted to new perspectives on the field, including the development of artificial intelligence agents to explore novel social phenomena.

Many fields are related to opinion dynamics but are not covered here, 
such as 
cultural dynamics, understood as the distribution of shared cultural attributes like values, norms, and practices \cite{axelrod_dissemination_1997}; language dynamics, encompassing the formation, evolution, and competition of languages \cite{abrams_modelling_2003};  and the broader domain of ``human dynamics'' \cite{barabasi_origin_2005}.
Further on, %
seminal models on the emergence of segregation \cite{schelling_models_1969}, collective action \cite{granovetter_threshold_1978}, and collective emotions \cite{schweitzer_agent-based_2010} closely relate to polarization and opinion spreading, but are not treated in detail here. 

Finally, while often studied distinctly, social contagion processes fundamentally belong to the field of opinion dynamics.
These processes, encompassing the diffusion of information, memes, innovations, and even the emergence of political riots, closely resembling opinion formation as they are mediated by social influence and driven by peer pressure within a network \cite{centola_complex_2007}. 
In its simplest form, social contagion can be interpreted as a binary opinion process where a specific state spreads among individuals through sustained or cumulative social interactions \cite{watts_simple_2002}. 
Even if it often focuses on irreversible states or sudden behavioral cascades, the underlying microscopic mechanisms, associated to social pressure, align social contagion with the core principles of statistical physics applied to opinion dynamics. 
Despite these close associations, the study of social contagion forms a well-identifiable, distinct domain and is not discussed in this review due to space constraints.

\section{Preliminaries: Concepts and tools}
\label{sec:building_blocks_od}

We first review essential concepts and definitions. 
In \sref{subsec:opinion-def} we define opinions by establishing a consistent theoretical framework used throughout the text. 
In \sref{subsec:micro_mech} we describe the microscopic mechanisms that drive opinion change, from agents assimilating new views to their preference for homophily.
Shifting to a broader picture, in \sref{subsec:macro-phen} we describe the macroscopic phenomena that emerge in societies, such as the formation of consensus or the emergence of polarization.

\subsection{Basic concepts} 
\label{subsec:opinion-def}

\subsubsection{Measures of opinions}
\label{subsubsec:opinion-mesures}

\noindent\emph{``Any clod can have the facts; having opinions is an art.''}\\Charles McCabe (American columnist, 1915-1983)
\medskip

An opinion is typically understood as a subjective view shaped by a person's or group's perception and judgment. This stands in contrast to an objectively ``true'' statement with factual content that remains independent of any individual.
Opinions are related to concepts of interest in many disciplines, including \emph{attitudes} in psychology \cite{eagly_psychology_1993}, \emph{ideologies} in political science \cite{jost_political_2009}, and \emph{preferences} in economics or social choice theory \cite{druckman_preference_2000}. 
Also, \emph{social norms} \cite{ullmann-margalit_emergence_2015} and \emph{individual values} \cite{schwartz_universals_1992} are often conceptualized and formalized similarly to opinions. 
All of these notions come with their own additional concepts and underlying disciplinary theories. Colloquially, opinions are also understood as pieces of text in newspapers, magazines, or online blogs expressing the views of the author. Such complex, extensive opinions have largely remained unaddressed by traditional opinion dynamics models. 
However, with the increasing integration of large language models (LLMs) as agents, future models may begin to process and interact with opinions in this textual format (for further discussion see \sref{sec:AI_op_dyn}).

\paragraph{Notation.}

As opinions are inherently subjective, they are also likely to change when individuals encounter others' opinions or new information, either from peers or their environment. 
\emph{Opinion dynamics} is how this process unfolds in groups of many individuals over time. 
Throughout this review, we indicate by $x_i(t)$ the opinion of individual $i$ at time $t$, unless otherwise noted, and study how it evolves for multiple individuals in continuous or discrete time.
Given a group of $N$ individuals $i = 1, \dots, N$, 
we indicate the opinions of all individuals at time $t$ by the vector $\boldsymbol{x}(t)$. 
The \emph{opinion space} $X$ is the set of all possible opinion values, such that $x_i(t) \in X$ at all times and $\boldsymbol{x}(t) \in X^N$.
Depending on the modeling purpose or data availability, opinion measures and formats may range from a binary ``Yes/No'' value to a multidimensional tuple, where each element in the tuple is either a number or a textual statement representing one of potentially many topics or issues regarding the opinion of an individual. 
In most models treated here, different topics are represented as dimensions in a vector space. %

\paragraph{Multiple opinion formats.}

Let us focus on the simplest case of a binary opinion space $X = \{\text{Yes}, \text{No}\}$, applicable, for instance, to a referendum, a political election with only two candidates, or the decision to buy a product or adopt a particular behavior. 
Often, binary opinions are represented as spins, $X = \{+1,-1\}$, or as bits, $X = \{0, 1\}$. 
The reasons for different representations are mathematical convenience or analogies to models in other domains, as discussed by \textcite{olsson_analogies_2024}. 
The two opinions can be symmetrical, meaning that there is no inherent bias toward either one, as in the Ising model without an external field \cite{brush_history_1967}. 
Alternatively, one opinion may have a special status, such as being the ``status quo,'' which introduces a bias into the system. 
Although binary opinions may seem like an oversimplification introduced by physicists, they are used in many psychological models of social response \cite{nail_proposal_2013} and often capture the essence of the opinions of individuals on important topics \cite{lewenstein_statistical_1992}.

Beyond binary opinion spaces, various extensions to multistate discrete opinions exist, via, e.g., multiple labels, ordered labels, or quasi-continuous intervals of integers, where a neutral opinion may or may not exist at the center or on the borders of the opinion space.
This resembles different levels of measurement as used to classify variables in the social sciences, such as nominal, ordered, interval, or ratio scales \cite{stevens_theory_1946}. 
What labels or numbers are used is often a mix of mathematical and conceptual arguments, plus convenience for quantifying opinions in empirical surveys. 
In the following, we list notable examples of opinion measures found in the literature (see \tref{tab:opinion_formats}): 

\begin{itemize}
    \item $X = \{\text{Yes, } \text{No}\}$ is the outcome of a simple referendum-like decision or political elections with just two candidates as typical under plurality voting systems, most prominently the US (here, candidate names replace Yes and No). In the form of $X = \{+1, -1\}$, it is the most common opinion space used for the voter,  $q$-voter, Sznajd, majority-vote, and social impact models in the physics literature as discussed in \sref{sec:models}.
    
    \item $X =$ \{Yes, Abstain, No\} or  $X =$ \{Yes, Neutral, No\} are outcomes of a referendum-like decision which takes abstention or neutrality into account. Depending on the model's purpose and the required level of realism, neutrality and abstention might not be exactly the same thing. 
    Models using this opinion state are multi-state models, and are often  used to describe fragmentation (\sref{subsec:opinion_fragmentation}) and polarization (\sref{subsec:polarization}).
    \item The \emph{five-point Likert scale} \cite{likert_technique_1932}, 
    $X =$ \{strongly disagree, disagree, neutral, agree, strongly agree\}, is an example of a typical scale used in surveys. 
    Likert scales are centered around a semantically meaningful ``neutral'' midpoint, i.e., they are represented by $\{-2, -1, 0, +1, +2\}$. 
    Likert scales without a midpoint are therefore called \textit{forced-choice} scales \cite{allen_likert_2007}. 
    Modern surveys often use Likert-type scales with more response categories (e.g., 7 or 11 points, see below).
    
    \item The \emph{11-point scale} $X = \{0, 1, \dots, 9, 10\}$ is a common scale for attitudes used by the \textcite{european_social_survey_european_research_infrastructure_ess_eric_ess11_2024}, where 0 and 10 are labeled with a statement. For example, a survey question might be: ``Thinking about the European Union, some say European unification should go further. Others say it has already gone too far. Using this card, what number on the scale best describes your position?''. The card then shows 0 with the label ``Unification already gone too far'', and 10 with the label ``Unification go further''. 
    Conceptually, this is similar to the Likert scale, but here numbers are not only used to code but also to communicate the scale. 
    The scale includes a midpoint, numerically identified as five. 
    A common issue with such numerical scales is the inherent bias created by associating one pole with a negative or lower number (e.g., zero), a situation ideally to be avoided. 
    In happiness measurements, the \emph{Cantril's ladder} uses a similar approach; respondents rate their life on a scale from 0 (worst possible) to 10 (best possible) \cite{cantril_pattern_1966}.

    \item $\bigstar$-ratings for products or cultural items is another similar discrete rating scale, where the opinion -- typically about the quality of some item -- is represented by a certain number of stars. 
    In the Internet Movie Database (IMDb), movies can be rated with $1\bigstar$ to $10\bigstar$, a forced-choice scale without a midpoint. However, five may be considered a midpoint by many respondents. This phenomenon is clearly visible in the comparison of political left-right self-placement scales 0-10 and 1-10 by \textcite{gestefeld_decomposing_2022}. 
    For quality scales, like star ratings, neutral midpoints are not so relevant \cite{lorenz_universality_2009}. 
    
    \item In theoretical models of continuous opinion dynamics, opinion spaces are usually the intervals $[0,1]$ or $[-1,1]$ (see, e.g., \sref{subsec:opinion_fragmentation} and \sref{subsec:coevolution}). Similarly, some models use the whole real line and assume initial opinions to be normally distributed. Continuous opinions can also be measured empirically in questionnaires using a visual analogue scale \cite{reips_interval-level_2008}.
    
    \item Multistate discrete opinion spaces are relevant in complex referendums with three or more proposals on the same legislative topic, and elections of more than two political parties or candidates as typical for proportional voting systems.
    A set of a certain number of options is also the classical setup of social choice theory \cite{arrow_social_1963, austen-smith_social_1998}. 
    Multistate opinions appear in the multistate voter model (\sref{subsec:opinion_fragmentation}) or in models where agents can create new labels \cite{baronchelli_sharp_2006, bornholdt_emergence_2011,choi_analysis_2025}.  
\end{itemize}

\begin{table*}[t]
\centering
\caption{Summary of opinion formats found in the literature.}  \label{tab:opinion_formats}
\begin{tabular}{cp{115mm}}
        \hline
        \textbf{Opinion space $X$} & \textbf{Description and examples} \\ 
        \hline\hline
        $\{\text{Yes}, \text{No}\}$ or $\{-1, +1\}$ & Binary choice or spins/bits. Referendums and 2-party elections. \\
        $\{\text{Yes}, \text{Abstain}, \text{No}\}$ or $\{-1, 0, +1\}$ & Binary decision with neutrality. Referendums with abstention. \\ 
        $\{-2, -1, 0, +1, +2\}$ & Likert scale or five-point scale. Social surveys. \\ 
        $\{0, 1, \dots, 9, 10\}$ & 11-point scale with midpoint at 5. Social surveys. \\ 
        $\{1\bigstar, \dots, \bigstar10\}$ & Discrete quality ratings, often without a midpoint. IMDb. \\ 
        $[0,1]$, $[-1,1]$, $[-M,+M]$, $\mathbb{R}$  & Continuous scale (bounded by parameter $M$). Models with continuous opinions. \\ 
        $\{A, B, C, \dots\}$ & Multistate opinions. Party preference or referendum with many options. \\ 
        Multidimensional & Tuples/vectors of any of the options above. Complex cultural preferences. \\ 
        \hline
        \end{tabular}
\end{table*}

Many of these opinion measures include a semantically meaningful midpoint. 
Indeed, any bounded numerical scale has a mathematical middle that is often considered by models. 
In some of them, it is important that the scale is centered around zero when the microscopic dynamics of opinion change takes neutrality into account. 
For example, in models of opinion reinforcement, agents tend to move away from zero in the direction previously taken by the agent (see \sref{subsubsec:directional}).

Multistates should not be confused with \emph{multidimensionality}. Multidimensional opinions encompass several opinion spaces $X_1, \dots, X_m$ for $m$ different topics (see \sref{sec:multidim_pol}).
An opinion can be represented by a tuple $x \in X_1 \times X_2 \times \dots \times X_m$. Many theoretical models of multidimensional opinion dynamics work with $m$ identical opinion subspaces: $\{0,1\}^m$ where opinions are bits \cite[an early example is in][]{deffuant_mixing_2000}; $\mathbb{R}^m$ for unbounded continuous opinions; or $[0,1]^m$ where opinions are bounded between zero and one in every dimension \cite[an early example is in][]{lorenz_mehrdimensionale_2003}.

\paragraph{Different representations.}

The opinion of an individual may also be represented as a probability density function over one of the above-mentioned non-probabilistic opinion spaces \cite{sobkowicz_opinion_2018, steiglechner_social_2023}. This is consistent with the Bayesian idea of modeling \emph{beliefs} as {subjective probabilities} \cite{martins_bayesian_2012}, which largely overlaps with definitions of opinions when they do not lack a connection to facts or evidence \cite{olsson_analogies_2024}. 
In decision-theoretic settings, it is often assumed that agents have to assess the truth of a statement (i.e. a binary opinion $x\in\{0,1\}$), about which they are not completely sure. 
Thus, they communicate the probability of how likely they consider the statement to be true. Likewise, their belief about an underlying continuous opinions in $[0,1]$ is a subjective probability distribution with support $[0,1]$. 

In models, individuals often change their opinions in reaction to the opinions of others. 
It is then relevant to formalize the \emph{opinion discrepancy} between two agents' views, $x_i$ and $x_j$. 
In one-dimensional continuous opinion spaces, the simplest measure is the distance $|x_i - x_j|$, but non-linear discrepancies can also be found in mathematical psychology \cite{hunter_mathematical_1984}. While the Euclidean distance is a mathematically tractable approach to defining discrepancy for multidimensional continuous opinions, other metrics are used as well \cite{ye_evaluating_2011}.
The choice of the most suitable measure usually requires a thorough assessment of how individuals are hypothesized to perceive and relate to the different dimensions of opinion.  

Individuals may distinguish between their \emph{private} and \emph{public} opinions, i.e., what they keep to themselves as opposed to what they share with others (see the review by \textcite{kaminska_impact_2025} on expressed-private opinions models, and the work by \textcite{iniguez_effects_2014} on the role of antisocial public opinion on network cohesion, or \cite{banisch_opinion_2019} for a model with private evaluations of opinions). 
Expressed public opinions might be more extreme than private views, e.g., they can be an exaggerated signal as part of a negotiation, or less extreme due to conformity pressure or a taboo. 
Private opinions are also known as \emph{concealed} views \cite{gastner_consensus_2018, garcia-millan_concealed_2020}. 
According to \textcite{noelle-neumann_spiral_1974}, concealed opinions may trigger a \emph{spiral of silence} where individuals suppress their private opinion in public \cite{juncosa_toxic_2024}. This can lead to unexpected results in elections with ballot secrecy. 
Different formal representations of opinions have been reviewed by \textcite{jedrzejewski_statistical_2019} (see Fig.~1 therein) and by \textcite{olsson_analogies_2024}.

\subsubsection{Social interactions}
\label{sec:soc_nets}

The dynamics of opinions is often mediated by social interactions. 
These interactions are the fundamental processes through which agents influence one another, exchange information, and update their views on an issue. 
Understanding the nature and structure of these interactions is paramount, as they mediate the microscopic mechanisms of opinion change and eventually determine the emergence of a macroscopic state. 
Consequently, models of opinion dynamics include assumptions on social interactions.

\paragraph{Social networks.}

Opinion dynamics is typically assumed to be mediated by an underlying social network, where agents are nodes and the edges between them represent social interactions that regulate information exchange \cite{newman_networks_2018}.
Networks are described by their \emph{adjacency matrix} $\mathbf{A} = \{ a_{ij}\}$. 
In the simplest case, the element $a_{ij}$ is equal to 1 if an edge exists from node $i$ to node $j$, and 0 otherwise. The edges $a_{ij}$ might be directed, weighted, multivariate, temporal, or stochastic, depending on the context. The specific topology of the underlying network impacts the emergent collective behavior. Researchers have explored various network structures, ranging from synthetic networks built from statistical models, such as Barab\'asi--Albert networks \cite{barabasi_scale-free_2009}, to networks directly reconstructed from empirical data, such as interactions on online social media \cite{peel_statistical_2022}.

The simplest structure is a complete graph, representing a homogeneous, all-to-all mixed population, where every agent can interact with everyone else, implying that influence propagates directly throughout the system. 
Complete graphs usually allow for full analytical tractability (sometimes referred to as a ``mean-field'' approximation, see \sref{sec:AMEs}), serving as a baseline for comparison with more realistic network topologies. 
Likewise, lattices (such as square lattices or one-dimensional chains) have also been used historically, due to the direct analogies and the application of methods from statistical physics models, such as the Ising model \cite{brush_history_1967}.

However, most opinion dynamics models unfold on {complex networks}, aiming to capture the heterogeneous nature of real-world social connections via diverse topologies. 
Erd\H{o}s-R\'enyi graphs, for instance, model networks where connections are formed randomly, leading to a homogeneous degree distribution \cite{newman_evolution_2011}. 
Scale-free networks, characterized by a few highly connected hubs and many sparsely connected nodes, provide a more realistic representation of empirical social networks \cite{barabasi_emergence_1999}. These topologies can be complemented by modular or community structure \cite{fortunato_community_2016}, further increasing the realism of the representation.

More recently, research has expanded to include richer representations of social interactions. 
For instance, temporal networks acknowledge that social ties are not static but change over time, with interactions forming and dissolving \cite{holme_modern_2015, braha_centrality_2006}. 
Multi-layer networks (or multiplex networks, in case the sets of nodes are identical across layers) capture situations where individuals are connected via different types of relationships (e.g., family, work, online acquaintances) simultaneously \cite{kivela_multilayer_2014}.
Finally, higher-order networks extend beyond pairwise interactions to represent group-based influence, where the collective state of a small group of agents might influence another, reflecting complex social dynamics \cite{battiston_networks_2020}.

\paragraph{Communication regimes.}

Social interactions may mediate opinion dynamics through various communication regimes. 
These specify how and when agents spread opinions to others in the network, and how they perceive the information they receive. 
The communication regime operates on a social network but also includes specifications of when agents act in sending, receiving, and processing opinions.
The basic version is the \textit{passive communication} paradigm, where one (passive) individual receives opinion(s) of others and processes them without necessarily sending their own opinions.
The dynamics in the classical voter model belongs to this case (see \sref{sec:voter_model}). A more complex type of communication regime is given by pairwise interactions, where both individuals discuss and adapt (like in the models of \sref{subsubsec:bounded_confidence}). 
In this case, we need to specify the order and selection of pairs communicating at a specific time, which can make a difference [see \cite{lorenz_about_2007} and \sref{sec:updating} for a detailed discussion]. 
Regular meetings of more than two individuals might be relevant \cite{urbig_opinion_2008}, or situations where each individual processes the opinions of all others \cite{hegselmann_opinion_2002}. 
Many more complex one-to-many or many-to-one interactions are also relevant \cite{keijzer_communication_2018}. For example, many current online social networks are not undirected social graphs (like Facebook's early friends network), but directed follower-following networks.

\subsection{Microscopic mechanisms of opinion change}
\label{subsec:micro_mech}

\begin{figure}
    \centering    \includegraphics[width=1\linewidth]{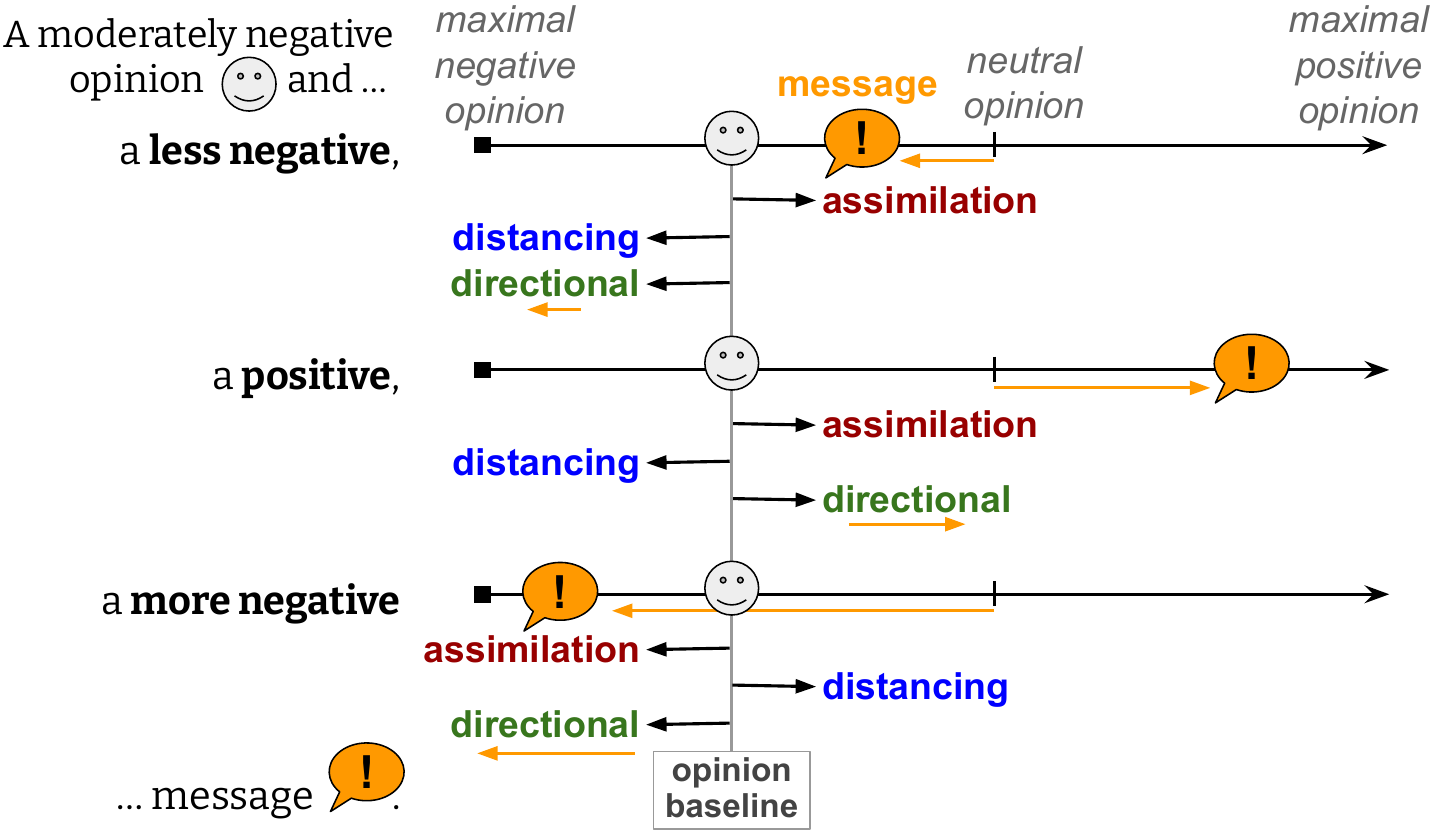}
    \caption{Examples of opinion change under a variety of micro-mechanisms and received messages. 
    Let us assume the focal individual has a negative opinion.
    If a message less negative than the opinion is received, distancing and directional update can trigger similar opinion changes, although the micro-mechanism builds on opposite valuations of the message. 
    If a positive message is received, assimilation and directional update can be identical as well. 
    If instead a more negative message is received, we observe the inverse behavior of the positive message, showing that distancing can also lead to more moderate opinions, while assimilation and directional update lead to extremization.}
    \label{fig:micromechanism}
\end{figure}

We catalog the various microscopic mechanisms of opinion change based on a \emph{passive communication} setting \cite{hunter_mathematical_1984}, where a focal individual receives a message and reacts to it, akin to experimental settings in psychology. 
The micro-mechanism underlying the interaction is characterized by the way the opinion of the focal individual changes after receiving the message. 
\fref{fig:micromechanism} depicts the differences between three common micro-mechanisms of opinion change (assimilation, distancing, and directional updating) in a stylized example of a one-dimensional continuous opinion space. 
The example emphasizes that the observed opinion change of an individual using different micro-mechanisms can appear indistinguishable in some situations.

Under both \emph{assimilation} and \emph{distancing}, the focal individual’s opinion update depends solely on the relative position between their current opinion and the incoming message: assimilation shifts the opinion \emph{toward} the message, whereas distancing shifts it away.  
In \emph{directional updating}, by contrast, the change depends on the opinion positions relative to the neutral point. 
The opinion change is in the direction of the message. 
A negative message makes the opinion more negative, a positive message makes it more positive, and a neutral message does not change it. 

The passive communication framework can be used to formalize the opinion dynamics unfolding between individuals, for instance, driven by pairwise interactions between individuals $i$ and $j$. 
Here, the opinion of agent $i$, $x_i$, is influenced by a message sent by agent $j$ with opinion $x_j$, and vice versa. For example, the messages may correspond to arguments derived from opinions $x_i$ and $x_j$, and therefore the process can be a simple model for a discussion or negotiation. 
In the following, we explain how these mechanisms are motivated and implemented, and then follow up with other important mechanisms of opinion change. 

\subsubsection{Assimilation}

Opinion assimilation assumes that individuals continuously influence each other, leading to a decrease in their opinion difference \cite{hovland_assimilation_1957}. 
In models of continuous opinion dynamics, this means that the opinion moves towards the message to decrease the discrepancy between them. 
These models often also consider that individuals process more than one message. 
Then, assimilation is typically implemented either through opinion averaging \cite{french_formal_1956,harary_criterion_1959,lehrer_rational_1981,degroot_reaching_1974} or a diffusive coupling scheme between agents \cite{abelson_mathematical_1967,taylor_towards_1968}. 
Forming a new opinion as the average opinion of others can be derived from the psychological theory of information integration of attitudes and social judgments \cite{anderson_integration_1971}. 
Other psychological reasons for assimilation can be normative pressure for conformity \cite{asch_opinions_1955}, or the reduction of cognitive dissonance regarding the opinions of others \cite{festinger_theory_1957}.

In binary or multistate opinion models, assimilation manifests as a tendency to adopt an opinion of a peer, such as in the classical (\sref{sec:voter_model}) and multistate (\sref{sec:MSVM}) voter models. 
When individuals process more than one binary opinion, assimilation can take the form of adopting the majority's opinion (\sref{sec:majority_rule}). In multidimensional models, opinions might align in one dimension \cite{axelrod_dissemination_1997}, or get closer to each other in a multidimensional vector space \cite{peralta_multidimensional_2024}.
Indeed, when the opinion space is a discrete ordered scale, assimilation can be the adoption of an opinion closer to the message, instead of the message itself. It is worth recalling that, as noted in \sref{subsubsec:opinion-mesures}, binary opinions are also used in models of social response. In social psychology, the tendency to adopt the opinion of a peer is referred to as conformity \cite{nail_proposal_2013}.

\subsubsection{Distancing}

Some models relax and even contradict the assumption that social influence always fosters opinion assimilation, introducing a mechanism of \emph{opinion distancing}.  
Within these frameworks, an interaction can be \emph{repulsive}: instead of converging, the opinions of the two agents diverge, pushing each other further apart in opinion space. 
This form of repulsive influence is also called contrasting or \emph{boomerang effect} \cite{hovland_assimilation_1957}, which describes cases in which the attempt to attract someone through social influence can have the opposite effect \cite{flache_models_2017}.
Typically, opinion distancing is implemented in models of continuous opinion dynamics, where the distancing mechanism is only activated when the opinion difference of agents already exceeds a given threshold \cite{jager_uniformity_2005,salzarulo_continuous_2006}. 

The mechanism of distancing has also been introduced in binary and multistate models. \textcite{macy_polarization_2003} conceptualize distancing as a tendency to change opinion when others have the same opinion, but the links to them are evaluated negatively. Analogously, \textcite{galam_contrarian_2004} characterizes agents with this behavior as \emph{contrarians} who adopt the opposite of the message, or the minority opinion of their peers. Another terminology for the same behavior is \emph{anticonformity}, as used by \textcite{nyczka_anticonformity_2013,jedrzejewski_statistical_2019}. This term is in line with psychological models of social response \cite{nail_proposal_2013}. \sref{sec:conformity_and_anticonformity} discusses the role of anticonformity for opinion fragmentation, while \sref{subsec:opinion_repulsion} presents distancing and repulsion as mechanisms driving polarization.

\subsubsection{Directional updating} \label{subsubsec:directional}

In one-dimensional opinion spaces with a notion of neutral opinion, directional updating refers to mechanisms where a message consistently induces opinion change towards the message, irrespective of the individual's current opinion. 
Directional updating appears in various forms and implementations, such as the explicit exchange of discrete arguments \cite{mas_differentiation_2013}, reinforcement \cite{banisch_opinion_2019}, Bayesian-like learning \cite{martins_continuous_2008}, and phenomenological differential equations \cite{baumann_modeling_2020, shin_tipping_2010}. 
In most models, directional updating is combined with similarity bias (see below), often leading to reinforcement and extremization of opinions at the individual level, and polarization at the collective level \cite{lorenz_individual_2021}.

Models based on directional updating can reproduce the phenomenon of ``group polarization'' studied in psychology: Members of a deliberating group all move towards a more extreme opinion in the direction of some of the members' pre-deliberation tendencies \cite{sunstein_law_2002, schkade_when_2010}.
\textcite{myers_group_1976} highlight other, more complex concepts where the term polarization refers to a split within a group of people.
Directional updating can also be associated with persuasive arguments theory, which explains changes in opinion through the exchange of arguments between individuals \cite{isenberg_group_1986}. Interestingly, directional updating is the only of the three basic mechanisms in \fref{fig:micromechanism} that leads to self-reinforcement: 
When an individual perceives their own opinion as a message, they might further move in the direction of their opinion, a relevant mechanism for polarization. In contrast, neither assimilation nor distancing would trigger opinion change in this case.

\subsubsection{Similarity bias}

The influence of a message on an individual's opinion may depend on the degree of similarity between them.
A similarity bias can be implemented as an opinion threshold, defining a range within which individuals can influence one another \cite{flache_models_2017}.
This bias serves as a fundamental prerequisite for social influence, governing whether interactions lead to convergence or other forms of opinion change. 
A prominent example is the class of bounded confidence models, where agents' opinions interact with a fixed strength, but only with messages whose discrepancy from their own opinion falls below a defined confidence bound (\sref{subsubsec:bounded_confidence}).
In other models, the discrepancy diminishes opinion influences smoothly \cite{abelson_mathematical_1964, mas_individualization_2010,kurahashi-nakamura_robust_2016, lorenz_individual_2021} or even changes the direction of influence \cite{vazquez_role_2020}. 

Similarity bias is a distinct mechanism in opinion dynamics, psychologically motivated by concepts like motivated reasoning \cite{lorenz_individual_2021}, confirmation bias \cite{nickerson_confirmation_1998}, or biased assimilation \cite{lord_biased_1979}. 
Sociologically, it reflects homophily \cite{mcpherson_birds_2001}, where individuals prefer communicating with similar peers. 
In models, this often means agents with similar opinions are more likely to influence each other, with further refinements possible through cognitive perception biases \cite{sobkowicz_opinion_2018}. \textcite{feliciani_how_2017} have asked, among other things, whether similarity between agents increases the chance of interaction or the persuasiveness of others’ arguments, and have studied the consequences of both assumptions in a series of agent-based models.

\subsubsection{Confining, heterogeneity, noise} %

Some mechanisms are about confining expansive dynamics in an explicitly or implicitly bounded opinion space. 
A psychological mechanism reflecting this dynamics is the \emph{polarity effect} \cite{hunter_mathematical_1984}. 
This states that individuals are less prone to changing their opinion the more extreme their opinion is (regardless of a message or discrepancy with a message). 
The same effect has been implemented by~\textcite{deffuant_how_2002} via initially placing individuals with tiny confidence bounds at the extremes of a continuous opinion space. 
A related mechanism is to assume that opinion movements slow down as opinions move away from neutrality  \cite{shin_tipping_2010, baumann_modeling_2020}. 
This also serves a more technical purpose of confining expansive dynamics (triggered, e.g., through distancing or directional updating) within the given opinion space. 
In this way, an implicit mechanism in some models is absorbing boundaries \cite{vazquez_ultimate_2004}.

Another aspect taken into account is the heterogeneity of individuals, which can occur for all mechanisms. 
Most studied is heterogeneity in confidence bounds \cite{lorenz_heterogeneous_2010, schawe_when_2020} where individuals can be more open- or close-minded. 
An important special case of this is full \textit{stubbornness}, where some agents never change opinion  \cite{galam_stubbornness_2016,baumann_laplacian_2020, botte_clustering_2022}. 
Those cases have also been studied under the label of constant signals \cite{hegselmann_opinion_2015} and are interpreted as charismatic leaders or media signals. 
A similar type of agent is a \textit{zealot} that either never changes their opinion or always returns to a specific opinion \cite{mobilia_does_2003,mobilia_role_2007,galam_role_2007}. Models implementing zealotry are reviewed in \sref{sec:zealotry}. We also point to the Friedkin-Johnsen model as an example in which an agent's initial opinion keeps influencing their later opinions \cite{friedkin_social_1990, friedkin_choice_1999}, as discussed further in \sref{sec:zealotry}.

It is natural to assume that not all opinion change is through social interaction, but can be through one's own reconsideration or other intrinsic processes. 
A typical way to incorporate such aspects, without modeling them explicitly, is through a form of noise.  
However, the term noise can mean many different things in opinion dynamic models. 
Noise is fundamental in many binary models of opinion dynamics, starting with the noisy voter model \cite{granovsky_noisy_1995}, by making adoption probabilistic. 
In continuous opinions, it can mean additive noise \cite{mas_individualization_2010, steiglechner_noise_2024}, where an individual adds a random number to their current opinion. 
Another often-used mechanism is to assume that an agent may sometimes choose a new opinion randomly \cite{pineda_noisy_2009, pineda_noisy_2013, kurahashi-nakamura_robust_2016, carro_role_2013,lorenz_individual_2021}, which has been described as free will or idiosyncratic opinion change.

\subsubsection{Network-based adaptivity}
\label{sec:net_adapt}

When exploring opinion dynamics over static networks, we usually make the assumption that links are persistent in time, or change on time scales longer than the dynamics of opinion. 
If connections are fixed, highly connected groups of individuals tend to have similar opinions, leading to polarized states where group structure and opinion segregation coincide (for a variety of models on the topic, see \sref{subsec:polarization}). 
On the other side of the spectrum of relevant time scales, we might consider a situation where social ties change, but opinions remain constant [becoming attributes or convictions that generate homophilous identity within groups \cite{mcpherson_birds_2001}]. 
Links are more likely to be rewired between people with similar opinions, due to homophilic or assortative mixing mechanisms \cite{soderberg_general_2002,newman_mixing_2003,boguna_models_2004,toivonen_comparative_2009,murase_structural_2019}. 
Examples include conviction-driven social segregation, based only on homophily \cite{teza_network_2019} or its dynamic interplay with triadic closure \cite{asikainen_cumulative_2020}, and the role of assortative and preferential attachment mechanisms in the emergence of core-peripheries \cite{urena-carrion_assortative_2023}.

In real-world opinion dynamics, arguably both node opinions and network structure evolve on comparable time scales and adapt to each other \cite{gross_adaptive_2008,zschaler_adaptive-network_2012,sayama_modeling_2013,peralta_opinion_2022,berner_adaptive_2023}. This network-based mechanism of adaptivity is considered in so-called coevolving, coevolutionary, or adaptive opinion dynamics, typically via a plasticity (or mutation) parameter measuring the relative probability or rate of link rewiring vs. opinion change \cite{castellano_statistical_2009}. 
The concept of coevolutionary networks implies an adaptive process in which the time-evolution of network edges and node opinion states are coupled to each other, in analogy to the notion of species coevolution in biology \cite{thompson_concepts_1989}. 
When social ties between agents with similar opinions are favored, adaptive dynamics self-organize into heterogeneous networks where groups of individuals sharing attributes are structurally distinguishable from each other, leading, e.g., to fragmentation--consensus transitions as a function of the rewiring rate \cite{holme_nonequilibrium_2006,iniguez_opinion_2009}. 
We describe models of adaptive or coevolving opinion dynamics in \sref{subsec:coevolution}.

\subsection{Macroscopic phenomena in opinion distributions}
\label{subsec:macro-phen}

The focus of opinion dynamics is to understand the mechanisms by which macroscopic phenomena emerge in the opinion distributions of the relevant population, most notably consensus, extremeness, fragmentation,  polarization, and opinion alignment. 
These phenomena cannot be assessed for individuals alone, but rather for the population as a whole. 
In the following, we outline the treatment of these phenomena in the social sciences and summarize methods for their formal conceptualization and measurement.

\subsubsection{Consensus} 

Consensus decision-making is a method of achieving a collective decision agreed on by all participating individuals through deliberation, collective judgment, negotiation, and compromise. 
Societal consensus refers to agreement in groups larger than those using collective decision-making, which may also emerge through indirect communication. 
Empirically, societal consensus does not mean strict consensus where all individuals have exactly the same opinion, just enough similarity according to a chosen metric of opinion distance. 

Formally, consensus denotes a state where all individuals share the same opinion, a condition straightforward to quantify in its strict form. 
However, assessing the degree of consensus in a population is often complex. 
For binary opinions, this is typically captured by the relative size of the group holding the majority opinion. 
When options for abstention or neutrality are introduced (e.g., opinion space $\{-1, 0, +1\}$), defining consensus becomes ambiguous. Is a distribution of $(0.85, 0, 0.15)$ across states $-1, 0,$ and $1$ respectively, more consensual than $(0.1, 0.8, 0.1)$? 
While the first shows an $85\%$ majority, the second, with $80\%$ neutrality, exhibits lower overall opinion discrepancy, demonstrating that the ``most consensual'' state lacks a unique definition without further criteria.

\begin{figure*}[tbp]
\centering
\includegraphics[width=1.8\columnwidth]{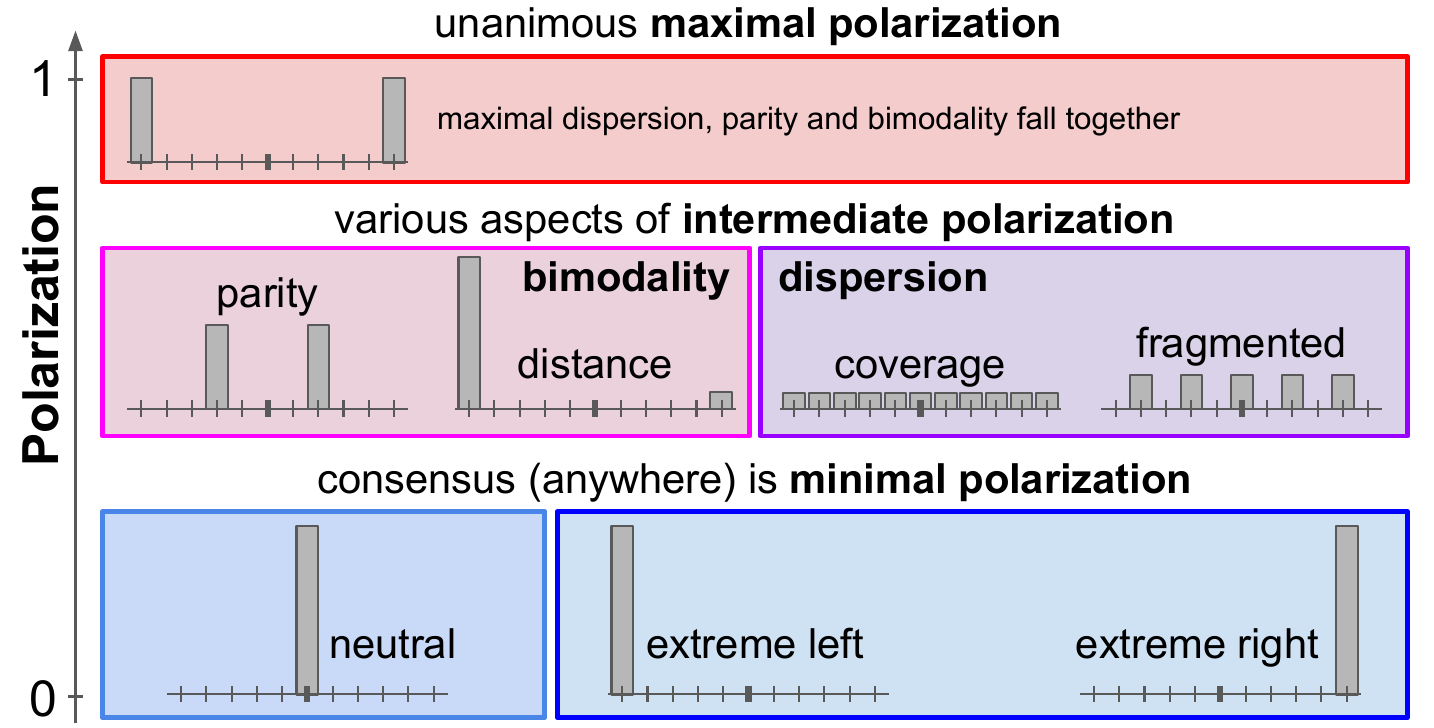}
\caption{Eight examples of opinion distributions on an eleven-point scale showing the conceptual measurement problem of polarization for a one-dimensional bounded opinion space:
Scholars agree on maximal polarization and minimal polarization (consensus). 
However, there are many different aspects of polarization between the maximal and the minimal. Figure adapted from \textcite{gestefeld_decomposing_2022} and extended.}
\label{fig:polarization1}
\end{figure*}

For the quantification of consensus, it is typically considered irrelevant if the consensus opinion is neutral, moderate, or extreme (see \fref{fig:polarization1}). 
Therefore, statistical measures of \emph{dispersion} such as the standard deviation can be used to measure the inverse of consensus: Minimal dispersion is the maximal degree of agreement. Dispersion measures can be used for continuous opinion scales as well as ordered categorical scales (when equal or other distances between categories are assumed). 
However, the standard deviation relies on the computation of the mean opinion, and on ordered categorical scales such as $\{0,1,\dots,10\}$, the mean opinion is usually not a valid opinion (because the mean may not be an integer). An \emph{agreement index} has been proposed by \textcite{van_der_eijk_measuring_2001} as an alternative that does not rely on the computation of a mean opinion. 
This measure was created to assess how much experts or lay persons agree on the ideological position of a party. 
Measures of consensuality can be taken from similar ideas from other fields. 

Consensus is also generally seen as the least polarized state, as outlined in \fref{fig:polarization1}. 
Before we treat the different facets of polarization, we consider two other phenomena to distinguish them from polarization: extremeness and fragmentation. 

\subsubsection{Extremeness} 
\label{sec:macro_extremeness}

Extremeness quantifies the degree to which opinions deviate from a neutral state. 
For a finite ordered opinion space such as \{0, 1, \dots, 10\} (with 5 as neutral state), extremeness can be measured by the fraction of extreme opinions. 
When an opinion discrepancy measure and a neutral opinion are specified, extremeness can also be measured by the mean distance from the neutral opinion. 
We note that extremeness is agnostic concerning the question of whether opinions are polarized into two or more camps of extremists, or if there is extreme consensus. 
\fref{fig:polarization1} shows some example opinion distributions, which can demonstrate that maximal extremeness can take different forms: The distributions ``extreme consensus'' as well as the distribution "maximal polarization" both have all opinions maximally far away from neutral and are thus all instances of maximal extremeness. 
However, they are fundamentally different in their degree of polarization, which we discuss later in \sref{sec:macro:polarization}. 

The increasing extremeness of a distribution, as observed in the group polarization phenomenon (\sref{subsubsec:directional}), is considered a form of polarization in psychological terminology. However, this definition diverges from the dominant understanding in modern political science literature, as outlined in \sref{sec:macro:polarization}.

The measurement of extremeness by the average distance to the neutral opinion can be extended to multidimensional opinions. 
\fref{fig:polarization2} shows examples of multidimensional opinion distributions, which show different types of fragmentation and polarization but all with a high level of or even maximal extremeness (assuming the opinion space is the bounded cube shown).

\subsubsection{Fragmentation} 
\label{sec:macro_fragmentation}

\begin{figure*}[tbp]
\centering
\includegraphics[width=1.8\columnwidth]{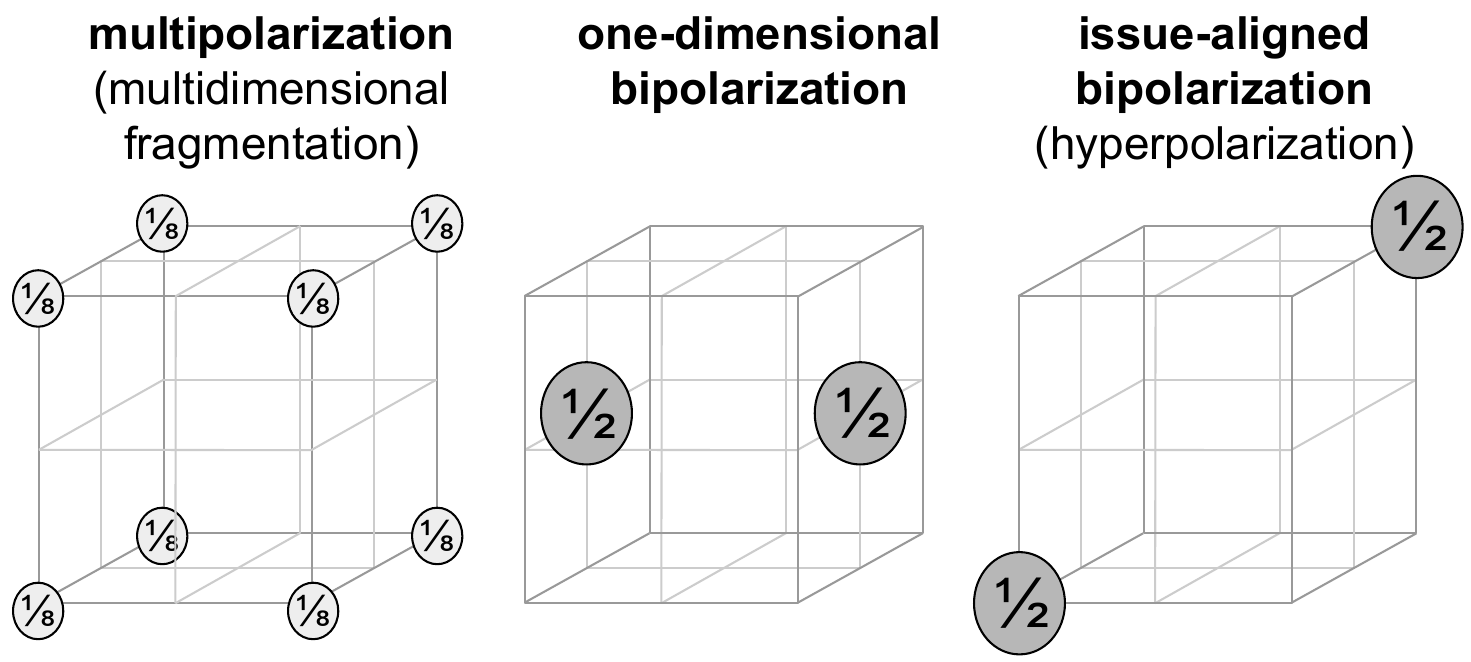}
\caption{Three examples of opinion distribution in a bounded three-dimensional opinion space. In all examples, all opinions are in some sense extreme and not consensual. So, they can all be considered polarized. However, there are three different versions of it: Fragmented (or multi-polarized), and two bipolarized versions, one essentially one-dimensional, the other issue-aligned (hyper-)polarized.}
\label{fig:polarization2}
\end{figure*}

A possible outcome of opinion dynamics is fragmentation, i.e., the separation of individuals into several groups (also called clusters, poles, modes, or fragments), where individuals hold similar opinions within groups, but dissimilar between groups \cite{bramson_disambiguation_2016}. 
When fragmentation occurs, there can be more than two clusters, such that the distance between all pairs of clusters does not have to be maximal. 
As we will describe in \sref{sec:models}, fragmentation can emerge in continuous models of opinion dynamics when several groups are formed along one dimension (as in the bounded confidence model under low bounds of confidence). 
Fragmentation can also occur in multidimensional models when clusters form in various combinations of values along the dimensions, for example, filling the corners of a hypercube (see \fref{fig:polarization2}) \cite{schweighofer_agent-based_2020}. 

Fragmentation has been understood as a form of ``polarization'', a concept which we will discuss in more detail below. Specifically, in the picture of \emph{community fragmentation} \cite{bramson_disambiguation_2016}, a society is considered more polarized the more it can be broken into several subgroups. 
The \emph{multi-polarized} opinion distribution in \fref{fig:polarization2} is therefore both fragmented and polarized. 
Some works consider polarized opinion distributions as those fragmented into two poles only, a state known as \emph{bipolarization} \cite{mas_differentiation_2013,velasquez-rojas_opinion_2018,saintier_model_2020}. 
Within this concept of polarization, more fragmentation indicates less polarization. 

Fragmentation is usually characterized by the number of groups in the opinion distribution, such as those arising in the Deffuant-Weisbuch model (\sref{subsubsec:bounded_confidence}). 
Even in this model, it can be non-trivial to assess what a cluster is, and how to deal with small groups of agents between major clusters \cite{ben-naim_bifurcation_2003, lorenz_repeated_2007}. 
Moreover, simply counting the number of different groups discards differences in the group sizes. 

For multistate discrete opinion spaces (with non-ordered labels), distinguishing groups is easier since every label defines a group. 
Simply counting the number of distinct groups in a population (i.e., the number of different opinions supported by at least one individual) can mask differences in group sizes. 
A way to measure fragmentation, borrowed from political science \cite{laakso_effective_1979}, adapts the concept of an effective number of parties to an effective number of opinions. 
If $p_1,\dots,p_n$ are the vote shares of $n$ political parties (with $p_i \geq 0$ and $\sum_{i=1}^n p_i = 1$), we may define the effective number of parties or opinions as $1/(\sum_{i=1}^n p_i^2)$. 
This measure is the inverse Simpson index, also known as participation ratio in physics, and the Herfindahl-Hirschman index in economics. 
It is used as a measure of concentration, e.g., market share, species in an ecosystem, or ethnic groups in a residential area. 
Adapted to opinion dynamics, the inverse participation ratio $\sum_{i=1}^n p_i^2$ represents consensuality, i.e., the probability that two randomly sampled individuals have the same opinion. 
The effective number of opinions $1/(\sum_{i=1}^n p_i^2)$ is equal to $n$ only when the shares across individuals are all equal ($p_i=1/n$ for all $i$), and goes to one when all agents hold one opinion ($p_{i_0}=1$ for one opinion $i_0$, and $p_i=0$ for all other $i$).

The effective number of opinions is closely related to Shannon's entropy, defined as $H = - \sum_{i=1}^n p_i \log p_i$, which quantifies the expected self-information.
Empirically, entropy is more sensitive to lower frequency values than the Simpson index.
Entropy has been used as a measure of global order in nonlinear multistate voter models \cite{ramirez_ordering_2024}, and as an order
parameter to characterize the transition between uni- and bimodal opinion distributions \cite{franco_shannon_2021}.

\subsubsection{Polarization} \label{sec:macro:polarization}

\paragraph{Definitions and aspects of polarization.}
Almost all definitions of opinion polarization fall under the abstract understanding that polarization is an \emph{accentuation of differences}. 
This broad idea relates to various concepts of polarization. Early works in political science \cite{dimaggio_have_1996} differentiate between four principles still used today: \emph{dispersion}, \emph{bimodality}, \emph{constraint}, and \emph{consolidation}. 
Dispersion and bimodality deal with one-dimensional opinion distributions (see \sref{sec:polarization_measures}): 
The more dispersed the opinions are, the more accentuated their differences become, while the more bimodal the distribution is, the more antagonistic the differences are between two groups (i.e., the bipolarization of \sref{sec:macro_fragmentation}).  

We discuss constraint and consolidation principles under \emph{opinion alignment} in \sref{sec:macro:alignment}.
The constraint principle involves polarization in multidimensional opinions, specifying that a society is more polarized when opinions regarding different topics are correlated. The consolidation principle deals with accentuated differences between defined groups (such as supporters of different political parties), stating that a society is more polarized when opinions become correlated with other characteristics of the individuals. 
In \sref{sec:polarization_measures}, we focus on conceptualizing and measuring polarization in a one-dimensional opinion distribution.

From a political science perspective, specifying the population under consideration is crucial when evaluating polarization. 
A typical distinction is drawn between mass polarization and elite polarization \cite{fiorina_political_2008}. 
Elite polarization refers to the ideological divergence and affective animosity observed among political actors, such as elected officials, party leaders, activists, and prominent political journalists. 
In contrast, mass polarization involves divisions within the broader electorate or general public. 
Scholars have extensively debated whether elite behavior primarily shapes public opinion or merely reflects it, with some studies suggesting that high elite polarization can exist even with limited mass polarization \cite{macy_polarization_2021}.

\textcite{lelkes_mass_2016} has distinguished four aspects of polarization in political science: ideological consistency, ideological divergence, perceived polarization, and affective polarization. 
The debate regarding increasing or stable mass polarization \cite{abramowitz_is_2008,fiorina_political_2008} largely centers on measurement, focusing on either consistency (increasing correlation across issues) or divergence (spread of opinion distributions). 
Perceived polarization refers to the public's perception of ideological distance between political parties, distinct from individuals' own opinions. 
Affective polarization, conversely, measures group-based sentiment, such as in-group liking and out-group disliking, relying on political or social identities rather than direct opinion content \cite{iyengar_opinion_2011,tornberg_how_2022}. 
These concepts, along with political sectarianism \cite{finkel_political_2020}, which bundles issue alignment, extremeness, affective polarization, and moralization, offer a nuanced understanding of political polarization.

\paragraph{Measures of polarization.} 
\label{sec:polarization_measures}

For a bounded numerical scale, most researchers agree that maximal polarization occurs when two groups of equal size have internal consensus at maximally discrepant opinions. 
Minimal polarization occurs when all individuals are in consensus on one opinion in the opinion space (which can be an extreme position, neutral, or any other opinion) \cite{gestefeld_decomposing_2022} (\fref{fig:polarization1}). 
For these extreme cases of minimal and maximal polarization, dispersion and bimodality are either both minimal or both maximal. Consequently, distinguishing between bimodality and dispersion becomes important for intermediate levels of polarization. For example, there is no agreed-upon standard measure to order the four intermediate opinion distributions in \fref{fig:polarization1} by their degree of polarization. 

The examples in the figure show opinion distributions on a discrete numerical scale. We can then represent the opinion distribution by the fractions of individuals $p_0,\dots,p_n$ (with $p_i \geq 0$ and $\sum_{i=0}^n p_i = 1$) holding the opinions $x_0 < x_1,\dots <x_n$. In this setting, the \emph{normalized standard deviation} is 
\begin{equation}
    \text{SD}(p,x) = \frac{2}{|x_n - x_0|} \sqrt{\sum_{i=0}^np_i(x_i-\bar{x})^2} 
\end{equation}
with $\bar x = \sum_{i=0}^np_i x_i$  the mean opinion. 
Versions for continuous opinions are omitted here, but are mostly straightforward to obtain (by replacing sums with integrals). The \emph{normalized mean absolute difference} is a special case of the polarization measure in \textcite{esteban_measurement_1994}:
\begin{equation}
    \text{Pol}_\alpha(p) =\frac{2^{1+\alpha}}{|x_n - x_0|} \sum_{i,j=0}^n p_i^{1+\alpha} p_j |x_i-x_j|,
    \label{eq:Pol}
\end{equation}
obtained by setting $\alpha=0$.
Here we focus on $\alpha=0$ and discuss general values of $\alpha$ further below.
Both of these measures ---normalized standard deviation and normalized mean absolute difference--- capture the definition of maximal and minimal polarization discussed above. A sensible threshold distinguishing between consensus and polarized opinion distributions can be based on the uniform distribution, which has a normalized standard deviation of $1/\sqrt{3} \approx 0.577$. Its normalized mean absolute difference is $2/3$ (in the limit of large $n$).   

The normalized mean absolute difference [obtained from Eq.~(\ref{eq:Pol}) for $\alpha=0$] is the (normalized) expected opinion distance between two randomly selected individuals from the opinion distribution.
It is more demanding to compute this than the normalized standard deviation, as the computational cost to calculate the mean absolute difference scales with $n^2$ instead of $n$. On the other hand, the normalized mean absolute difference has a conceptual advantage, as it is more versatile and can be adapted to other opinion spaces, such as, for example, text-based opinions. 
The absolute difference only relies on a numerical quantification of the distance [discrepancy \cite{hunter_mathematical_1984} or antagonism \cite{esteban_measurement_1994}] between two opinions, while it does not require the computation of an average opinion. 
For instance, one can define a distance between two strings, but it is not always meaningful to average strings. Normalized mean absolute difference is also very similar to the Gini index, which can be understood as the mean absolute difference normalized by the mean of the distribution (and not by the maximal discrepancy as here for polarization). 
 
The quantity in Eq.~\eqref{eq:Pol} (for general $\alpha$) was originally inspired by questions in economics (such as income polarization) \cite{esteban_measurement_1994}, and was later extended to continuous distributions \cite{duclos_polarization_2004}. Choosing $\alpha>0$ allows one to distinguish polarization from inequality, as measured by the Gini index. 
The idea of polarization ``beyond'' inequality is that for $\alpha>0$, antagonism is weighted more strongly when a higher share of individuals is involved, as captured by the term $p_i^{1+\alpha}$. 
This is built on the conception that polarization appears when there is high homogeneity within ``groups'', defined as individuals with the same opinion. Significantly sized groups contribute more than proportionally with respect to isolated groups. 
Versions of the measure for continuous opinion spaces require some more technical specification.

Generally, there remains conceptual ambiguity for the middle ground between maximal and minimal polarization as outlined in \fref{fig:polarization1}. 
Measurement can emphasize different aspects, and it does not look like a one-measure-fits-all number for polarization will emerge. \textcite{flache_small_2011} for example use the variance of the distribution of pairwise dissimilarities as a measure of polarization.
\textcite{martino_quantifying_2025} compare several ways of measuring bimodality. \textcite{koudenburg_new_2021} generalize the concept of antagonism ($|x_i-x_j|$) and pair appearance probability ($p_ip_j$) and compose a measure based on expert judgments about example distributions. \textcite{bramson_disambiguation_2016} alone distinguish nine different aspects. 
Here, we take a pragmatic approach and exclude extremeness and fragmentation as aspects of polarization and treat them separately in \sref{sec:macro_extremeness} and \sref{sec:macro_fragmentation}.
We focus on bimodality and dispersion, as these seem to be the most important quantities to assess intermediate levels of polarization. For interdisciplinary communication on polarization, it is important to be aware of its different and sometimes even contradictory verbal conceptions, and specify clearly what measures are used and what aspect of polarization they measure.

\subsubsection{Alignment}\label{sec:macro:alignment}

Up to this point, we have treated polarization solely based on the one-dimensional opinion distribution $P(x)$. This description is unable to describe further phenomena of polarization that can arise when opinions become aligned with other salient entities of the system. 
For instance, opinions on one issue can be correlated to opinions on another topic in multidimensional opinion spaces (issue alignment), \cite{baldassarri_partisans_2008,kozlowski_issue_2021}, or individuals' opinions can be aligned with the opinions of their neighbors in the social network, which gives rise to echo chambers \cite{sunstein_republiccom_2009,cinelli_echo_2021}.

\paragraph{Constraint principle.} The constraint principle \cite{dimaggio_have_1996} of polarization quantifies polarization as high when the opinion of an individual on one issue is highly predictive of their opinions on other dimensions. Maximal constraint occurs when the multi-dimensional opinions of individuals align along one direction in the multi-dimensional opinion space. Other terms for this aspect of polarization are \emph{issue alignment} \cite{baldassarri_partisans_2008, baumann_emergence_2021} and attitude consolidation \cite{dellaposta_pluralistic_2020}. 
\emph{Hyperpolarization} has been defined as the appearance of issue alignment together with extremeness of opinions \cite{schweighofer_raising_2024}, as shown in the right panel of \fref{fig:polarization2}.

\paragraph{Consolidation principle.} 

\begin{figure}
    \centering
    \includegraphics[width=\linewidth]{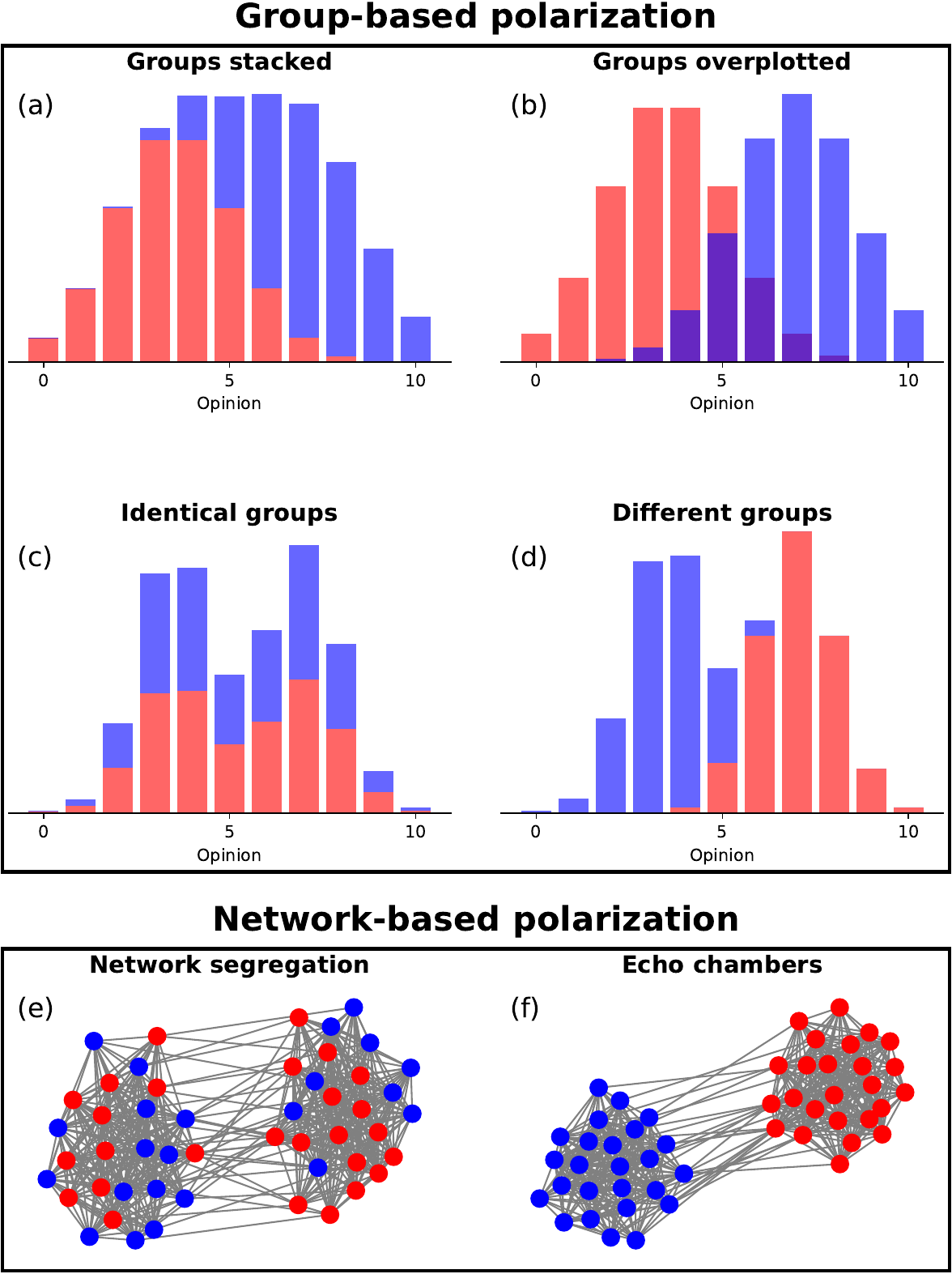}
    \caption{Differences in group-based polarization, network segregation and echo chambers. (a) and (b): Groups have exactly the same opinion distributions. (a): stacked, visualizing the total distribution is unimodal. (b): Overplotting, visualizing group-based polarization with little overlap (the purple area). (c) and (d): Same total opinion distribution with visible bimodality. (c): Driven by within-group polarization. (d): Driven by between-group polarization. (e) and (f): Same total opinion distribution (color coded). (e): Segregated network structure but no correlation between the opinion of an agent and their neighbors. (f): Echo chambers. In addition to a segregated network, there is a strong correlation between agents' opinion and the opinion of their neighbors.}
    \label{fig:polarization_groupbased}
\end{figure}

The consolidation principle asserts that polarization is high when large opinion differences exist between groups originally defined by characteristics other than the opinions themselves \cite{dimaggio_have_1996}. Measurements of such \textit{group-based polarization} therefore rely on the existence of exogenously given group labels, such as the affiliation to a political party or groups formed around identities.
This phenomenon has also been referred to as \emph{issue partisanship} \cite{baldassarri_partisans_2008} or \emph{between-group polarization}.
Most commonly, group-based polarization considers two opposing groups, such as voters in the US leaning towards the Democratic or Republican parties, having different opinions. 
The importance of the concept of group-based polarization is schematically depicted in \fref{fig:polarization_groupbased}, which shows some examples of two equally sized groups. In panels (a) and (b), the overall opinion distribution (sum of the two group distributions) is unimodal; however, group-based polarization emerges if group information is taken into account and the opinion distribution of each group is plotted separately (b). Likewise, a bimodal opinion distribution in the overall population [panels (c) and (d)] can either be driven by groups [panel (d)] or not [panel (c)].

\paragraph{Echo chambers.}

A further kind of opinion alignment arises in the form of echo chambers; 
environments where individuals are primarily exposed to information that aligns with their preexisting beliefs, due to, for instance, selective exposure 
\cite{sunstein_republiccom_2009} or algorithmic curation \cite{pariser_filter_2012}. 
Echo chambers have frequently been associated with both the emergence of polarization \cite{barbera_tweeting_2015} and the spread of misinformation \cite{stein_network_2023}. 
The reinforcement of similar views within segregated communities can amplify intergroup differences, fostering polarization, while the lack of exposure to dissenting perspectives may reduce a group's ability to distinguish between accurate and false information.

Formally, echo chambers can be characterized by an increased level of homophily (defined based on agents' opinions) in the interaction networks \cite{cinelli_echo_2021,hartmann_systematic_2025}. 
This can be measured by the correlation between the opinion of individual (or node) $i$, $x_i$, and the average opinion of their neighborhood $\langle x_i\rangle^{NN}= {1}/{{k}_i}\sum_j A_{ij} x_j$, where ${k}_i$ is the out-degree of node $i$. 
Here, the out-degree is used because a directed link from node $i$ to node $j$ indicates that $i$ is influenced by $j$.
Echo chambers are defined by a high correlation between $x_i$ and $\langle x_i\rangle^{NN}$.
It is important to distinguish echo chambers (panel C2 of \fref{fig:polarization_groupbased}) from pure network segregation, where there is a clear community structure in the network but no significant opinion correlation between neighboring individuals (panel C1).
See \fref{fig:echo_chambers_pnas} for a measurement of echo-chambers across social media.

\section{Quantifying opinion dynamics in empirical data}
\label{sec:empirical_data}

Many opinion dynamics models aim at demonstrating \textit{generative sufficiency} \cite{epstein_generative_2012}, i.e., the identification of a small set of principles that suffice to generate a given macro-level behavior, such as opinion polarization or consensus. 
However, for opinion dynamics models to be relevant for the social sciences and to motivate public policies or interventions, they also need to be based on empirically tested assumptions and their predictions have to be verified \cite{flache_models_2017}. 
To this end, researchers in opinion dynamics consider empirical data of various types to support models in terms of \textit{calibration} and \textit{validation} \cite{matteo_richiardi_common_2006}. 

While the usage of these terms has been inconsistent in the literature \cite{banisch_validating_2024}, an emerging consensus is to denote as calibration the task of empirically estimating plausible values for parameters in the model, often linked to micro behaviors or other assumptions. 
Validation, on the other hand, is the empirical testing of model predictions, often as observable statistical patterns and stylized facts \cite{flache_models_2017}. 
Opinion dynamics research has several decades of history and there have been several iterations between empirical observation and model building, blurring the line between predictions and motivating examples. 
Still, opinion dynamics models only represent society if they are constantly tested against empirical data \cite{sobkowicz_modelling_2009,peralta_opinion_2022}.

This section gives an overview of notable research in the empirical analysis of opinions.
Controlled experiments, both lab and online, provide internal validity for testing causal assumptions in microscopic dynamics (\sref{sec:micro}). 
For macroscopic phenomena, empirical approaches using surveys and large-scale digital traces on social media (\sref{sec:macro}) are crucial for model validation.

\subsection{Microscopic behaviors}
\label{sec:micro}

Behavioral experiments are a rich source of empirical data to validate idealized models of opinion dynamics, particularly to identify which influence mechanisms are the most relevant in emulating human social behavior \cite{mercier_majority_2019}. 
In \textit{controlled experiments}, both offline and computerized, individuals choose between a select set of behavioral options after being exposed (\textit{treatment}) or not (\textit{control}) to information from other subjects or the researchers themselves. 
These experiments are controlled in the sense that many aspects of decision-making of the subjects are constrained in advance by the design of the experiment, apart from selected treatment variables.

The conformity experiments of \sref{sec:exps_discrete} study how likely it is that a subject gives, for example, a manifestly wrong answer to an obvious question, as a function of the number of people who give the same wrong answer \cite{asch_opinions_1955}. 
Conformity in this sense is a type of imitation behavior that can be formalized into a theory of social impact \cite{latane_social_1981}. 
In controlled experiments of opinion revision and social influence of \sref{sec:exps_cont}, a typical quantity to be measured is the percentage of subjects who change their perspective (or opinion) as a function of various controlled variables, such as the absolute and relative number of people holding that opinion, their competence or opinion strength (i.e. status, power, or ability), or the prior beliefs of the individual. 
In addition to their controlled counterparts, the \textit{field experiments} of \sref{sec:field_exps} explore human behavior in more natural scenarios subject to some intervention by the researchers, both in social media and offline.

\subsubsection{Experiments on social impact and conformity}
\label{sec:exps_discrete}

A typical output of controlled experiments on social impact is an estimate of the functional form of the transition rate $F$ between a pair of discrete opinion states (for details on the mathematical treatment of these rates, see \sref{sec:methods}). 
There is debate among social psychologists on the shape of $F(y)$ as a function of the relative number of neighbors of a node that hold a given opinion. 
Social impact theory \cite{latane_social_1981,stauffer_social_2001,eriksson_are_2009} suggests a sub-linear trend, $F(y) \sim y^{\alpha}$ with $\alpha<1$, implying that minority opinions have a relatively strong influence on individuals. 
This is known as the `psychosocial law' or the principle of marginally decreasing impact in social impact theory: if an agent is exposed to the opinions of its acquaintances sequentially, each additional opinion is, on average, less influential than previous ones. 
Conversely, a super-linear trend $F(y) \sim y^{\alpha}$ with $\alpha>1$, implying that the minority opinion has lower influence, has also been proposed \cite{tanford_social_1984,morgan_evolutionary_2012} as the so-called `conformist' transmission of opinions \cite{boyd_why_2005,henrich_evolution_1998}. 
Integer values of $\alpha$ larger than 1 can be interpreted as a situation in which an individual needs opinion unanimity (agreement) between its information sources in order to change its opinion \cite{castellano_nonlinear_2009}. In \textcite{eguiluz_bayesian_2015} Bayesian theory applied to experimental data shows conformist social behavior, although it is difficult to distinguish from a linear trend, i.e. $F(y) \sim y^{\alpha}$ with $\alpha \approx 1$. Based on observational data, there is also a discussion on whether the absolute number of neighbors or their ratio is the relevant variable for opinion change \cite{karsai_complex_2014}.

Given the variability and sometimes conflicting results in experimental setups \cite{mercier_majority_2019}, empirical estimates depend on the social context and experimental design. For instance, social impact experiments typically involve individuals without strong prior opinions \cite{eriksson_are_2009}, whereas conformist bias experiments test subjects with existing opinions, often in scenarios where sensory information is contradicted by peers \cite{morgan_evolutionary_2012}. This latter scenario is more comparable to opinion dynamics models, which commonly feature stationary states of consensus, fragmentation, or polarization.

Beyond the number of influencing peers, social and situational factors significantly impact individual adoption. 
Sources of influence vary in strength [e.g., teachers vs. students causing stress \cite{latane_cross-modality_1976}] and proximity [physical or social distance, as shown by the length of disaster news article correlating with proximity \cite{latane_psychology_1981}]. 
Crucially, a single dissenting individual can drastically reduce majority influence \cite{asch_effects_1951}, implying that even minimal agreement can alter perceived social forces.

\subsubsection{Experiments on revising and influencing opinions}
\label{sec:exps_cont}

Research on voter behavior in political elections is based on experiments that aim to identify how individuals choose who to vote for, based on a dimensional representation of issue positions \cite{merrill_iii_unified_1999}. This embedding is a mixture of proximity and preference for the perceived opinion of the candidate, in itself not an opinion change but an expression of agreement towards the opinions of others.
In continuous opinion experiments, subjects interact over rounds, stating opinions and confidence \cite{moussaid_social_2013,chacoma_opinion_2015}. 
\textcite{chacoma_opinion_2015} identified three behaviors: 'keep' (opinion unchanged), 'adopt' (copying a reference), and 'compromise' (approaching a reference). 
'Keep' was most common, followed by 'compromise' and 'adopt', with confidence difference, not opinion difference, determining the choice. 
This suggests a negligible effect of metric opinion discrepancy on response to influence, contrasting with the bounded confidence modelling framework (see \sref{subsubsec:bounded_confidence}). 
A similar distribution of influence types was reported by \textcite{vande_kerckhove_modelling_2016}, whose model successfully predicted observed opinion evolution.

A survey experiment on wealth redistribution \cite{balietti_reducing_2021} used artificial social media interactions to show that extended discussions, unlike short texts, can reduce polarization by narrowing opinion gaps. 
The study found that perceived closeness to a peer predicted opinion convergence, supporting the bounded confidence principle.
Recently, a series of controlled experiments have been conducted on strategic conformity and anti-conformity with discrete choices, binary or multiple \cite{dvorak_strategic_2024}. 
In one of the experimental settings, subjects were divided into groups of three and first made binary choices without knowing the choices of others, then chose again after receiving such information. 
This scheme is almost identical to the $q$-voter model of opinion dynamics \cite{castellano_nonlinear_2009} (see \sref{subsec:q_voter}), and thus the experimental results reported by \textcite{dvorak_strategic_2024} can serve as a basis for further developing the model.

\begin{figure}
    \centering
    \includegraphics[width=\linewidth]{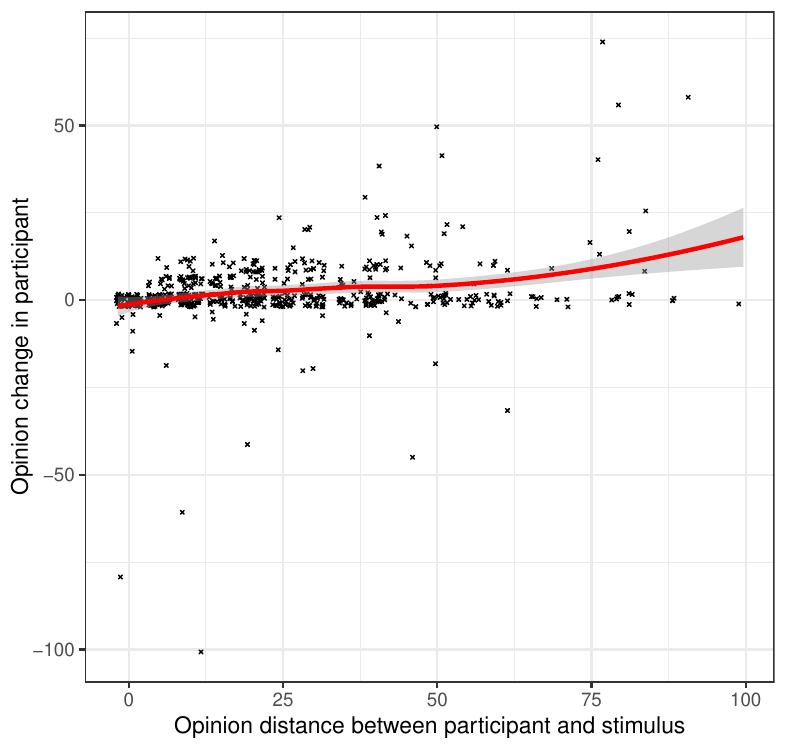}
    \caption{Opinion changes in a controlled experiment. Jittered scatter plot of the results of the first experiment of \textcite{takacs_discrepancy_2016}, where participants were shown opinions at a distance from them, recording their opinion changes. The red line shows a LOESS fit (locally estimated scatterplot smoothing), which reveals a positive tendency to move towards the opinions of others.
    }
    \label{fig:experiment}
\end{figure}

Experiments also suggest repulsive forces or ``\emph{backfire effects}'' that drive individuals away from a stimulus, such as the opinions of others.
This effect was initially discovered as an unintended consequence in setups where participants were subjected to a stimulus advocating for a change in behavior. Unexpectedly, people showed instead a change in the opposite direction. This observation was initially called the \emph{``contrast effect''} \cite{hovland_assimilation_1957} or the \emph{``boomerang effect''} \cite{cohen_dissonance_1962}.
More recently, this effect has been studied in experiments of political misperception in the context of misinformation, where researchers attempt to change a participant's opinion with factual information, but the intervention provokes a move in the opposite direction \cite{nyhan_when_2010}.

The backfire effect has also been identified in field experiments on Twitter (\sref{sec:field_exps}), although it is far from universal. Predicting when and in which conditions it occurs is a complicated task \cite{wood_elusive_2019}. For instance, experiments have tested whether a prior distance between opinions or disliking a source of information explain why individuals change their opinion away from the source, but have not found a reliable effect yet \cite{takacs_discrepancy_2016}. The results of one of these experiments are shown on Fig. \ref{fig:experiment}, where changes in opinions towards a source are more likely than away from it. A follow-up study finds weak evidence for negative influence, but only if strong disagreement and ideological difference combine~\cite{keijzer_polarization_2024}.
Based on these observations, it seems that models based on the backfire effect need to include clear explanations for its existence, apart from merely phenomenological observations. 

Argument-communication theory goes beyond scalar opinion models to propose that individual positions are sets of pro- and contra-arguments. The experiment of \textcite{mas_differentiation_2013} tests opinion-only, argument-only, and combined interactions (with varied matching), showing that argument exchange can explain polarization even without negative influence. Instead, some form of biased directional updating seems to be relevant, where individuals preferably pick up arguments in favor of the side they already lean towards.

\subsubsection{Field experiments}
\label{sec:field_exps}

Field experiments are a complementary approach to test effects influencing human behavior, based on natural scenarios rather than highly controlled laboratory situations \cite{duflo_handbook_2017}. 
This is a challenging approach, due to the risk of influencing human behavior in the wild and the difficulty of isolating the effect of a treatment under the full variance of behavior outside of a lab. Concerning opinions, field experiments in psychology have been used to test treatments against prejudice or inter-group conflict. 
For example, field experiments in schools have shown that fostering inter-ethnic contact leads to higher levels of linking and respectful behavior across groups than in a control condition \cite{cook_experimenting_1985}.
Another field experiment with data from a radio show in Rwanda suggests that a reconciliation-focused program, as opposed to a control condition concentrating on health, leads to a significant reduction in prejudice towards other ethnic groups \cite{paluck_reducing_2009}.
Experiments inducing friendships between individuals of different ethnic groups have tracked the effects of this intervention in diaries and physiological measurements of stress \cite{page-gould_little_2008}, finding significant reductions in stress in inter-group settings for treated participants.
Beyond these studies, there is a wider set of examples of field experiments on prejudice and inter-group opinions (see \textcite{paluck_prejudice_2009} for a review on the subject).

Social media offer new opportunities for field experimentation, allowing the application of treatments at scale, especially in the form of bot interventions or through changes in feed algorithms.
A notable experiment used bots to reply to users who have previously posted racist tweets, finding the conditions of the bot profile that lead to a decrease in racism \cite{munger_tweetment_2017}.
Other experiments with social media bots identify how hashtags spread on social media in a manner akin to social contagion.
\textcite{monsted_evidence_2017} applied treatments through bots to control the exposure of users to different sources of a hashtag, finding that the users' tendency to adopt the hashtag is explained more accurately by a threshold effect than by independent exposure events.
With respect to opinions, bot experiments show how the political position of conservatives can become more extreme after being exposed to cross-cutting content reshared by a bot \cite{bail_exposure_2018}.

Field experiments with feed algorithms and interface design also explore the ways interactions might change behaviors related to opinions. A large-scale experiment on Facebook tested how seeing the voting behavior of friends motivates other users to vote in US elections \cite{bond_61-million-person_2012}.    
Recent experiments changing the feed algorithms of Facebook and Instagram showed that reducing the exposure to sources of similar political alignment does not produce the predicted reductions in polarized or extremist attitudes \cite{nyhan_like-minded_2023}. 
Another experiment changed the algorithm to reduce the exposure to reshared content \cite{guess_reshares_2023}, also finding no effects on reducing polarized opinions as measured through surveys.
We note that these experiments took place over three months centered around the 2020 US presidential election, while the company Meta had in place special mechanisms to control misinformation during the run-up of the election \cite{bagchi_social_2024}. This situation differs from the usual operation of the platform and thus affects the potential generalization of these results. Changing feed algorithms does not affect individuals in isolation, but alters the way people interact. This calls for agent-based models that can better predict the effects of these interventions \cite{garcia_influence_2023}.

More recently, lab-in-the-field experiments on opinion dynamics have combined the control of lab settings with the realism of field experiments. An experiment in the UK \cite{hobolt_polarizing_2024} recruited participants from a representative sample to have Zoom discussions about immigration policy with groups of the same or mixed political ideology, providing minimal steering to the discussion dynamics. Results support previous hypotheses of echo chamber formation: opinion extremity and affective polarization increase after interactions in like-minded groups, while affective polarization decreases after interacting with mixed groups.

\subsection{Macroscopic phenomena in societies}
\label{sec:macro}

Opinion dynamics models often make predictions about macro-level behavior, requiring empirical approaches that consider data at the scale of entire communities or societies. 
For example, some predictions have been tested against election results (\sref{sec:elections}). While election outcomes offer macroscopic data, models often address behaviors not reflected in them, require testing on unidentifiable groups or attitudes, or need higher-frequency measurements. 
A general approach for capturing these phenomena is conducting surveys with questionnaires on representative population samples (\sref{sec:surveys}).
Surveys, though established, face challenges like declining response rates and ``expressive responses.'' 
An alternative, large-scale digital trace data from social media (\sref{subsec:opinion_data_sm}), offers higher temporal resolution and structural information but has its own sampling and measurement limitations.

\subsubsection{Election results}
\label{sec:elections}

Elections are large-scale efforts to capture and aggregate the opinions of a society in order to make collective political decisions. 
As a byproduct, they create publicly available datasets that can be used to study macroscopic phenomena in opinion dynamics. 
While the topic of election research is too broad to be considered in detail here, there are points of contact between statistical physics and election research that provide insights into opinion dynamics.

Within this framework, votes are discrete multistate opinions showing a preference for a certain candidate or political party. \textcite{chatterjee_universality_2013} have analyzed voting results in open-list proportional electoral systems, finding a universal pattern that explains the relative votes of candidates with respect to their parties across countries and years. Their analysis shows that electoral systems and voting rules play an important role in the shape of voting distributions, the key components of the data generation process when using voting data to study opinion dynamics. Deviations from the universal pattern, especially in the right tail of vote counts, are consistent with voting cascades generated through social influence mechanisms \cite{granovetter_threshold_1978,watts_simple_2002,centola_complex_2007,ruan_kinetics_2015}.
Analyzing statistical patterns of election results is also useful to detect deviations from hypotheses underlying opinion dynamics, such as how excessive kurtosis in vote distributions implies potential voter fraud \cite{klimek_statistical_2012}.

More directly related to opinion dynamics, election results in Brazil have been emulated by the Sznajd model over a preferential attachment network \cite{bernardes_election_2002}. The voter model is also able to reproduce data on vote-share distributions at the county level in the 1980–2012 US
presidential elections \cite{fernandez-gracia_is_2014}. Similar data has been used to explain the vote-share distribution as a combination of external influence and peer contagion \cite{braha_voting_2017}, amounting to a model that combines behavioral spreading with voting. 
Opinion dynamics models have led to predictions of the US presidential elections of 2020 \cite{galam_will_2021} and 2024 \cite{kaufman_application_2025}. They also attempt to explain discrepancies between popular and electoral college results in the US \cite{biswas_critical_2017,biswas_block_2021}, and the eventual result of the Brexit referendum \cite{biswas_social_2023,mukherjee_long_2020}.
Beyond political elections, voting networks in the Eurovision song contest have been understood as a polarization dynamics that oscillated during the 2008 Euro crisis \cite{garcia_measuring_2014}.

\subsubsection{Survey data}
\label{sec:surveys}

Representative surveys provide a vital mechanism for validating emergent opinion patterns at the macro level, covering topics from ideological self-placement to migration and political values. 
These surveys utilize various answering options, most notably the 0-10 scale, which yields quasi-continuous opinion distributions that are well-suited for comparison against the outcomes of continuous opinion dynamics models. 
International programs, such as the European Social Survey (ESS) and World Value Survey (WVS), are conducted regularly with new representative samples across multiple countries, allowing for the empirical study of opinion evolution in the general population over time. 
While less common panel surveys track the same individuals, their usefulness for validating micro-mechanisms in opinion dynamics is severely limited by the typical absence of social network and contact information.

\textcite{gestefeld_decomposing_2022} demonstrated that opinion distributions remain remarkably stable over time on an extensive macro-level study using ESS data, showing less variation annually than they do between different countries or topics. 
This finding reveals a crucial empirical regularity: macroscopic stability persists despite microscopic changes, as separate panel surveys indicate that individual opinions frequently shift. Additionally, empirical opinion distributions often exhibit complex, non-trivial shapes, frequently displaying a pentamodal structure—one central peak, two peaks at the extremes, and two off-center non-extreme peaks—a shape more common for political topics. 
Since empirical distributions are rarely found in states of pure consensus or maximal polarization, \textcite{gestefeld_decomposing_2022} concluded that these detailed real-world data are currently underutilized for exploring and validating advanced opinion dynamics models.

Political opinion surveys have been collecting so-called affective thermometers since the 1970s, where participants are asked about their feelings for political parties, partisans, and candidates in a scale between 0 and 100. 
\textcite{iyengar_fear_2015} have measured affective polarization, i.e. the tendency to have positive attitudes with people that mostly agree with us, and negative attitudes towards the rest (see \ref{sec:macro:polarization}). 
The authors conducted survey experiments with an implicit association test for parties and partisans, as well as other bias tests (including selecting the winner for scholarships and dictator games with financial rewards). 
The US is a prominent example of increasing affective polarization over time \cite{webster_ideological_2017}. 
Comparative survey data shows that affective polarization is increasing in other countries too, but it is far from a worldwide phenomenon \cite{boxell_cross-country_2024}. 
Metrics of multi-party affective polarization have been developed and applied to multiple European countries \cite{reiljan_fear_2020}, finding that there is a wide range of affective polarization in Europe above and below the level of the US. 
Research on macroscopic opinion measurements beyond the US is becoming prominent to calibrate opinion dynamics models \cite{gestefeld_calibrating_2023}, especially as a way to cover a wider range of possible outcomes and systems than just one country \cite{falkenberg_towards_2025}.

Causes for the rise in affective polarization in the US are debated by \textcite{abramowitz_polarized_2013}, primarily revolving around ideological polarization and partisan sorting hypotheses.
\textcite{webster_ideological_2017} suggested that affective polarization may be primarily driven by ideological polarization, i.e., individuals having increasingly divergent beliefs on ideological issues, such as abortion or affirmative action. 
\textcite{fiorina_political_2008} argue that, while political elites are clearly polarized, ordinary Americans are less so. 
Instead, according to the partisan sorting hypothesis, the electorate has become more ``sorted'', meaning party affiliation and ideology are more strongly aligned than in the past, leading to consistent opinions across issues. 
This point is mostly illustrated by analyzing mean differences by self-reported partisan sorting in the General Social Survey (GSS), American National Election Studies (ANES), and Gallup. Research on ANES survey data has studied issue alignment in the US and found a steady increase since the early 2000s \cite{kozlowski_issue_2021}.

\textcite{ojer_charting_2025} have also used ANES data to disentangle the alternative hypotheses of ideological polarization or partisan sorting. 
By mapping the opinions of ANES respondents within a two-dimensional ideological space, the authors show that the ideological gap between Democrats and Republicans on key issues has indeed widened over the past 30 years.
At the same time, parties became more ideologically heterogeneous, especially Democrats after 2010, thus contradicting the partisan sorting hypothesis for the increase in affective polarization. 
Further research on ANES has identified an increase in hyperpolarization, signalled by the simultaneous presence of issue alignment and opinion extremeness \cite{schweighofer_weighted_2020,schweighofer_raising_2024,baumann_emergence_2021}.
Studies also show that opinion-influenced behaviors can create feedback loops that drive opinion extremism. 
An identified mechanism is acrophily \cite{goldenberg_homophily_2022}, the tendency to form links with similarly or more extremely minded individuals, resulting in selective exposure that reinforces opinions within echo chambers (see \ref{subsec:echo_chambers}).

\begin{figure}
    \centering
    \includegraphics[width=\linewidth]{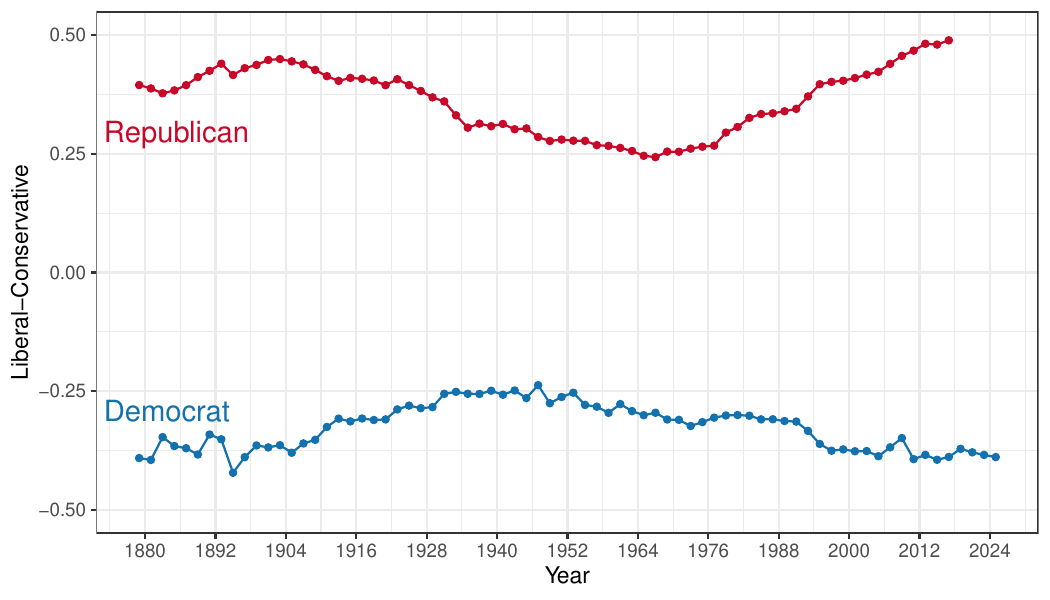}
    \caption{Mean ideological positions of parties in the US House of Representatives. Using roll-call vote data, which is generated when votes are done verbally and publicly calling the name of each member of the US Congress, each member can be mapped to a space from liberal to conservative using the DW-Nominate (Dynamic Weighted NOMINAl Three-step Estimation) method. This shows periods where the representatives of both parties were closer in their opinions and an increasing division across party lines.}
    \label{fig:polarization_congress}
\end{figure}

Other observational data in the social sciences can complement survey data to contextualize the measurement of opinion dynamics.
\cite{feezell_exploring_2021} analyzed the relationship between consumption of news through algorithmic curation and other means in the Youth Participatory Politics Survey (YPP) and did not find an effect on algorithmically-curated news consumption on polarized opinions.
Other notable sources of opinion data is voting records in parliaments where individual votes are publicly recorded, where the position of parties can be compared and tracked over time. 
Fig. \ref{fig:polarization_congress} shows the mean positions of Republicans and Democrats in the US House of Representatives, which reveals the increasing pattern of elite polarization in the US.
Polarization can also be measured through text, for example, as a metric of how well a tweet can be classified for ideology based only on its text \cite{green_elusive_2020}. 
That approach gives a timeline of polarization based on the degree of predictability of partisanship in the state of political discussion. 
Text analysis of congressional records can be used to analyze polarization, for example, with respect to the topic of immigration \cite{card_computational_2022} or to identify how components of populist rhetoric lead polarization in votes \cite{aroyehun_computational_2025}. 
Beyond the US, text analysis of Swiss parliamentary records has also been used to measure multi-party polarization over time \cite{schlosser_textual_2024}.

\subsubsection{Online social media}
\label{subsec:opinion_data_sm}

The scale, resolution, and pervasiveness of digital traces from social media allow new approaches to the observational analysis of opinions and their dynamics. 
Early examples of social media research have analyzed the spread and adoption of behavioral patterns akin to research on social impact (see \ref{sec:exps_discrete}).
The adoption of informal conventions can be traced through social influence in online networks, such as the spreading of norms on how to retweet someone without the ``retweet'' button \cite{kooti_emergence_2012}. 
Behavioral data traces from microblogging platforms allow the tracking of the co-evolution of two layers of the network: the follower relationships between users and the spreading of information through reshares \cite{weng_role_2013}.
Focusing on the adoption of a hashtag or the resharing of content, multiple exposures increase the probability of a user spreading content \cite{romero_differences_2011,lerman_information_2010}, but the effects of these exposures are not independent. 
Indeed, an assumption of independence between exposures actually overestimates the intensity of spreading \cite{lerman_information_2016}.
Research on online discussion dynamics has introduced generative models that simulate the structural and temporal patterns of conversation growth \cite{aragon_generative_2017}.
Several other methods have been developed to measure opinion-related phenomena in social media, which we review in this section.

\paragraph{Opinion and ideology estimation of individuals.}

The vast majority of research on opinions using social media data builds on at least one of two methods: i) text analysis to infer ideology or opinions of users; and ii) network analysis to infer political positioning or diagnose polarization (e.g., studies of who follows whom on a given social medium).
Among the first studies using text to identify the political orientation of social media users, \textcite{conover_partisan_2012} detected the political alignment of Twitter users based on the use of ideology-signalling hashtags. 
They also found high correspondence with the location of users in two communities in their retweet network.
Hashtags have been used to identify the political leaning of Twitter users \cite{cota_quantifying_2019}, in relation to the impeachment of former Brazilian President Dilma Rousseff.
They showed that pro-impeachment users transmitted information to a larger audience on average, and that a user's spreading capacity correlates with the political diversity of their audience.

The use of hashtags to identify ideology is sensitive to the time period of analysis and self-selection bias. This means that measurement is only possible on individuals who actively choose to express their views, potentially leaving a silent majority outside the study.
One way to overcome such an issue is to train supervised classifiers on real tweets \cite{conover_predicting_2011}, but this is only accurate for users highly engaged in politics and performs substantially worse when analyzing more representative samples of Twitter users \cite{cohen_classifying_2013}.
When a focus on individuals is not necessary and researchers want to measure an overall aggregate of content polarization for a topic, expressions of positivity and negativity can both be used as a metric of polarization. An example is the analysis of tweets about Bitcoin, where the latter serves as a predictor of trading volume \cite{garcia_social_2015}.

An alternative to text analysis is the use of social network information, especially the lists of followed accounts by individuals on Twitter. 
This is the main data source of methods identifying a scale of ideology from conservative to liberal in the US \cite{barbera_birds_2015}, a metric that correlates with the degree of partisanship across regions in the country.
This method has been extended to capture ideology in more than one dimension in multi-party systems \cite{ramaciotti_morales_inferring_2022,peralta_multidimensional_2024}, particularly when considering more dynamic retweet data \cite{martin-gutierrez_multipolar_2023}. 
Such approaches combine the time sensitivity of text analysis with the robustness of network analysis, offering a promising approach for future research using social media data in opinion dynamics.

\paragraph{System-level measurement of polarization.}

Beyond individuals as units of analysis, a major motivation for the use of social media data is its potential to reveal system-level structures that signal various types of polarization. 
Among the first studies to follow this approach, \textcite{conover_political_2011} collected tweets related to the 2010 U.S. congressional midterm elections through a hand-curated hashtags selection. 
The authors reconstructed both the retweet network (i.e., two users are connected if one retweets the other) and the mention network (i.e., two users are connected if one mentions the other), 
finding that the former is highly segregated, while the latter is not. 
This study opens the way to using the structure of Twitter interactions as a tool to infer the opinions of users, as retweets can often be seen as an endorsement of the opinion of the retweeted user.
For example, retweet and follow networks can be partitioned to quantify the level of controversy, or polarization, in a conversation over a certain topic \cite{garimella_quantifying_2018}, without relying on a priori assignments to opinion groups.
Alternatively, network polarization was operationalized as a metric of fragmentation or modularity with respect to party identity on networks of links with a positive meaning, such as likes \cite{garcia_ideological_2015} or retweets \cite{barbera_tweeting_2015}.
Polarization manifested as partisan sorting also in other contexts, such as scientific book purchases \cite{shi_millions_2017} and other preferences like restaurants or music \cite{shi_cultural_2017}.

Different metrics aim to capture more nuanced forms of polarization beyond modularity \cite{guerra_measure_2013}, for example, by using positive and negative links between individuals.
Text classifiers that measure toxicity, defined as rude or harmful content that drives users away from online interaction, are used to measure a signal of disagreement or negative expressions across party divides, as a first approach to measuring affective polarization \cite{lerman_affective_2024}, as well as fragmentation in signed networks \cite{pougue-biyong_debagreement_2021}. 
This type of social media data with positive and negative relationships in discussions can be used to identify alignment in networked structures \cite{fraxanet_unpacking_2024}, thus revealing fault lines that divide the population across various issue dimensions.
We highlight the relevance of emotional expression as a manifestation of affective polarization in online social networks. This can be used to identify correlations between user activity and negativity
\cite{del_vicario_echo_2016}, the possible association between moral emotional content and its spreading \cite{brady_emotion_2017,burton_reconsidering_2021}, and how negative expressions about the out-group motivate user reactions \cite{rathje_out-group_2021}.

\begin{figure}[tbp]
\centering
\includegraphics[width=\columnwidth]{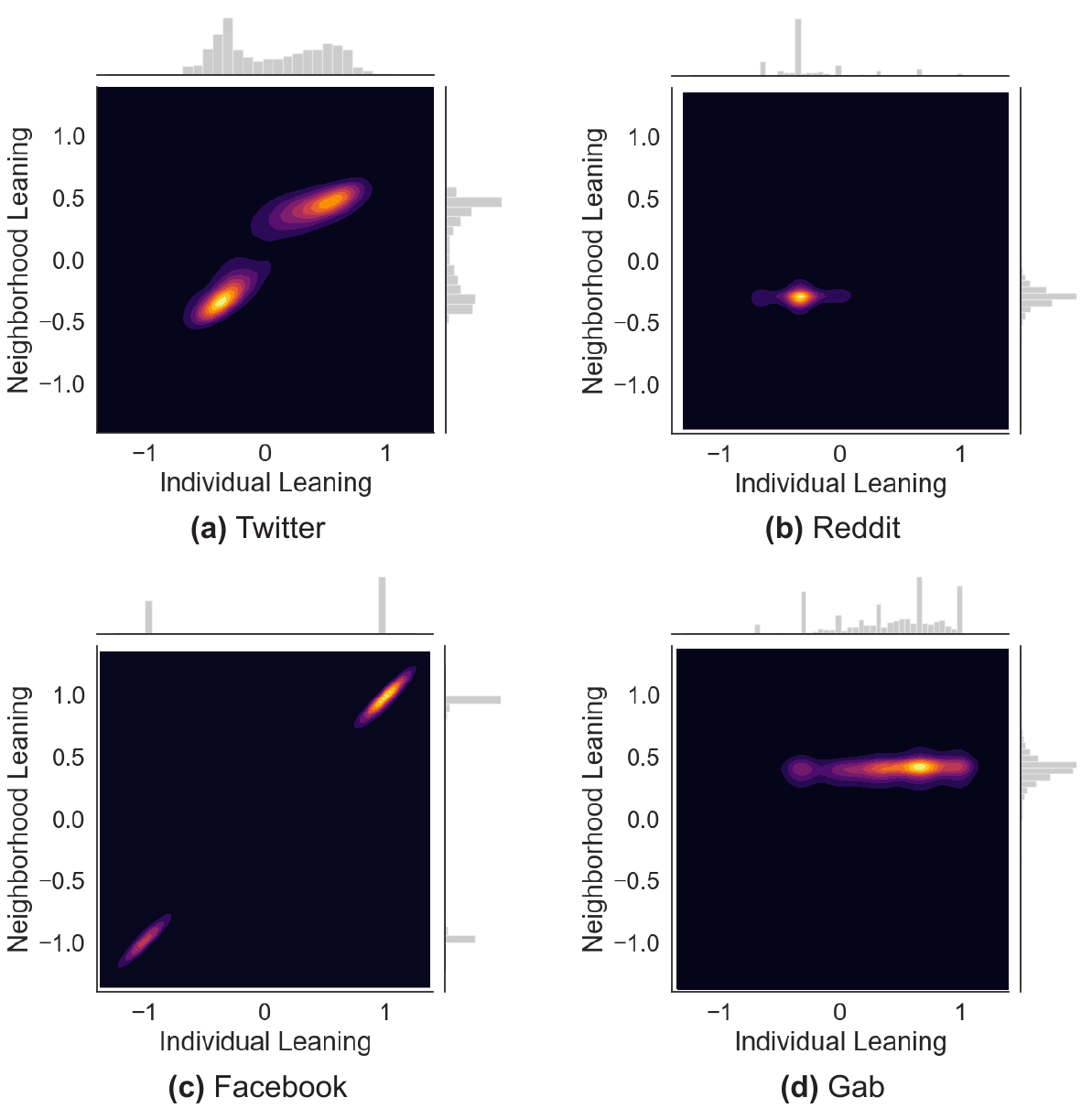}
\caption{
Joint distribution of the opinion of users, $x$, and the average opinion of their neighborhood $x_{NN}$ for different topics and datasets: 
(a) Abortion on Twitter, (b) Politics on Reddit, (c)
Vaccines on Facebook, and (d) Gab as a whole. 
Colors represent the density of users: The lighter the color, the larger the number of users. 
Marginal distributions $P(x)$ and $P_N(x)$ are
plotted on the x and y axes, respectively. 
Figure from \textcite{cinelli_echo_2021}.
\label{fig:echo_chambers_pnas}}
\end{figure}

When opinion polarization combines with network segregation, echo chambers may emerge \cite{cinelli_echo_2021}.
These are situations where individuals are primarily exposed to opinions and beliefs that align with their own. 
\textcite{cinelli_echo_2021} quantified echo chambers across social media platforms by inferring the opinion of users on a specific topic, for example, vaccines, and their exposure to content from their peers.
Crucially, these two elements must be inferred independently. 
Echo chambers emerge when the opinion of a user is correlated with the opinions of their neighbors, as showed in Fig. \ref{fig:echo_chambers_pnas}.
Results indicate that platforms such as Facebook and Twitter tend to exhibit echo chambers for various topics, while others, such as Reddit, do not. 
\textcite{de_francisci_morales_no_2021} confirmed this finding on Reddit, by showing that Trump and Clinton supporters during the 2016 US election debate preferred cross-cutting political interactions rather than solely engaging with like-minded peers. 
\textcite{morini_toward_2021} develop a general framework that treats echo chambers as meso-scale structures in debate networks.  Using label-guided community detection, their method uncovers multiple echo chambers not only across opposing camps but also within the same ideological leaning.
Echo chambers are relevant for polarization research due to their role for the spreading of misinformation \cite{del_vicario_spreading_2016}. Like-minded communities also provide a breeding ground for low-quality information sharing without proper engagement across various opinions.

\paragraph{Polarization across topics.}

The metadata available on online discussions allows researchers to identify the context in which opinions manifest on social media. 
For example, retweeting across the spectrum of ideology on Twitter shows that some topics, such as elections or government shutdowns, create more polarized discussions than others, such as sports and entertainment \cite{barbera_tweeting_2015}. 
This approach can also show how some topics can become polarized over time, for example, the Newtown school shooting went from a national-scale tragedy to a politicized topic.
In order to study political polarization \textcite{morales_measuring_2015} quantified the degree of polarization in the Twitter conversation about Hugo Chávez, by employing a model that estimates opinions based on how influential minorities propagate their views through the network.

Several studies focus on well-defined cases where opinion polarization matters, for example, on the topic of vaccination.
\textcite{cossard_falling_2020} 
found the Italian Twitter debate on vaccines to be polarized and segregated into echo chambers. 
They achieved this by partitioning the retweet and follower networks [similar to \cite{garimella_quantifying_2018}] and then predicted a new user's echo chamber alignment based on the content of their previous tweets, unrelated to vaccination. 
The vaccination debate on Facebook is also polarized,
as revealed by a quantitative analysis of 2.6 million Facebook users interacting with nearly 300,000 vaccine-related posts over seven years \cite{schmidt_polarization_2018}. 
The authors found that users predominantly consumed content either supporting or opposing vaccines, rarely both, leading to well-segregated online communities.
The relation between opinions regarding vaccines and political leaning has also been explored. 
\textcite{paoletti_political_2024}
investigated the link between political interest and exposure to vaccine-hesitant content on Twitter across 17 European countries, finding that users following right-wing, authoritarian, or anti-EU politicians were more likely to endorse vaccine-hesitant content, while followers of left-wing, pro-EU, or liberal parties were less so. 

Focusing on climate change discussions on Twitter reveals that polarization manifests around the release of COP reports \cite{falkenberg_growing_2022}, with an increasing trend that aligns with the growth of right-wing activity on Twitter on climate-related topics.
The topics of climate change and health overlap when it comes to issues of food consumption. 
In particular, the consumption of meat has emerged as a polarizing topic around the release of the EAT-Lancet report \cite{garcia_eatlancet_2019}. 
This is another example of how choosing a topic can highlight other dimensions of opinion dynamics, as that case revealed how users in a moderate community reoriented themselves to a pro-meat community in their choice of who to retweet.

\paragraph{Polarization over time.}

Long trends of polarization can be identified with large-scale social media data, as illustrated by the analysis of more than seven years of Twitter data that showed that polarization was consistently increasing \cite{garimella_long-term_2017}. This has been measured more recently for Twitter influencers, who display increasing polarization between the 2016 and 2020 US presidential elections \cite{flamino_political_2023}.
Evolution in the polarization about vaccines can also be observed, especially related to the onset of the COVID-19 pandemic.
\textcite{lenti_global_2023} analyzed tweets across 28 countries from October 2019 to March 2021, to quantify global anti-vaccination misinformation flows and content moderation effects. 
They found that during the pandemic, anti-vaccination communities became more central in their national debates and their international connections strengthened, forming a global network. 
Notably, Twitter's content moderation, especially post-January 6th US Capitol attack suspensions, significantly reduced global vaccine misinformation spread.
The time resolution of social media data also allows the measurement of the reaction of online polarization to political events, such as how elections relax polarized structures across parties \cite{garcia_ideological_2015} and how positive and negative links appear across fault lines in online political discussions \cite{fraxanet_unpacking_2024}.

\paragraph{Polarization correlates.}

Studies have shown the effects and correlates of online polarization, for example ``the wisdom of polarized crowds'', as evidenced by the improved quality of Wikipedia articles edited by ideologically polarized groups \cite{shi_wisdom_2019}.
In online interactions about Bitcoin, polarization in text correlates with trading volume  \cite{garcia_social_2015}. This has been described as ``differences of opinion make a horse race'' \cite{harris_differences_1993}, i.e., trading can happen among individuals with access to exactly the same information as long as they hold different subjective opinions.
Polarization can also appear as a function of how content spreads on social networks, for example in the scaling of likes and dislikes with audience size, which shows that larger audiences are more negative and create polarization in the presence of positive biases \cite{van_mieghem_human_2011}. This leads to a ``breaking the bubble'' effect in which content attracts polarization once it reaches a level of attention beyond the immediate social context of the user who posted it \cite{abisheva_when_2016}.

While polarization is an important topic, the focus on controversial topics and highly engaged users has led to some overestimation of polarization and to inadequate generalization across platforms.
Reddit is an example where echo chambers are not evident in political interactions \cite{de_francisci_morales_no_2021} and where demographics are stronger predictors of interaction than ideological segregation in discussions \cite{monti_evidence_2023}.
Beyond polarization, social media data has been used 
to measure opinion trends \cite{oconnor_tweets_2010}, track presidential polls \cite{bovet_validation_2018} and even to predict election outcomes \cite{tumasjan_predicting_2010}. 
Some of these early results stood the test of time with robust correlation methods, such as the case of opinions about the state of the economy \cite{pasek_stability_2018}, but in general social media data tends to lead to biased predictions \cite{jungherr_why_2012} and overstatements when building on simple aggregates of opinionated expression or popularity of parties and candidates \cite{gayo-avello_meta-analysis_2013}.
However, a new line of work appears to be promising in this area \cite{cerina_2024_2025}. This combines Twitter data with demographic inference and generative language models to correct sampling biases towards a more reliable estimate of voting intention in US presidential elections.

\section{Models of opinion dynamics}
\label{sec:models}

The development of mathematical and computational models of opinion dynamics goes back more than 50 years to the works of \textcite{french_formal_1956,abelson_computer_1963, abelson_mathematical_1964,degroot_reaching_1974}. It also finds roots in statistical physics with the models of polarization of \textcite{weidlich_statistical_1971}, sometimes called the founder of sociophysics \cite{batey_wolfgang_2023}, the synergetics of \textcite{haken_cooperative_1975}, and the mathematical theory of social imitation by \textcite{callen_imitation_1974}. The voter model was initially proposed as a model of spatial conflict \cite{clifford_model_1973}, attracting first the interest of probabilists \cite{holley_ergodic_1975} to later become one of the most studied opinion dynamics models \cite{redner_reality-inspired_2019}. \textcite{galam_sociophysics_1982} used ideas from critical phenomena to model strikes in collective worker behavior.

Further early work on opinion dynamics by \textcite{nowak_private_1990} and \textcite{lewenstein_statistical_1992} focused on agent-based models and the statistical physics of social impact \cite{latane_psychology_1981}. The Kirman model \cite{kirman_ants_1993}, describing herding and animal decision making, is a variation of the original voter model. \textcite{helbing_boltzmann-like_1993} proposed models inspired by Boltzmann and Fokker--Planck equations to describe human behavior, as well as early models of pedestrian movement \cite{helbing_mathematical_1991}. Another influential work is the model of cultural dynamics by \textcite{axelrod_dissemination_1997}, who hypothesized that future online media, despite creating more connections between people, would not lead to global consensus in one culture. Instead, he predicted a trend towards polarization by selective exposure. 
Work from that period also includes the Sznajd model \cite{sznajd-weron_opinion_2000}, as well as the models by \textcite{deffuant_mixing_2000} and the Hegselmann--Krause model \cite{hegselmann_opinion_2002}.

Although the use of statistical physics to model the dynamics of opinions, beliefs, or attitudes might appear oversimplified from a psychological perspective, 
recently, some psychologists 
adopted classical models from statistical physics to describe social phenomena \cite{olsson_analogies_2024}. 
For example, by modeling individual attitudes as Ising networks of interconnected attitude elements, \textcite{van_der_maas_polarization_2020} developed the hierarchical Ising opinion model, 
in which agents behave either discretely or continuously depending on their level of attention to the issue. 
Building on this framework, \textcite{dalege_networks_2025} proposed a theory for networks of beliefs, which also incorporates the 
attitudinal entropy framework \cite{dalege_attitudinal_2018}.
More recently, \textcite{van_der_maas_statistical_2025} used the Spin-1 Blume-Capel model \cite{capel_possibility_1966} to represent responses with uncertain or neutral categories (e.g., “don’t know” or “not relevant”), which often appear in psychological datasets. 
Furthermore, opinion dynamics models are the basis of Social Sampling Theory, which explains the emergence of polarization from a combination of mechanisms of social extremeness aversion and authenticity preference when exchanging opinions \cite{brown_social_2022}.

Here we give an overview of the literature on models of opinion dynamics, focusing mostly on statistical physics papers. However, it should be noted that the study of opinion dynamics is of interest to many other disciplines as well \cite{bernardo_bounded_2024,forster_anonymous_2013,grabisch_model_2013,grabisch_model_2019}.
We broadly structure this along the macroscopic phenomena they describe (see \sref{subsec:macro-phen}). Whenever possible, we organize the material according to the microscopic mechanisms of opinion change implemented (see \sref{subsec:micro_mech}).
\sref{subsec:emergence_of_consensus} is dedicated to consensus formation, \sref{subsec:opinion_fragmentation} to opinion fragmentation, \sref{subsec:polarization} to polarization, and \sref{subsec:echo_chambers} to the emergence of echo chambers. 
Finally, in \sref{subsec:coevolution}, we summarize work on models in which the agents' opinions and the underlying social network co-evolve. 
Of course, the division into these topics is not always clear-cut. For example, the phenomena of fragmentation and polarization are related to one another (see \sref{subsec:macro-phen}). Additionally, a given model dynamics can lead to consensus formation for some values of the model parameters, and to more polarized outcomes in other regimes. 
Likewise, a model may involve co-evolving opinions and networks, and result in fragmentation or consensus. 

\subsection{Consensus}
\label{subsec:emergence_of_consensus}

Consensus, broadly understood as some sort of general agreement within a group, plays a crucial role in social dynamics, as many situations in daily life require collective decision-making \cite{baronchelli_emergence_2018}. 
Moving animal groups need collective decision-making as a type of consensus that emerges from local interactions \cite{couzin_effective_2005} and that can be improved by the presence of uninformed individuals \cite{couzin_uninformed_2011}.
In this field, researchers model the opinion states of a population and identify the basic processes that drive changes in these states, aiming to understand the mechanisms behind opinion formation.

Mechanisms leading to consensus in stylized discrete opinion models fall into two classes: by randomness or by deterministic drift. 
In the first class, consensus arises solely from fluctuations in finite populations, emerging by chance. 
These models typically feature absorbing consensus states and lack deterministic flow towards them, as seen in the conventional voter model, where consensus time can be long. 
Conversely, consensus via deterministic drift occurs when it is a locally attracting fixed point of the deterministic rate equations, leading to relatively rapid consensus in both finite and infinite populations, exemplified by variants of the nonlinear voter model \cite{redner_guide_2001, ramirez_ordering_2024}

In models of continuous opinions, the formation of consensus is seen as a gradual process: when two agents interact, their opinions shift toward one another.  
Discrete-time formulations capture this by updating each agent’s opinion in proportion to the difference between its current view and that of its interaction partner.  
In continuous-time this process is usually replaced by a diffusive coupling scheme, encoded by the graph Laplacian of the underlying network.

\subsubsection{The conventional two-state voter model}
\label{sec:voter_model}

\begin{figure}
    \includegraphics[width=\columnwidth]{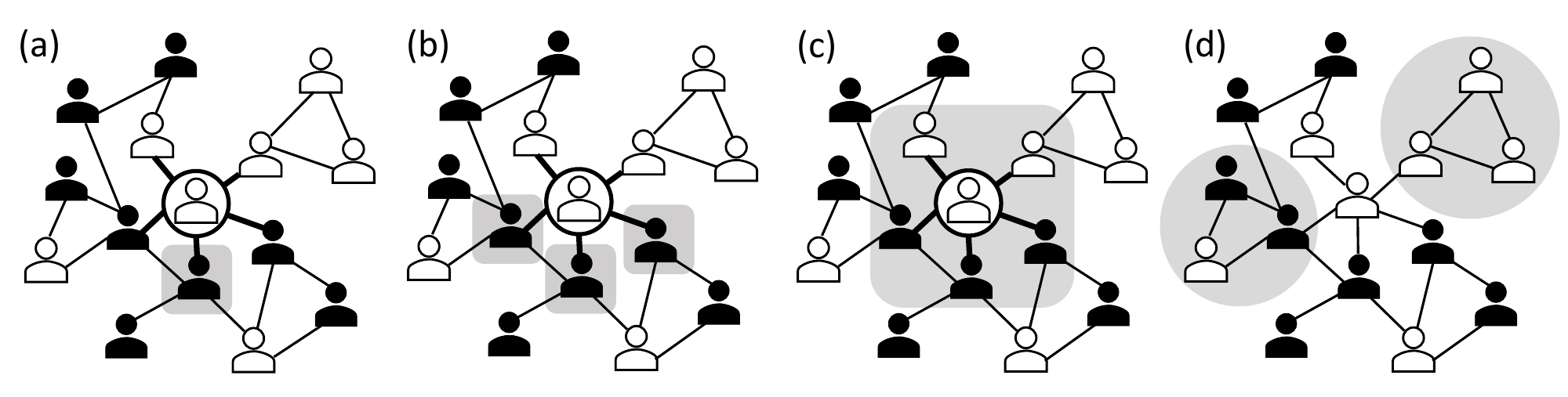}
    \caption{Schematic representations of (a) the voter model, (b) the $q$-voter model, with $q=3$, (c) the nonlinear voter, the majority vote or the threshold model, and (d) the Galam majority model and the majority rule model, with size of the discussion group $G=3$. 
    In (a)-(c), only a single agent (in the circle) updates its state in a given time under the influence of agents in the shaded fields, while in (d), the whole group within a shaded circle changes its state. 
    These diagrams can also represent multistate models, or even continuous opinion models, if we ignore the black-and-white representation of agents and focus only on their interactions.}
    \label{fig:binary_models}
\end{figure}

The voter model is probably the simplest and most popular model of opinion dynamics. 
Its simplicity is a key factor to its success, as it is one of the rare non-equilibrium stochastic processes that can be solved exactly in any dimension \cite{redner_guide_2001}. 
In its basic formulation, first proposed by \textcite{holley_ergodic_1975}, individuals in a population hold one of two possible opinions and are connected in a lattice. We write the opinion of agent $i$ at time $t$ as $x_i(t)\in\{-1,1\}$.
At each time step, a randomly selected individual adopts the opinion of a randomly chosen neighbor [see Fig. \ref{fig:binary_models}(a)]. 
It is equally likely that an agent changes from one opinion to the other and vice versa. 
This means that the stochastic process is a martingale where the average magnetization across realizations is conserved [usually defined as $m(t)=1/N\sum_ix_i(t)$, i.e. the opinion average over all $N$ agents]. When considering the exit probability $E(x)$ (defined as the probability that the system ends up with $x_i = 1$ for all agents, when starting from a homogeneous initial condition where a fraction $x$ of agents has $x_i = 1$), the conservation law implies the linear relation $E(x)=x$ \cite{sood_voter_2005}. The conservation law does not hold on heterogeneous interaction networks, as we discuss below.

Starting from a disordered initial condition, voter dynamics tends to increase the order of the system by coarsening. 
Because spontaneous opinion changes are not allowed, any state where all individuals share the same opinion (consensus) is absorbing. 
In a finite system, consensus is always reached asymptotically. 
On a complete network, the mean time to consensus can be computed exactly for any value of the initial magnetization, scaling linearly with the population size $N$ \cite{slanina_analytical_2003}.
However, in the thermodynamic limit (infinite population), reaching consensus depends on the underlying network topology or lattice dimension.
In the following we distinguish between different types of networks.

\paragraph{Regular lattices.}

One of the first studies of the voter model on regular lattices in the physics community goes back to \textcite{frachebourg_exact_1996} [for earlier work in mathematics see \cite{holley_ergodic_1975, cox_diffusive_1986, cox_coalescing_1989}]. \textcite{frachebourg_exact_1996} show that the asymptotic behavior of the density of active interfaces (links connecting opposing opinions) depends strongly on the dimension $d$. 
For $d < 2$, the density decays polynomially to zero. 
For $d = 2$, decay is logarithmic, but consensus is still reached asymptotically. 
For $d > 2$, the density remains positive, indicating indefinite coexistence of opinions; consensus is never achieved in an infinite system.

This behavior can be understood via a duality with coalescing random walkers, i.e., random walkers who coalesce and thus move together whenever they meet  \cite{holley_ergodic_1975}.
In the duality, two voters sharing the same opinion because they copied from a common neighbor in the past corresponds to a coalescing path (i.e, two random walkers met and became one) if looked backward in time. 
In $d > 2$, random walks are transient, meaning diffusing interfaces have a finite chance of never meeting and annihilating. 
For $d \leq 2$, random walks are recurrent, and this probability approaches zero, ensuring the interfaces vanish.
In finite lattices of size $N$, fluctuations lead to absorption at finite times. 
The mean time to consensus, $T_N$ \cite{cox_coalescing_1989}, scales as: $T_N \propto N^2$ for $d = 1$, $T_N \propto N \ln N$ for $d = 2$, and $T_N \propto N$ for $d > 2$.

The voter model has been studied as a member of a family of spin models with up/down symmetry by \textcite{de_oliveira_isotropic_1992,drouffe_phase_1999, molofsky_local_1999}. This family includes the conventional Ising model and the majority vote model. We also point to the work by \textcite{dornic_critical_2001}, who introduced the voter universality class. Langevin descriptions of models with $\mathbb{Z}_2$ symmetry and two absorbing states have been developed in \cite{al_hammal_langevin_2005}.

For $d=2$, \textcite{dornic_critical_2001} and \textcite{al_hammal_langevin_2005} further showed that, unlike the Ising model, the voter model does not show surface tension. This means that “bubbles” of individuals with a given opinion do not slowly shrink to consensus, but rather regions with agents of one opinion gradually disintegrate as their boundaries become rougher in a diffusive process.
When noise-reduction is introduced, the surface tension emerges and drives the ordering dynamics \cite{dallasta_effective_2007}. Similarly, the coarsening of voter models with inertia or multiple states can also be driven by surface tension \cite{stark_decelerating_2008, dallasta_algebraic_2008}. 

We note that the noisy voter model, i.e., a two-state voter model with the possibility for spontaneous opinion changes, remains exactly solvable on hypercubic lattices in all dimensions \cite{de_oliveira_linear_2003}. The model turns into the conventional voter model as the noise rate $\varepsilon$ approaches zero. \textcite{de_oliveira_linear_2003} obtained the corresponding critical exponents $\beta=0$, $\gamma=1$, and $\nu=1/2$ for $d>2$, with logarithmic corrections at the upper critical dimension $d_c=2$. For $d=1$ de Oliveira finds $\beta=0$, $\gamma=1/2$, and $\nu=1/2$.

\paragraph{Static networks}

On regular lattices, the order of agent updates and the direction of opinion copying are irrelevant. Specifically, if an agent A is chosen randomly and a random neighbor B of A is selected, the dynamics remain the same whether A copies the state of B or B copies the state of A.
On a network with a heterogeneous degree distribution, the detailed sequence is important, as high-degree nodes are more likely to be selected as neighbors than low-degree nodes. This is akin to the distinction between birth-death and death-birth processes in evolutionary dynamics (see e.g. \cite{fu_evolutionary_2009,kaveh_duality_2015}).
Three choices are possible: i) in the direct voter model \cite{wu_information_2004,suchecki_conservation_2004}), a randomly chosen agent copies one of its neighbors; ii) in the reverse voter \cite{garrido_modeling_2005} a randomly chosen agent transmit its opinion to one of its neighbors; and iii) in a link-update dynamics \cite{suchecki_conservation_2004}, a link is selected at random and then one of the two nodes (chosen at random) copies the state of the other. 
These different definitions result in different dynamics. 
For example, the conservation of overall magnetization is guaranteed for dynamics ii) and iii), but not for the direct voter process. However, a different conservation law holds for a degree-weighted density of up- (or down-) spins  \cite{sood_voter_2005,sood_voter_2008}.

The mean consensus time of the direct voter model on a heterogeneous network can be computed analytically for networks with scale-free degree distribution with exponent $\gamma$ \cite{barabasi_emergence_1999}. 
When $\gamma>3$, the mean consensus time scales linearly with the system size $N$, and sublinearly for $\gamma \le 3$. Extensive numerical simulations confirm this behavior \cite{castellano_comparison_2005,sood_voter_2005,suchecki_conservation_2004}.  The dependence on the exponent $\gamma$ can be related to the fact that the second moment of the degree distribution diverges for $\gamma \le 3$. This impacts many dynamical processes on scale-free networks, from epidemics to synchronization \cite{barrat_dynamical_2008}.  
For link-update dynamics, the mean consensus time scales linearly with $N$ for any degree distribution. Linear scaling is also found for the reverse-voter dynamic on scale-free networks with $\gamma > 2$ \cite{sood_voter_2008}, while such a scaling is logarithmic for $\gamma = 2$ and sublinear for $\gamma < 2$
\cite{castellano_effect_2005}.

On small-world networks \cite{watts_collective_1998}, one finds a metastable state with coexisting domains of opposite opinions, whose typical length scales as $1 / p$ \cite{vilone_solution_2004}, where $p$ is the fraction of long-range connections of the small-world rewiring. In the thermodynamic limit, such a state prevents the complete ordering of the voter model. 
\textcite{suchecki_voter_2005} carried out a comprehensive study of the voter model as a function of the network topology, showing how the ordering dynamics depend on the effective dimensionality and degree distribution of the underlying networks.
The consensus time, in particular, has been shown to be independent of the network's structure \cite{baxter_fixation_2008}.

While the voter model offers simplicity and analytical tractability, its realism is limited (see however \cite{fernandez-gracia_is_2014}). 
The next sections will explore numerous extensions designed to capture diverse mechanisms that drive consensus or opinion fragmentation, together with other modeling approaches. 
For a more specific examination of voter dynamics, readers are directed to the comprehensive reviews by   \cite{redner_reality-inspired_2019,jedrzejewski_statistical_2019}, while \textcite{zschaler_adaptive-network_2012} provides an in-depth analysis of adaptive voter dynamics

\subsubsection{\texorpdfstring{Nonlinear and $q$-voter models}{Nonlinear and q-voter models}} %
\label{subsec:q_voter}

The spreading dynamics of opinions is often more complex than that of viruses, rumors, or information, as spreading opinions requires not only informing people, but also persuading them. 
This, in turn, frequently requires interactions with multiple people rather than a single individual, as in the original voter model. 
This mechanism is often modeled by complex contagion \cite{granovetter_threshold_1978,watts_simple_2002,centola_complex_2007,ruan_kinetics_2015},
in contrast to simple contagion, where a single individual can be sufficient to induce change.

To model complex contagion in opinion dynamics,  \textcite{castellano_nonlinear_2009} proposed the `nonlinear $q$-voter model'. 
In this model, a focal (or target) agent is influenced by a group of $q$ agents chosen at random from all its neighbors. The influence process only affects the focal agent if all $q$ agents hold an opinion different from that of the focal agent, as illustrated in Fig. \ref{fig:binary_models}. In particular this requires the $q$ neighbors to agree with one another.
The elementary update consists of the following steps: 
(1) selecting a random agent; 
(2) selecting $q$ of their neighbors (either with repetition \cite{castellano_nonlinear_2009} or without \cite{nyczka_phase_2012}); 
(3a) If all $q$ neighbors are in the same state, then the focal agent copies this state, or 
(3b) If the $q$ neighbors are not in agreement, the focal agent flips state with probability $\epsilon$. 
In the limit of $N\rightarrow\infty$ such a model can be defined by the flip rate
\cite{castellano_nonlinear_2009}
\begin{equation}
f(x)=x^q +\epsilon \left[1-x^q-(1-x)^q \right],
\label{eq:qvm_flip}
\end{equation}
where $x$ is the fraction of neighbors in the state opposite to that of the focal agent (the term multiplying $\epsilon$ is the probability that $q$ agents, chosen at random with repetition among the neighbors of the focal agent, are not all in agreement).
The standard voter model is obtained for $q=1$. In this latter case, the value of $\varepsilon$ is immaterial, as case (3b) is never activated (because a single neighbor is always in agreement with itself). In such a case, the flip rate reduces to $f(x)=x$. We note that the flip rate in Eq.~(\ref{eq:qvm_flip}) preserves the absorbing states $x=0$ and $x=1$, for all $\epsilon$. Therefore, the exit probability can be determined as in the linear voter model. The result is a nonlinear, S-shaped curve that depends on system size: the larger the system, the steeper the slope at $x=1/2$ \cite{castellano_nonlinear_2009}.

When $\epsilon>\epsilon_c$, the population quickly reaches consensus, with the average consensus time scaling logarithmically with the population size. In contrast, for $\epsilon<\epsilon_c$, the system can become trapped in a polarized state, where one group favors the $+1$ opinion and the other favors $-1$. 
In this regime, the time required to escape polarization and reach full consensus grows exponentially with the population size \cite{castellano_nonlinear_2009}.

Confusingly, different versions of this model travel in the literature under names such as `nonlinear voter model', `$q$-voter model', and `nonlinear $q$-voter model'. 
The following remarks aim to distinguish between these different models and clarify the terminology.
 
\paragraph{Nonlinear \texorpdfstring{$q$}{q}-voter, \texorpdfstring{$q$}{q}-voter, and nonlinear voter models.}

In the conventional voter model, a focal agent is selected for potential update. 
The agent changes state if and only if a randomly selected neighbor is in the opposite opinion state. 
Thus, once the focal agent is selected, the probability for it to switch state is given by the fraction of neighbors, $x$, who are in the opposite state. 
This is the `linear' voter model (the dependence of the flip rate on $x$ is linear). 
The term `nonlinear voter model' was proposed by  \textcite{durrett_nonlinear_1991} to describe extensions of the voter model in which the propensity for an agent to change states depends on the fraction of neighbors in the opposite state via a nonlinear function $f(x)$. 
This includes, e.g., threshold models, in which changes are only possible if the fraction of neighbors in the opposite state exceeds a given value \cite{watts_simple_2002,watts_influentials_2007,nowak_symmetrical_2020,nowak_threshold_2022}.  It also includes models in which the propensity to change depends polynomially on the fraction of neighbors in the opposite state. Therefore, the term `nonlinear voter model' encompasses a large class of extensions of the original voter model. 

\textcite{castellano_nonlinear_2009} coined the term ``nonlinear $q$-voter model'' for dynamics defined by the flip rate in Eq.~\eqref{eq:qvm_flip}. 
In the original formulation, the parameter $q$ is an integer, as it represents the number of agents influencing the focal agent \cite{castellano_nonlinear_2009}. The authors note though that the flip rate in Eq.~\eqref{eq:qvm_flip} is well-defined for any real value of $q>0$, and thus it can be regarded as a general definition of the nonlinear $q$-voter model \cite{castellano_nonlinear_2009}. 

Despite this observation, there seems to be a tendency in the literature to refer to models with real values of $q>0$ as ``nonlinear voter models''~\cite{min_fragmentation_2017,peralta_analytical_2018,min_multilayer_2019,ramirez_ordering_2024,schweitzer_nonlinear_2009, jedrzejewski_statistical_2019}. 
The choice of non-integer $q$ is partly motivated by social impact theory \cite{latane_psychology_1981}, see also \cite{nowak_private_1990}. 
Latan\'e, in particular, uses empirical data to conclude that social impact depends sublinearly on group size (i.e., $q<1$) in a number of different situations.
The special case $\varepsilon=0$, which is equivalent to $f(x)=x^q$, is sometimes referred to as the ``$q$-voter model'', see e.g. \cite{nyczka_phase_2012,jedrzejewski_pair_2017,cox_complete_2025}.

The related term ``noisy nonlinear voter models'' can also be found in the literature. \textcite{peralta_analytical_2018}, e.g., uses this term to describe a model in which $f(x)=x^q$, but where spontaneous opinion changes are {\em always} possible with a certain rate, even if all neighbors of a focal agent agree with one another. 
Such a model then does not have any absorbing states, in contrast with the model by \textcite{castellano_nonlinear_2009}. 
The same model was also introduced as the $q$-voter with independence \cite{nyczka_phase_2012,jedrzejewski_pair_2017}.

In summary, there is some variation in the use of the terms ``$q$ voter model'', ``nonlinear voter model'', and ``nonlinear $q$-voter model''.  
Therefore, when reading a particular contribution to the field, we recommend not to rely too much on this terminology, but to carefully examine the precise form of propensities for updating, the circumstances under which spontaneous changes in opinion are possible, and whether absorbing states are present or absent.

\paragraph{Nonlinear \texorpdfstring{$q$}{q}-voter model (with integer \texorpdfstring{$q$}{q}).}
\textcite{castellano_nonlinear_2009} studied their model on the complete graph using rate equations. 
On random graphs, the model has been analyzed with so-called heterogeneous mean-field theory in \cite{moretti_mean-field_2013}. 
It has been shown that there is a paramagnetic-ferromagnetic phase transition driven by $\epsilon$, and that the nature of this phase transition depends on $q$. 
When a complete graph is replaced by a random regular graph, the phase transition changes significantly, especially for $q \ge 4$. 
In the mean-field case, a mixed phase separates ordered and disordered states, but on random regular networks even minimal disorder eliminates this mixed phase, leading to a single transition point \cite{moretti_mean-field_2013}. 
In more complex networks with different degree distributions, the influence of the interplay between $q$ and the network structure is not yet understood. 

The binary $q$-voter model was later generalized to the threshold $q$-voter model \cite{nyczka_anticonformity_2013, vieira_threshold_2018}, in which a minimum of $q_0$ agents out of a group of $q$ ($0 \leq q_0 \leq q$) must share the same opinion to influence the focal agent. 
In this model, consensus is still possible, but only for certain parameter values. 
As shown by \textcite{vieira_threshold_2018}, the model exhibits rich phase diagram, including both continuous and discontinuous order-disorder phase transitions, as well as transitions from fluctuating phases to absorbing states.

In the original binary $q$-voter model and most of its modifications, positive and negative opinions are symmetric, but recently two asymmetric versions introducing a bias toward one of the opinions have been proposed and studied on the complete graph \cite{doniec_modeling_2025,mullick_social_2025}. In both cases, the exit probability is no longer linear or S-shaped and exhibits a highly nontrivial form.

\paragraph{Nonlinear voter model (real valued \texorpdfstring{$q$}{q}).}
Voter models with real valued exponents $q$ have been studied for example in \cite{min_fragmentation_2017,min_multilayer_2019, raducha_emergence_2020, peralta_analytical_2018, vazquez_agent_2010, llabres_universality_2025}. Different types of behavior are seen in nonlinear voter models with $q<1$, $q=1$ and $q>1$, respectively \cite{ramirez_ordering_2024}. This is the case both with all-to-all interaction and on networks. For $q<1$, the dynamics favor minority opinions, and thus promote co-existence, leading to long-lived states. The extinction of opinions driven by fluctuations works against this deterministic pull towards coexistence. 
For $q=1$, one has the standard voter model, in which there is no underlying deterministic drive, i.e., in infinite populations, the dynamics favor neither coexistence nor extinction. For $q>1$, the deterministic flow in infinite populations favors majority opinions and therefore moves the system towards extinction. Noise in finite populations then plays virtually no role in the ordering.

\paragraph{Sznajd model.}
The nonlinear $q$-voter in the sense of \cite{castellano_nonlinear_2009} is based on the principle of unanimity, inspired by the experiments on conformity, which showed that unanimous opinion of a group is significantly more influential than any form of majority  \cite{asch_opinions_1955}.
A similar unanimity-driven mechanism had already appeared earlier in what is now known as the Sznajd model \cite{sznajd-weron_opinion_2000}. The core idea of the Sznajd model is that “\textit{united we stand, divided we fall}”: a group of neighboring agents that agree can collectively convince their neighbors, whereas a group that disagrees has no persuasive power. In its original one-dimensional formulation, the influence group consists of a pair of adjacent agents \cite{sznajd-weron_opinion_2000}; if the two agents share the same opinion, they attempt to impose it on their neighbors. When the model is implemented on more complex networks or higher-dimensional lattices, various generalizations of this rule have been proposed, and there is no single, universal definition of how the influence group should be selected on arbitrary graphs. This is also the primary difference between the $q$-voter model and the Sznajd model, namely the way the influence group of $q$ agents is chosen. In the $q$-voter model, the $q$ agents are randomly sampled from the set of all agents directly connected to the target agent, as illustrated in Fig. \ref{fig:binary_models}(b). By contrast, the Sznajd model typically requires the influence group to form a contiguous cluster of neighboring agents, such as an adjacent pair in one dimension or a local plaquette or clique in higher dimensions. For an overview of these variants, we refer the reader to the comprehensive review \cite{sznajd-weron_review_2021}.

\paragraph{\texorpdfstring{$q$}{q}-voter models with mechanisms other than conformity.}

The original $q$-voter and nonlinear voter models include only conformity (imitation) as a possible response to social influence. However, social psychologists have identified other types of social responses, such as anticonformity and independence. Therefore, the (nonlinear) $q$-voter model has been extended in these directions (see, e.g., \cite{jedrzejewski_statistical_2019}). In these extended models, consensus is no longer an absorbing state, and thus they will be presented in \sref{subsec:opinion_fragmentation}.

\subsubsection{Majority models}
\label{sec:majority_rule}

There are two families of models that rely on the majority principle, according to which individuals update their opinions in response to the majority opinion within their local neighborhood. The first family is based on the Galam majority model \cite{galam_majority_1986,galam_social_1990}, whereas the second is based on the majority vote model  \cite{de_oliveira_isotropic_1992}. 
While the first family leads to consensus, at least in its basic version, the second does not and will therefore be discussed in \sref{sec:conformity_and_anticonformity}.

\paragraph{Galam model.}

The original Galam model was developed to describe a hierarchical election process \cite{galam_majority_1986}. In its original form, agents do not change their opinions. Inspired by the real-space renormalization group, the model represents an iterative process in which the number of agents is reduced at each step until a single representative of the entire population is selected \cite{galam_social_1990}. Interestingly, an almost identical idea appears in a paper by Shimbel from 1952 \cite{shimbel_communication_1952}, who drew an analogy between a hierarchical communication network and a pyramidal election system. Later, Galam reformulated his original model to describe opinion dynamics using two different approaches: repeatedly enlarging group sizes \cite{galam_fashion_2005} and reshuffling agents \cite{galam_minority_2002}.

The original model with enlarging groups was first presented in the French newspaper \textit{Le Monde} under the title \textit{Les réformes sont-elles impossibles?}, to illustrate how broad democratic consultation can paradoxically lead to the rejection of reforms, even when a large majority initially supports them. Suppose that a society must decide whether or not to introduce a reform. One faction favors maintaining the status quo (opinion $-1$), while the other favors the reform (opinion $+1$). In the first step, the population is divided into groups of size $r$. All individuals within each group update their opinions by adopting the majority opinion in that group. If no majority emerges, the status quo (option $-1$) is chosen. The process then continues by merging the previously formed groups, each consisting of $r$ individuals. After the groups are merged, a new majority decision is made among the resulting $r^2$ individuals. This process is repeated until a consensus is reached. A similar approach was later applied to fashion trends \cite{galam_fashion_2005}, with one difference: when there is no majority, a novelty (option $B$) is chosen instead of the status quo.

\paragraph{Majority rule model.}

The model now widely known as the Galam majority model was originally introduced in the context of competing species dynamics \cite{galam_competing_1998}, and later reformulated to describe the spread of minority opinions in public debates \cite{galam_minority_2002}. 
Independently, an almost identical model was introduced by \textcite{krapivsky_dynamics_2003}. 
In this model, at each time step, a randomly chosen group of $G$ individuals adopts the state of the majority within the group  \cite{galam_minority_2002,krapivsky_dynamics_2003}. 
The original majority rule model includes only conformity, but was modified to also cover other types of agents, such as contrarians \cite{galam_contrarian_2004,toth_deviations_2022} and inflexibles (stubborn individuals) \cite{galam_role_2007}. 
In his models, Galam considers both odd and even values of $G$ \cite{galam_sociophysics_2008_1}. 
In contrast, Krapivsky and Redner assume that $G$ is always odd \cite{krapivsky_dynamics_2003}, which ensures that a local majority always exists. 

This seemingly minor difference has significant consequences due to the rule that Galam applies when there is no local majority, which can occur when $G$ is even. For instance, to model political choices, Galam assumes that the status quo is more likely to prevail \cite{galam_minority_2002}. 
Again, let $-1$ represent the current law and $+1$ represent a proposed reform, then in the case of an equal number of agents who hold opinions $-1$ and $+1$ within a group, all agents in that group will adopt the opinion $-1$ \cite{galam_minority_2002}. 
In such a scenario, the minority opinion can easily spread throughout the society. For instance, when $G=4$, achieving consensus in which all agents hold opinion $+1$ requires the initial proportion of agents with opinion $+1$ to be greater than $0.77$. Another idea was proposed in \cite{galam_heterogeneous_2005}. In this model, in the case of a tie, the group adopts one opinion or the other with complementary probabilities, representing a social bias.

For odd $G$, if there are no contrarians and no inflexible agents, the system reaches a consensus in which all agents have opinions $+1$ if the initial fraction of agents in state $+1$ is more than half of all agents, and to a consensus in which all agents have opinion $-1$ otherwise. This consensus is reached in a time that scales with the system size $N$ as $\ln{N}$ \cite{krapivsky_dynamics_2003}. 
The majority rule model with odd $G$ was later generalized to the majority-minority model, where agents in the group adopt the local majority state with probability $p$ or the local minority state with probability $1-p$ \cite{mobilia_majority_2003}. Thus, the majority rule model corresponds to $p=1$. The probabilistic choice between adopting the majority or minority opinion is reminiscent of the majority-vote model, which will be discussed in  \sref{sec:conformity_and_anticonformity}. 

\paragraph{Homophilous majority rule.}

Another interesting and more recent modification of the majority rule model is the homophilous majority rule \cite{krapivsky_divergence_2021}, which incorporates homophily, the tendency of individuals to ignore the opinions of those who are not like them. 
In this model, the majority rule dynamics is investigated in a population with two classes of people, each with two opinion states $\pm 1$. 
This means that each agent is described by two binary variables: a dynamic one, which is an opinion, and a static one, which is a class. 
An elementary update consists of the following steps: 
(1) a group of agents is randomly selected from the entire population; 
(2a) if all agents in the group belong to the same class, they adopt the majority opinion; 
(2b) if the group includes agents from different classes, they adopt the majority opinion with probability $\epsilon$, otherwise no opinion change occurs. 
It is seen that if there is only one class, then this algorithm would be identical to the one in the majority rule model \cite{krapivsky_dynamics_2003}. 

In the homophilous majority rule \cite{krapivsky_divergence_2021}, the rule that requires unanimity and allows random changes in the lack of unanimity is the same as in the original $q$-voter model \cite{castellano_nonlinear_2009}. 
However, the crucial difference is that in the $q$-voter model, this rule is related to opinions that change over time, while in the homophilous majority rule, it is related to classes that do not change. 
Nevertheless, the main result is the same as in the $q$-voter model. 
Consensus is achieved in a time that scales logarithmically with the system size only when $\epsilon>\epsilon_c$. 
For $\epsilon<\epsilon_c$, the population may become trapped in a polarized state, with one group favoring the $+1$ opinion and the other favoring $-1$. 
In this case, the time required to escape polarization and reach consensus increases exponentially with system size. Furthermore, the critical value $\epsilon_c=1/9$ for the size group $3$ in both models is significant, although the update rules of the opinion dynamics are significantly different. The larger groups were not considered in \cite{krapivsky_divergence_2021}, but they were in \cite{castellano_nonlinear_2009}. It would be interesting to see in the future if, for larger groups, the critical value of $\epsilon$ is also the same in both models.

Recently, majority rule models have been investigated on hypergraphs \cite{noonan_dynamics_2021}. Focusing on interaction groups of size three, \textcite{noonan_dynamics_2021} consider various hypergraph models, from tripartite hypergraphs to modular and heterogeneous hypergraphs. The majority rule dynamics is recast into Fokker-Planck equations, allowing for the prediction of transient dynamics toward consensus, with numerical simulations showing strong agreement with theoretical predictions for large populations.

\paragraph{Neighboring consensus domains.}

Majority-like models have also been studied within the context of the size of the neighboring domains.
For instance, in a one-dimensional lattice, an individual's decision to change opinion may depend on the size of neighboring consensus domains, i.e., the number of neighboring nodes sharing the same opinion, which quantifies the strength of social pressure \cite{biswas_model_2009}.
Either individuals at the boundary between opposing domains follow the opinion of the larger domain, or individuals sandwiched between two domains of the same, opposing polarity flip their opinion, regardless of domain size.
In its simplest, disorder-free form, the model spontaneously reaches a homogeneous state (full consensus), but exhibits coarsening exponents drastically different from other known one-dimensional models. 
When disorder is introduced, the system undergoes a phase transition from a completely homogeneous to a heterogeneous, disordered state. 
A stochastic variation of the aforementioned model has been introduced by \textcite{biswas_opinion_2013}, where the probability of opinion change is determined by the size of neighboring consensus domains, weighted by a factor that biases the dynamics toward one opinion. This model yields a step function behavior in the exit probability, a pattern previously unobserved in one-dimensional opinion dynamics models, and exhibits lower exponent values for the persistence function during coarsening.

\subsubsection{Social impact theory}

Consensus in a population of agents with binary opinions can be modeled using frameworks based on social impact theory \cite{latane_psychology_1981}. The foundational work \cite{nowak_private_1990} extended the original concept from a single act of influence to a continuous group process studied through simulations. In this model, each agent is characterized by four attributes: attitude, persuasive strength, supportive strength, and position in the social structure. Attitudes, represented as binary variables, evolve from random initial states toward near-unanimous consensus. Reformulated in the language of statistical physics \cite{lewenstein_statistical_1992}, the model attracted considerable attention among physicists \cite{holyst_phase_2000,stauffer_social_2001,bordogna_statistical_2007}.
Recently, renewed interest in social impact models has led to several extensions \cite{bancerowski_multi-choice_2019,mansouri_phase_2020,malarz_phase_2023,malarz_fine_2025}. These include generalizations to multiple opinions \cite{bancerowski_multi-choice_2019} and studies of fully differentiated societies where each agent initially holds a unique opinion \cite{malarz_phase_2023,malarz_fine_2025}. Introducing a Müller’s ratchet–like mechanism, where extinct opinions cannot reappear, shows that consensus is eventually reached for any positive temperature, though convergence can be extremely slow. In the deterministic limit (zero social temperature), systems may instead freeze into clusters of differing opinions or enter oscillatory limit cycles \cite{malarz_fine_2025}.

\subsubsection{Opinion assimilation}
\label{subsubsec:opinion_assimilation}

So far, we have discussed the emergence of consensus in the context of models with discrete, spin-like opinions. 
However, consensus dynamics has also been extensively studied in models with continuous opinions \cite{abelson_mathematical_1967,french_formal_1956,degroot_reaching_1974}.
In these models, the dynamics does not occur via direct copying of another agent's state. Instead, agents change their opinions gradually, and as discussed further in \sref{subsubsec:bounded_confidence}, the population can converge towards a real-valued consensus opinion within the range of the initial opinions in the population. 

Models with continuous opinion states can be categorized into two types: discrete-time and continuous-time models. 
Discrete-time models typically rely on an opinion-averaging process. At each time step, agents update their opinions by taking a weighted average of their own opinions and those of their interaction partners. 
In continuous time, this translates into a scheme of \emph{diffusive coupling}, $\dot{x}_i=\sum_j A_{ij}(x_j-x_i)$, which leads to steadily shrinking opinion differences between interacting individuals.

\paragraph{Discrete time.} 
Building on the earliest formalizations of assimilative social influence, introduced by \textcite{french_formal_1956} and \textcite{harary_criterion_1959}, the classic DeGroot model described opinion assimilation as a social learning process through repeated opinion averaging in discrete time \cite{degroot_reaching_1974}. 
Specifically, in the DeGroot model, individuals are represented as nodes in a social network, with edges weighted to quantify the influence that neighbors exert on each other’s opinions.
The edge weight $A_{ij}\geq0$ indicates the confidence level that individual $i$ puts in $j$, and $A$ is a row-stochastic matrix, i.e., $\sum_j A_{ij} = 1 \quad \forall i$.
Individual $i$ updates their opinion as a linear averaging process according to $x_i(t+1) = A_{ii}x_i(t)+ \sum_{j\neq i}A_{ij}x_j(t)$, where $A_{ii}$ is sometimes referred to the self-weight. 
In the DeGroot model, consensus is always reached on \textit{strongly connected networks}, where there is a path of influence between any two individuals. 
The speed of convergence is governed by the second-largest eigenvalue $\lambda_2$ of $A$ where $|\lambda_2 |<1$ for stochastic matrices. 
More precisely, one finds $|x_i(t)-x_c|\leq C|\lambda_2|^t$ for all $i$, and where $x_c$ is the final consensus opinion and $C$ is a constant. 

A closely related assimilative opinion update rules proposed in the economics literature is the asymmetric gossip model \cite{acemoglu_opinion_2011,jo_finite-size_2021}. Here, the interaction between two agents $i$ and $j$ only leads to an update of opinion $x_i$, while $j$'s opinion stays constant. Namely, $x_i(t+1)=(1-\epsilon) x_i(t)+ \epsilon x_j(t)$, where $\epsilon$ defines the learning rate of agent $i$. Note that for $\epsilon=0$ the opinions never change, and for $\epsilon=1$ the model reduces to the multistate voter model. For $0 < \epsilon < 1$ the population still converges to consensus, but the path to the steady state differs from that of the multistate voter model.

Kinetic exchange models [see \textcite{biswas_social_2023} and \textcite{toscani_kinetic_2022} for a more detailed review]  are also based on the assimilation principle, but formulated by drawing an analogy with the kinetic theory of ideal gases and models of wealth exchange in economics \cite{ispolatov_wealth_1998, chakraborti_statistical_2000}. 
These models simplify social interactions by assuming that the resulting change in an individual's opinion, $x_i(t)$, follows a linear exchange relation with the opinion of another agent. 
The opinion values are typically continuous and bounded (e.g., $|x_i(t)| \leq 1$), with non-linearity introduced by enforcing these bounds. 
Unlike wealth exchange, opinion exchanges are generally not conserved; rather, the interaction causes opinions to either converge or drift apart.

The Lallouache-Chakrabarti-Chatterjee-Chakrabarti (LCCC) model \cite{lallouache_opinion_2010}, featuring non-negative interaction, exhibits a spontaneous symmetry-breaking transition as a function of the opinion fraction $\lambda$ that each agent retains in an exchange (also called the conviction parameter), with a very small critical exponent $\beta \simeq 0.1$.
Above the critical point [$\lambda_c \approx 2/3$, which can be demonstrated at the mean field level \cite{biswas_phase_2011}], one can observe consensus, while below it one obtains a neutral/indifferent absorbing state (all opinions decay to zero), which contrasts with expected social fragmentation. 
An extension of the LCCC model introduces stochasticity in the self-retention (or conviction) parameter $\lambda$ \cite{biswas_mean-field_2011}. 
Also this case typically results in consensus or the neutral absorbing state, with the stochasticity only modifying the system's critical behavior rather than altering the fundamental ordering tendency.
Another interesting variation considers the distinction between public expression and private belief, linked by a tolerance $\delta$ \cite{roy_opinion_2022}. While the core dynamics of this model favor consensus, the private layer introduces persistence and complexity, often leading to long transient periods of coexistence. The distinction between public expression and private belief (i.e., between public and private opinions) can also be introduced within the family of the so-called expressed-private opinions (EPO) models, as recently reviewed by \textcite{kaminska_impact_2025}.

\paragraph{Continuous time.}
\textcite{abelson_mathematical_1967} introduced a continuous-time counterpart to the models proposed by \textcite{french_formal_1956} and \textcite{degroot_reaching_1974}. 
The linear Abelson model captures opinion assimilation through a diffusive coupling scheme, resulting in a system of coupled ordinary differential equations of the form $\dot{x}_i = \sum_j A_{ij} (x_j - x_i)$, where $A_{ij}$ is a non-negative, but not necessarily row-stochastic, matrix of influence weights~\cite{proskurnikov_tutorial_2017}. The model can be written more compactly using the graph Laplacian $L_{ij}$, as $\dot{x}_i = \sum_j -L_{ij} x_j$, where \(L_{ij} = -A_{ij}\) for \(i \neq j\), and \(L_{ii} = \sum_k A_{ik}\).

The Abelson model exhibits convergence properties similar to those of the DeGroot model. First, the model always converges to a specific opinion distribution $\mathbf{x}^\infty=\mathbf{V}^\infty \mathbf{x}^0$, where we have the linear operator $\mathbf{V}^\infty=\lim_{t\rightarrow\infty}e^{-\mathbf{L}t}$ and $\mathbf{x}^0$ denotes the initial opinion distribution. Second, consensus is reached if and only if the underlying social graph is strongly connected \cite{proskurnikov_tutorial_2017}.
The Abelson model has been extended to non-linear dynamics of the form $\dot{x}_i = \sum_j A_{ij} g(x_i,x_j) (x_j - x_i)$, to capture more complex types of social influence \cite{matveev_stability_2013, lin_state_2007}. 

\subsubsection{Effects of temporal interaction patterns and network structure}
\label{subsubsec:consensus-temp}

Over the past few decades, it has become increasingly clear that the simplest network representations fail to capture key aspects of complex systems in general, and social dynamics in particular. 
Social interactions can evolve over time, vary in nature, and extend beyond simple pairwise relationships. 
To better reflect these additional levels of complexity, more advanced network models—such as temporal, multi-layered, and higher-order networks—have been developed. 
These enriched representations influence the dynamical processes that unfold on them, giving rise to new emergent phenomena. 
In this section, we review the emergence of consensus in networks that evolve over time, span multiple layers, and incorporate higher-order interactions.

\paragraph{Temporal networks.}

Temporal networks are networks in which the nodes and/or edges can change over time: 
They play a crucial role in modeling social dynamics, as they capture the time-dependent nature of human interactions \cite{holme_temporal_2013, braha_centrality_2006}.
These networks can be characterized by different probability distributions for node and link durations, or equivalently, by distributions for the time intervals between consecutive activations, i.e., inter-event time distributions.

One of the earliest and most popular models for temporal network generation is the activity-driven model \cite{perra_activity_2012}.
In this model, nodes are assigned an activity rate that determines how often they create connections over time. 
At each time step, active nodes establish links with randomly chosen peers, forming a temporal network.
The degree distribution of the time-integrated network has the same functional form as the activity distribution \cite{starnini_topological_2013}.
Due to its simplicity, the activity-driven model has been the chosen substrate to study several dynamical processes on temporal networks, from information diffusion to epidemic spreading \cite{liu_controlling_2014, rizzo_effect_2014}.
For instance, \cite{zino_consensus_2020} studied the formation of consensus on activity-driven networks, where the update of a node's state occurs by averaging the states of the nodes with which the node is connected at the time of the update. 
Using stochastic stability theory, Zino et al. show that the expected consensus state is primarily influenced by low-activity nodes. 
By applying eigenvalue perturbation techniques, they derive a closed-form expression for the mean-square convergence rate, revealing that moderate heterogeneity can slow down consensus in large networks. 

Similarly, \textcite{moinet_generalized_2018} investigated a generalized voter dynamics ---where a randomly selected agent can either copy the neighbor's opinion (with probability $p$) or impose their opinion to them  (with probability $1-p$)--- on activity-driven networks, where nodes are also endowed with attractiveness, affecting connection reception. 
Using a mean-field approach, they derive equations for state dynamics and evaluate consensus time and exit probabilities, highlighting a symmetry between voter and Moran dynamics and the significant impact of temporal networks on dynamic processes compared to static ones.
\textcite{wang_phase_2022} also studied the behavior of the majority vote model on activity-driven networks with attractiveness. 
They consider either a single-directed (SD) process, where individuals adopt opinions based on actively interactive neighbors, or an undirected (UD) process, with both active and passive neighbors. 
They derive critical noise thresholds via a mean-field approach, showing that UD process achieves a higher consensus level than SD process under the same noise level.

\paragraph{Temporal interaction patterns and aging.}

\textcite{fernandez-gracia_update_2011} introduced methods to simulate voter-like dynamics with arbitrary inter-event time distributions. 
They explore models in which activation probabilities of nodes depend inversely on the time since the last action, distinguishing between exogenous and endogenous updates.  
They find slow ordering for endogenous updates, with a power-law decay in interface density and power-law tails in interevent time distributions, suggesting undefined mean times to reach absorbing states in the thermodynamic limit. 
In further work by \textcite{fernandez-gracia_timing_2013}, each node is updated with a probability that depends on the time since the last event of the node. 
Events can be update attempts (exogenous update) or an actual state change (endogenous update). 
The authors find that these update rules can lead to power-law interevent time distributions. 
For exogenous update and standard voter updated rules the systems fails to reach consensus in large systems. 
Using an endogenous update rule leads to a coarsening process toward consensus.  

\textcite{li_impact_2019} studied how temporal patterns affect consensus formation speed in several continuous opinion dynamics using four empirical datasets, including human and social insect interaction networks. 
They find that static, aggregated networks overestimate consensus speed, weight heterogeneity inhibits consensus relative to unweighted networks, and nodal lifetimes significantly influence outcomes, while temporal patterns like interevent burstiness and edge lifetimes have negligible effects. 
\textcite{artime_dynamics_2017} studied interevent times from Twitter data and how the competition between temporal and network correlation affects the dynamics of voter processes on the network.

The term aging describes situations in which change becomes less likely the longer an agent has been in its current state. 
This type of inertia was first introduced in voter models by \textcite{stark_decelerating_2008}, who found that the time to reach an ordered state can be reduced by slowing down the microscopic dynamics.  
As shown by \textcite{artime_aging-induced_2018}, aging changes the finite-size transition of the noisy voter model to a phase transition of the Ising type that remains in the thermodynamic limit. In the voter model without noise coexistence of opinions is found when the activation probability increases with age \cite{peralta_ordering_2020}. For decreasing activation the authors find that the system can reach consensus or remain trapped in frozen states, depending on the asymptotic value of the activation probability and the rate with which this limiting value is approached. Analytical and numerical treatments of models with aging are further discussed in \cite{peralta_reduction_2020} and \cite{baron_analytical_2022}; in particular simulations of models with continuous-time aging can be carried out using Lewis' thinning algorithm (see \sref{para:thinning}). A linear model with aging is shown to be equivalent to a non-linear model without aging \cite{peralta_reduction_2020}. Aging in threshold models was studied in \cite{abella_aging_2023, abella_ordering_2024}, and a comparative study of aging in opinion formation models can be found in \cite{llabres_aging_2024}. \textcite{min_aging_2025} carried out an analysis of aging in coevolving voter models, and \textcite{llabres_complete_2025-2} showed that "complete aging" (aging affecting both imitation, and spontaneous opinion changes) can enhance consensus formation. A noisy kinetic exchange model with aging is studied by \textcite{vieira_noisy_2024}.

\paragraph{Multilayer networks.}
In complex systems, interactions among entities can take various forms. 
For example, in social networks, individuals may connect through face-to-face interactions, phone calls, text messages, or online communication. 
This added complexity can be effectively captured by multilayer networks, where nodes and edges are distributed across multiple interconnected layers, each representing a different type of interaction or relationship \cite{kivela_multilayer_2014}.
When the same set of nodes is present in all layers, but each layer features different types of edges, the network is referred to as a multiplex network.
Examples of multilayer and multiplex networks include urban transportation networks, where metro lines, bus routes, and bicycle paths form different layers with transfer points connecting them; biological interaction networks, where proteins interact through physical binding, and financial networks, where interbank lending, equity markets, and insurance relationships create multiple interconnected layers.
Such a multilayer structure can profoundly impact the dynamical processes that unfold within it, in particular spreading phenomena \cite{de_domenico_physics_2016}.

A central question in the study of multilayer networks is the extent of redundancy in their representation---specifically, whether certain layers can be effectively merged into others, without loss of information. \textcite{diakonova_irreducibility_2016} studied whether the voter dynamics on a multilayer network can be reduced to an equivalent process on a single-layer network. 
They compute the interface density and the consensus time in the thermodynamic limit, and compare numerical results on multilayer networks with analytical predictions on equivalent single-layer networks, obtained through various possible aggregation procedures.
They show that these aggregation methods fail to capture key nonlinear effects, particularly the prolonged lifetime of finite-size multiplexes at low interlayer connectivity, indicating fundamental differences between voter dynamics on multilayer networks and single-layer approximations.

Multilayer networks can be used to model coupled dynamical processes, one different process in each layer, interacting through intra-layer connections.
For instance, \textcite{amato_opinion_2017} explored opinion competition in a two-layered network where agents hold potentially conflicting opinions in each layer. 
While each layer individually tends toward consensus on a dominant opinion, interlayer interactions can stabilize a mixed state where both opinions coexist. 
Although finite systems eventually reach consensus due to finite-size effects, a long transient period of coexistence is observed, particularly in sparse, positively correlated networks which facilitate the formation of opinion-sharing groups across layers.
Multilayer networks can also model the coupling of different dynamical processes, such as the spreading of an epidemic in a population and the information awareness to prevent its infection \cite{granell_dynamical_2013}. \textcite{peng_multilayer_2021} proposed a similar model, where the disease spreads in one layer, and two opinions — pro-physical-distancing and anti-physical-distancing compete in the other layer. 
 Individuals adopting pro-distancing opinions are less likely to become infected, while those holding anti-distancing opinions are more susceptible.  
 The authors found that longer opinion durations and increased correlations in network structure, both between layers and within layers, can effectively reduce disease transmission. 
Similarly, \textcite{zino_two-layer_2020} proposed a model where opinion formation is coupled with a collective decision process.
In this model, a communication layer represents how individuals share their opinions, and an influence layer captures the social influence due to observing the actions of others.
The authors found that the diffusion of innovation in the population is strongly influenced by the network structure and the coupling strength between the two coevolutionary dynamics.
Depending on how such factors combine, they observed a range of different complex real-world phenomena, such as 
a paradigm shift with a rapid, widespread adoption of the innovation; the establishment of an unpopular norm where adoption fails despite widespread positive sentiment; and persistent community resistance to innovation, favoring the status quo. 

\paragraph{Higher-order networks.}
While dyadic networks have been the standard modeling workhorse for decades, real complex systems sometimes are better described by  \emph{higher-order networks}, or hypergraphs, capturing relations between more than two nodes.
The presence of higher-order interactions can give rise to novel emergent phenomena.
For instance, they can promote abrupt synchronization switching \cite{skardal_higher_2020} or 
discontinuous transition in social contagion models \cite{iacopini_simplicial_2019}. There are multiple review papers summarizing the knowledge base on higher-order networks and dynamical processes evolving on them \cite{battiston_networks_2020,torres_why_2021,majhi_dynamics_2022,bick_what_2023,wang_epidemic_2024}.

In opinion dynamics modeling, higher-order social interactions can impact consensus formation; for example, group interactions may exert a stronger influence on an individual's opinion than a simple dyadic relationship. 
The work by \textcite{kim_competition_2025} addresses this framework by introducing the group-driven voter model, a polyadic voter process that accounts for group interactions through nonlinear interactions constrained by group sizes. 
This model is an extension of the nonlinear voter model (see \sref{subsec:q_voter} for a full description) to higher-order interactions, where the focal node samples $q$ opinions from other members of the hyperedge and, if all sampled opinions are unanimous, it adopts the observed opinion.
Analytical and numerical results show that the group-driven voter model reaches consensus faster than the standard voter model, with the exit time scaling as $A\ln N$, where $N$ is the network size, and $A$ depends on both the group size and the degree of nonlinearity.

\textcite{neuhauser_multibody_2020}
develop opinion dynamics models for consensus on hypergraphs, showing that nonlinear interaction functions are essential for observing truly higher-order effects beyond simple pairwise scaling.  
They propose a nonlinear three-body consensus model, incorporating peer pressure, showing that, unlike network-based consensus, group effects can shift the system away from the average state, depending on the initial states, structure, and interaction function.  
In modular hypergraphs, these multi-agent interactions create asymmetric dynamics, leading to one polarized cluster dominating the other.
Similarly, \textcite{sahasrabuddhe_modelling_2021}
study nonlinear consensus dynamics on hypergraphs, where different model ingredients are associated with different sociological mechanisms.
Through mathematical analysis and simulations, they explore how nonlinearities influence consensus, on both synthetic block hypergraphs and real-world structures.

\paragraph{Simplicial complexes.}
The voter model has been studied also on simplicial complexes, i.e., structures that include all lower-dimensional faces of a higher-order interaction (e.g., if a triangle is present, its edges must also be included). 
\textcite{horstmeyer_adaptive_2020}
introduce an adaptive voter model on simplicial complexes, incorporating peer pressure among three-node groups in addition to classical voter and rewiring rules. Numerical simulations show that while peer pressure does not alter the fragmentation transition, it accelerates consensus below the transition and speeds up fragmentation above it. Additionally, the model exhibits a multiscale hierarchy, where higher-dimensional simplices deplete before lower-dimensional ones, suggesting similar behavior in other dynamic network models on simplicial complexes.
In pairwise networks, the consensus formation process can be studied using Laplacian matrices. 
On simplicial complexes, \textcite{ziegler_balanced_2022}
examine consensus dynamics by using a generalized Hodge Laplacian, that allows different strengths for higher- and lower-order interactions. 
By leveraging algebraic topology, the analysis reveals that collective dynamics converge to a low-dimensional subspace corresponding to the homology space. 
The authors show that optimally balancing higher- and lower-order interactions accelerates convergence, with the fastest consensus occurring when two-simplices are well dispersed rather than clustered.
A generalized Laplacian for simplicial complexes has also been studied in a paper by \textcite{deville_consensus_2021},
which introduces a generalized dynamical model on a simplicial complex of several consensus and synchronization processes.
The model reveals higher-dimensional analogs of structures found in traditional network-based models.

\paragraph{Temporal higher-order networks.}
Networks can evolve over time and simultaneously incorporate higher-order interactions.
In this case, both the temporal and higher-order nature of the network substrate influence the continuous dynamics of the consensus process.
\textcite{neuhauser_consensus_2021}
address this framework, examining consensus dynamics on temporal hypergraphs with time-dependent, multiway interactions, and comparing them to projections that simplify these features. 
For linear consensus, convergence is slowest in the temporal hypergraph, faster in its pairwise projection, and fastest in the static network. 
In the nonlinear case, the final consensus value can differ significantly from the static system, with a "first-mover advantage" where early-active groups disproportionately influence the outcome.

In the adaptive voter model (\avm), an individual either adopts the opinion of a neighbor or drops this connection to create a new one with an individual having the same opinion \cite{vazquez_generic_2008,chodrow_local_2020}. 
This update rule generates a temporal network in which links are rewired until consensus is reached (see \sref{subsec:coevolution} for more details on the coevolution between opinions and the underlying network, and \sref{subsubsec:aVM} for the \avm in particular). 

\textcite{papanikolaou_consensus_2022} have extended this modeling framework to group interactions (i.e., higher-order networks), distinguishing between influence and split-merge processes; the former assumes the minority adopts the majority opinion with a certain probability, while the latter posits that a minority group breaks away and joins another group with the same majority opinion. 
A threshold parameter dictates whether the influence or the split-merge process is triggered based on the minority group's size.
By assuming heterogeneous mean-field interactions, where at each time step agents are randomly chosen to form a group with a size sampled from a given distribution, the authors show that group interactions enhance the impact of small initial opinion biases, reduce the mean consensus time, and induce a shift in the average magnetization, violating the conservation of initial magnetization, a fundamental characteristic of classic voter models. 

\textcite{papanikolaou_fragmentation_2023} have studied the same model from the point of view of opinion fragmentation (given its close resemblance to the work of \textcite{papanikolaou_consensus_2022}, we discuss its findings here rather than in \sref{subsec:opinion_fragmentation}).
Instead of assuming a preserved group size distribution on an annealed higher-order network, they consider a quenched network where groups actually split and merge. 
In this case, the authors show that a larger threshold (stronger group influence) and higher initial connectivity generally promote opinion fragmentation. 
They observe fragmentation bands---states with varying levels of fragmentation depending on the threshold, showing discontinuous transitions between these states, a different behavior from the phenomenology of the \avm without group interactions (see \sref{subsubsec:aVM}). 
They also propose an analytical explanation for these bands and their abrupt changes, in the limit of sparse hypergraphs.

\subsection{Fragmentation}
\label{subsec:opinion_fragmentation}

As introduced in \sref{subsec:macro-phen}, opinion fragmentation refers to a state where a population evolves into multiple distinct and often non-interacting clusters of opinions. 
This situation is in contrast to both consensus, where all individuals converge to a single, shared opinion, and polarization, where the population splits into opposing, extreme opinion groups. 
Fragmentation, instead, describes a more diverse outcome with two or more dominant viewpoints, and individuals within each cluster may interact more with those sharing similar views.

To discuss fragmentation, it is useful to distinguish three social responses to others' opinions: conformity, independence, and anticonformity \cite{nail_proposal_2013}. 
Conformity, seen in voter or Axelrod models \cite{axelrod_dissemination_1997}, is aligning to someone else's opinion. 
Anticonformity is actively opposing others' opinions. 
Both are dependent behaviors. 
In continuous opinion models, these are typically termed assimilation and distancing, respectively (see \sref{subsec:micro_mech})

An individual may also exhibit independent behavior.
This can manifest in two forms: either an agent spontaneously changes opinion, independent of others [sometimes called "noise" in voter models \cite{granovsky_noisy_1995}], or an agent resists external influence, maintaining its current opinion regardless of interaction partners. 
The latter is also known as zealotry \cite{mobilia_role_2007}, or stubbornness \cite{galam_role_2007}.

Multiple social responses can be modeled by assigning agents fixed types (quenched disorder) or dynamically assigning types at each step with fixed probability (annealed randomness). 
Stubbornness, for instance, can be implemented as fixed "zealots" or as a probability of temporary stubbornness. 
Competition between these response types can induce order-to-disorder phase transitions in opinion dynamics models, where disorder signifies fragmentation and order implies consensus. 
The nature of these transitions and phases depends on whether the quenched or annealed modeling approach is chosen.

Here, we first describe multistate voter models (\sref{sec:MSVM}), showing transient fragmentation despite an imitation mechanism (conformity), then fragmentation as a consequence of spontaneous opinion change and noise (\sref{sec:noise}). 
Models that describe fragmentation as a result of mixed conformity and anti-conformity are discussed in \sref{sec:conformity_and_anticonformity}. 
We then discuss the effects of zealotry in models with discrete and continuous opinion states (\sref{sec:zealotry}).
Finally, in \sref{subsubsec:bounded_confidence} we review how fragmentation arises as a consequence of so-called `bounded confidence'.

\subsubsection{Fragmentation despite conformity: Multistate voter model} 
\label{sec:MSVM}

Opinion fragmentation with conformist imitation can be observed for one-dimensional, yet multistate systems.
In the multistate voter model (\msvm), each agent is on one of $G>2$ different opinions at any time 
\cite{bohme_fragmentation_2012, starnini_ordering_2012}. 
This bears some resemblance to the Potts model \cite{wu_potts_1982}, however, there is no energy function and the dynamics is intrinsically an out-of-equilibrium process.
While the \msvm generally tends towards consensus, it can exhibit transient or quasi-stationary states where multiple distinct opinion clusters coexist. 
At various stages of the dynamics, groups of agents coalesce around different opinions, forming discernible clusters that interact minimally or only at their boundaries. 
The investigation of these coexisting clusters—their number, size distribution, stability, and the time it takes for them to either merge into a larger consensus or persist as fragmented entities—becomes central to understanding the phenomenon of opinion fragmentation.

For all-to-all interactions, \textcite{starnini_ordering_2012} derived the consensus time of the  \msvm in the thermodynamic limit analytically, as well as the mean number of surviving states as a function of time, showing that in the limit $G \to \infty$ the mean consensus time is only $1/ \ln 2$ times bigger than in the binary voter model.
Such behavior also holds on square lattices, 
while the model behaves differently in one dimension.  
\textcite{pickering_solution_2016} also studied the \msvm at the mean-field level, %
finding the eigenvalues and eigenvectors of the Markov transition matrix. 
They found the expected time for the system to transition from $G$ to $G-1$ states, the moments of consensus time, and the expected number of states over time.

The three-state case of the \msvm has been extensively studied by \textcite{vazquez_constrained_2003}, in a model where the two extreme opinions do not interact with each other, but they do interact with the middle opinion. \textcite{vazquez_constrained_2003} solved the model exactly in one dimension, by mapping the dynamics onto a spin$-1$ Ising model with zero-temperature Glauber kinetics.
\textcite{ramirez_local_2022} studied the average density of active links (a measure of local order) in the \msvm, and the average entropy of the distribution of agents across opinions (a measure of global order). 
They showed that the density of active links decays exponentially, with a timescale set by the size and geometry of the underlying network, but independently of the initial number of opinion states. 
The average entropy's decay is instead exponential only at long times, when only two opinions remain in the population. 

The \msvm has also been studied on temporal networks. 
\textcite{bohme_fragmentation_2012}, for instance, studied the multistate voter model unfolding a network where links are rewired depending on the opinion dynamics
They show that such a model can display a fragmentation transition, where the underlying social network breaks into disconnected components. 
In particular, they analytically studied the three-opinion case.  
A \msvm unfolding on a coevolving network has also been studied by \textcite{shi_multi-opinion_2013}. 
In this case, a link connecting two disagreeing agents can either lead to agreement or be rewired, with a certain probability. 
\textcite{shi_multi-opinion_2013} showed that, in the limit of infinitely large networks and initial opinions, the model has infinitely many phase transitions (see \sref{subsec:coevolution} for coevolving opinions and networks).

Fragmentation, as a long-lived metastable state, is also present in the multistate nonlinear voter model. 
There, the predominant transient behavior is the co-existence of multiple opinions, although ultimately the system reaches consensus. 
The ordering dynamics in such models have recently been studied on graphs using the pair approximation and Monte Carlo simulations \cite{ramirez_ordering_2024}. 
We also note that the so-called `$q$-deformed dynamics' have recently been studied in evolutionary games \cite{kitching_q_2024}. 
It is also interesting to note that multistate linear voter models can often be reduced to effective two-state models \cite{redner_reality-inspired_2019, herrerias-azcue_consensus_2019, ramirez_local_2022} using ideas first developed in population genetics \cite{kimura_random_1955, littler_loss_1975}. 
This reduction is not applicable for nonlinear dynamics ($q\neq 1$) \cite{ramirez_ordering_2024}.

\subsubsection{Independent behavior as spontaneous opinion change: The noisy Voter model}
\label{sec:noise}

Independent behavior in the form of spontaneous opinion change has been implemented in what is known as the `noisy voter model' \cite{granovsky_noisy_1995}. 
Mathematically similar models have been proposed (mostly independently) in strongly correlated systems \cite{lebowitz_percolation_1986}, chemical reaction systems \cite{fichthorn_noise-induced_1989,considine_comment_1989}, herding in financial markets and groups of animals \cite{kirman_ants_1993}. 
We stress that the term `noisy' in `noisy voter model' is somewhat misleading, as the original (`non-noisy') voter dynamics in finite populations also defines a stochastic process \cite{holley_ergodic_1975, liggett_interacting_2005}. 
In the context of the voter model, the term `noise' is used specifically to refer to the possibility of spontaneous opinion change.

\paragraph{Transition between near-consensus and polarized phases.}
\label{para:caution}

One main phenomenon brought about by the introduction of spontaneous opinion change in the voter model is a transition between a state in which the population is mostly near consensus and another phase where both opinions coexist, leading to fragmentation. 
The former phase, at low noise rates, shows random switches between near-consensus states, characterized by a bimodal distribution of the number of agents holding a specific opinion. 
At high noise rates, the stationary distribution of agent counts becomes unimodal, with half the agents holding each opinion. 
This transition is a finite-size phenomenon, vanishing in the infinite population limit \cite{kirman_ants_1993, fichthorn_noise-induced_1989, considine_comment_1989, biancalani_noise-induced_2014}. We note that nonlinearities in the dynamics can result in {\em bona fide} phase transitions in the Ising universality class in the thermodynamic limit \cite{dornic_critical_2001,al_hammal_langevin_2005,artime_aging-induced_2018, peralta_analytical_2018, llabres_universality_2025}.

\begin{figure}[tbp]
\centering
\includegraphics[width=1.\columnwidth]{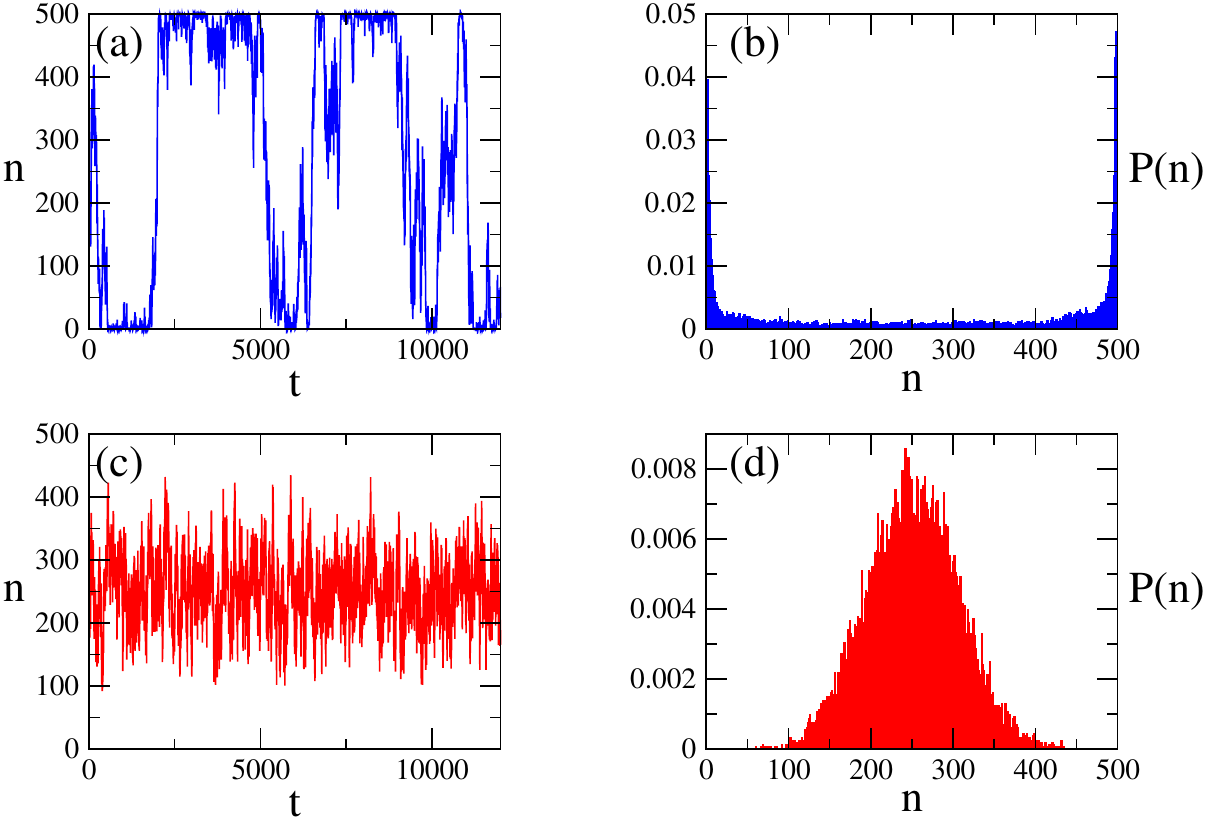}
\caption{
Simulation of the noisy voter model with two opinion states, with $N=500$ agents and all-to-all interaction [for details of the model see \sref{para:noisy_vm}, and/or \cite{biancalani_noise-induced_2014}]. 
The variable $n$ is the number of agents in opinion $1$. 
The left-hand plots (a) and (c) show $n$ as a function of time, for two different parameter choices. The right-hand panels (b) and (d) show the distribution of $n$. Panels (a) and (b) are for a low rate of spontaneous opinion changes ($\epsilon=1/2000$), panels (c) and (d) for a high noise rate ($\epsilon=1/50$). \label{fig:noisy_vm}}
\end{figure}

It is crucial to distinguish this usage of ``unimodal'' and ``bimodal'' from their application to opinion distributions in other contexts. 
While a polarized state is typically associated with a bimodal distribution in the opinion space (as discussed in \sref{subsec:macro-phen}), in the statistical physics literature on noisy voter models, the terms refer to the distribution of the number of agents $n$ in a particular opinion state, $P(n)$. 
The quantity $P(n)$ can be obtained from simulations, and it can be calculated analytically using methods from statistical physics (see \sref{sec:methods}).

As illustrated in \fref{fig:noisy_vm}, a bimodal $P(n)$ reflects a population predominantly found in one of two consensus states, despite occasional switches. 
The overall population, if observed at a random time, will mostly be found in (or near) consensus at one of the two opinion states.
Conversely, a unimodal $P(n)$ signifies stable coexistence of both opinions.
At any one point in time, roughly half of the agents hold opinion 0 and the other half holds opinion 1, representing a polarized state. 
This distinction is vital for interpreting results in noisy voter models \cite{khalil_noisy_2019, khalil_zealots_2018, khalil_zealots_2021, biancalani_noise-induced_2014, llabres_universality_2025, carro_noisy_2016, peralta_analytical_2018}.

This is not a contradiction, but rather a terminological difference: ``bimodal'' and ``unimodal'' describe distributions of opinion states in one context, and distributions of agent counts in another. 
Readers should note the specific type of distribution being referenced by these terms throughout the review.

\paragraph{Networks, and nonlinear noisy voter models.}
Noisy voter models on networks have been studied by \textcite{carro_noisy_2016}. The authors used an annealed approximation for the network structure, and showed that degree heterogeneity affects the location of the noise-induced finite-size transition occurring in the model, the local ordering of the system and correlations in time. The effects of nonlinearity in noisy voter models have been studied by \textcite{peralta_analytical_2018} and \textcite{artime_herding_2019}. The former explicitly confirm that mean-field critical exponents are valid in nonlinear noisy voter models on networks. The latter considers `ageing' in the dynamics, a type of memory effect in which the propensity of an agent to change depends on how long the agent has held its current opinion.  

\paragraph{Noise in models with conformists and anti-conformists.}
The possibility for spontaneous opinion change was also studied in $q$-voter models with conformist and anti-conformist behavior \textcite{nyczka_phase_2012}. Two-state and multistate $q$-voter models with quenched and annealed noise were studied and compared in \cite{jedrzejewski_pair_2022, nowak_discontinuous_2021}. Quenched disorder here indicates that a fixed subset of agents shows independent behavior, whereas annealed randomness indicates a situation in which all agents can undergo spontaneous opinion change with a non-zero rate. Quenched disorder was found to exclude discontinuous transitions when there are two opinion states. In the multistate model discontinuous phase transitions are possible even for quenched disorder. A particularly interesting case is the $q$-voter model with annealed independence \cite{nyczka_phase_2012}, which is equivalent to the non-linear noisy voter model \cite{peralta_analytical_2018}. In such a model, there is a tricritical point at $q=5$: the phase transition is continuous for $q<5$ and discontinuous for $q>5$. Such a model was studied on 
a complete graph \cite{nyczka_phase_2012}, random graphs \cite{jedrzejewski_pair_2017,jedrzejewski_pair_2022}, and multiplex networks \cite{gradowski_pair_2020}. In the latter two cases, the pair approximation and computer simulations were used to analyze the model. For the $q$-voter model with independence on scale-free networks, the ratio between critical exponents has been evaluated using finite-size scaling methods. For example, for $q = 2$ and an average degree of $10$, the following values have been obtained: $\beta/\bar{\nu} = 0.157$ and $1/\bar{\nu} = 0.314$. Thus, the model belongs to a different universality class than the Ising model \cite{jedrzejewski_pair_2017}.

\paragraph{Multistate noisy voter models.}
\textcite{khalil_zealots_2018} showed that zealotry in noisy voter models enhances the effective noise strength and that zealots can remove the quasi-consensus state. The authors find a discontinuous transition from an asymmetric bimodal phase to an extreme asymmetric phase, and a continuous transition from the extreme asymmetric phase to an asymmetric unimodal phase. In multistate voter models with noise, one observes a transition from consensus to a stable multistate stationary distribution \cite{herrerias-azcue_consensus_2019}. In populations with all-to-all interaction, this can be studied analytically via a mapping to an effective death-birth process for agents holding a selected opinion. \textcite{peralta_stochastic_2018} introduced a stochastic pair approximation (see also \sref{sec:AMEs}) to capture leading-order finite-size corrections in noisy voter models on networks.

\paragraph{Imperfect copying and transmission errors.}
Noise can also be introduced as imperfect copying, which leaves the states of the two interacting individuals similar but not exactly equal. \textcite{vazquez_multistate_2019} found that the ordering of the system, quantified by a quantity similar to the ordering parameter of flocking dynamics \cite{vicsek_novel_1995}, in the stationary state decreases quadratically with the amplitude of the copying error, and vanishes in the thermodynamic limit. Voter-like models with spontaneous state changes and transmission errors were used to model the evolution of language features in \cite{kauhanen_geospatial_2021} and \cite{kitching_estimating_2025}. 
\textcite{fernandez-gracia_is_2014} used a noisy voter model with geographically mobile agents to capture statistical features of presidential elections in the US and to model the logarithmic decay with distance of correlations of vote share fluctuations across counties.

\subsubsection{Models with conformity and anticonformity}
\label{sec:conformity_and_anticonformity}

Fragmentation can arise as a competition between conformity and anticonformity, as in the majority-vote model \cite{de_oliveira_isotropic_1992}.
Although in its original formulation it does not explicitly refer to the concepts of conformity or anticonformity, the underlying mechanisms correspond to these social responses \cite{nail_proposal_2013}. 
In the majority-vote model, these responses are implemented within the annealed approach, as will be clarified in the paragraph on the majority-vote model.

The same idea of competition between conformity and anticonformity within the annealed framework was adopted in other models, such as the majority-minority model \cite{mobilia_majority_2003} and in the $q$-voter model with anticonformity \cite{nyczka_phase_2012}. 
The key differences between these models lie in the system update rules and the construction of the group of influence (see \fref{fig:binary_models}). 
In the majority-vote model, the group of influence consists of all neighbors of the focal agent, whereas in the $q$-voter model, only $q$ neighbors are randomly selected from the entire neighborhood. 
However, in both models, only a single agent, the focal agent, is updated at a time. 
In contrast, the majority-minority model updates an entire group of $G$ agents simultaneously \cite{mobilia_majority_2003}.
A slightly different mechanism, which will be described in one of the next paragraphs, is used in the Galam model with contrarians \cite{galam_contrarian_2004}. 
Despite these differences, all these models yield qualitatively similar macroscopic behavior, showing a continuous agreement–disagreement phase transition, where the disordered state (fragmentation) is referred to as a mixed state \cite{mobilia_majority_2003} or disagreement \cite{jedrzejewski_statistical_2019}.

Anticonformity has also been incorporated into threshold models, both within the annealed approach \cite{grabisch_anti-conformism_2020,nowak_threshold_2022} and within the quenched approach, where it appears under the term hipsters \cite{juul_hipsters_2019}. While in binary opinion models anticonformity is most commonly implemented as adopting an opinion opposite to that of others, in continuous opinion models it typically takes the form of opinion distancing and is often used to explain polarization (see \fref{subsec:opinion_repulsion}). For example, \textcite{mas_individualization_2010} studied a continuous-opinion model with an element of anticonformity (“striving for uniqueness”) to describe opinion clustering. In the next paragraphs, we focus on models with discrete opinions in which anticonformity has led to interesting behaviors.

\paragraph{The majority-vote model.}

The model traces its origins to the classical majority voting process, one of the first interacting particle systems studied in mathematics \cite{liggett_interacting_2005,bramson_majority_2021}. 
However, in statistical physics, it was introduced in the early 1990s \cite{de_oliveira_isotropic_1992}. As in voter models and the Ising model with Metropolis or heat-bath dynamics, the majority-vote model updates a single target agent at a time (see \fref{fig:binary_models}). 
The target agent adopts the majority opinion of its neighbors with probability $1 - p$, and the minority opinion with probability $p$.  

In the physics literature, the majority vote model was initially studied on a square lattice using Monte Carlo simulations. 
By applying finite-size scaling to estimate critical exponents, it was shown that the model belongs to the same universality class as the equilibrium Ising model \cite{grinstein_statistical_1985, de_oliveira_isotropic_1992}. 
Many extensions of the original model were later proposed and briefly reviewed in \cite{vieira_phase_2016}, so here, we will focus only on more recent works.
\textcite{vieira_phase_2016} introduced additional noise to the majority-vote model in the form of independence. 
In this model, a randomly chosen agent chooses one of the two possible opinions $\pm 1$ with probability $p$, and with the complementary probability $1-p$, they apply the original rule of the majority-vote model. 
The introduction of independence did not change the universality class of the model. 

Later, the majority-vote model was extended by introducing inertia into the spin-flipping dynamics, allowing an individual's probability of changing state to depend not only on the opinions of its neighbors, but also on its own previous state \cite{chen_first-order_2017}. 
This modification fundamentally altered the nature of the phase transition, shifting it from continuous to discontinuous when inertia exceeded a critical threshold.

\paragraph{The \texorpdfstring{$q$}{q}-voter model with anticonformity.}

Anticonformity was initially introduced into the binary $q$-voter model within the annealed approach \cite{nyczka_phase_2012}. 
As in the original $q$-voter model \cite{castellano_nonlinear_2009}, an agent is influenced by a group of $q$ agents randomly selected from its neighborhood, but only if all members of the group share the same opinion. 
A key difference from the original model is that these $q$ neighbors are selected without repetition, whereas the original version allowed repetitions. Moreover, if the group is not unanimous, the focal agent does not change its state, unlike in the original model, where it flips with probability $\epsilon$.

In the model proposed by \textcite{nyczka_phase_2012}, if the influence group is unanimous, the focal agent adopts the opposite opinion with probability $p$ (anticonformity) or the same opinion with the complementary probability (conformity). This model exhibits a continuous agreement--disagreement phase transition for any value of $q$, both on complete graphs \cite{nyczka_phase_2012} and on random graphs, under both the quenched and annealed approaches \cite{jedrzejewski_pair_2022}. However, \textcite{jedrzejewski_pair_2022,abramiuk-szurlej_discontinuous_2021} showed that, in the annealed formulation of the $q$-voter model with anticonformity, the pair approximation may incorrectly predict a discontinuous phase transition where the true behavior is continuous. In a generalized version of the model, where the influence-group sizes for conformity ($q_c$) and anticonformity ($q_a$) are treated as independent parameters, discontinuous transitions appear only when $q_c$ is sufficiently larger than $q_a$ \cite{abramiuk-szurlej_discontinuous_2021}.

\begin{figure}[tbp]
\centering
\includegraphics[width=1\columnwidth]{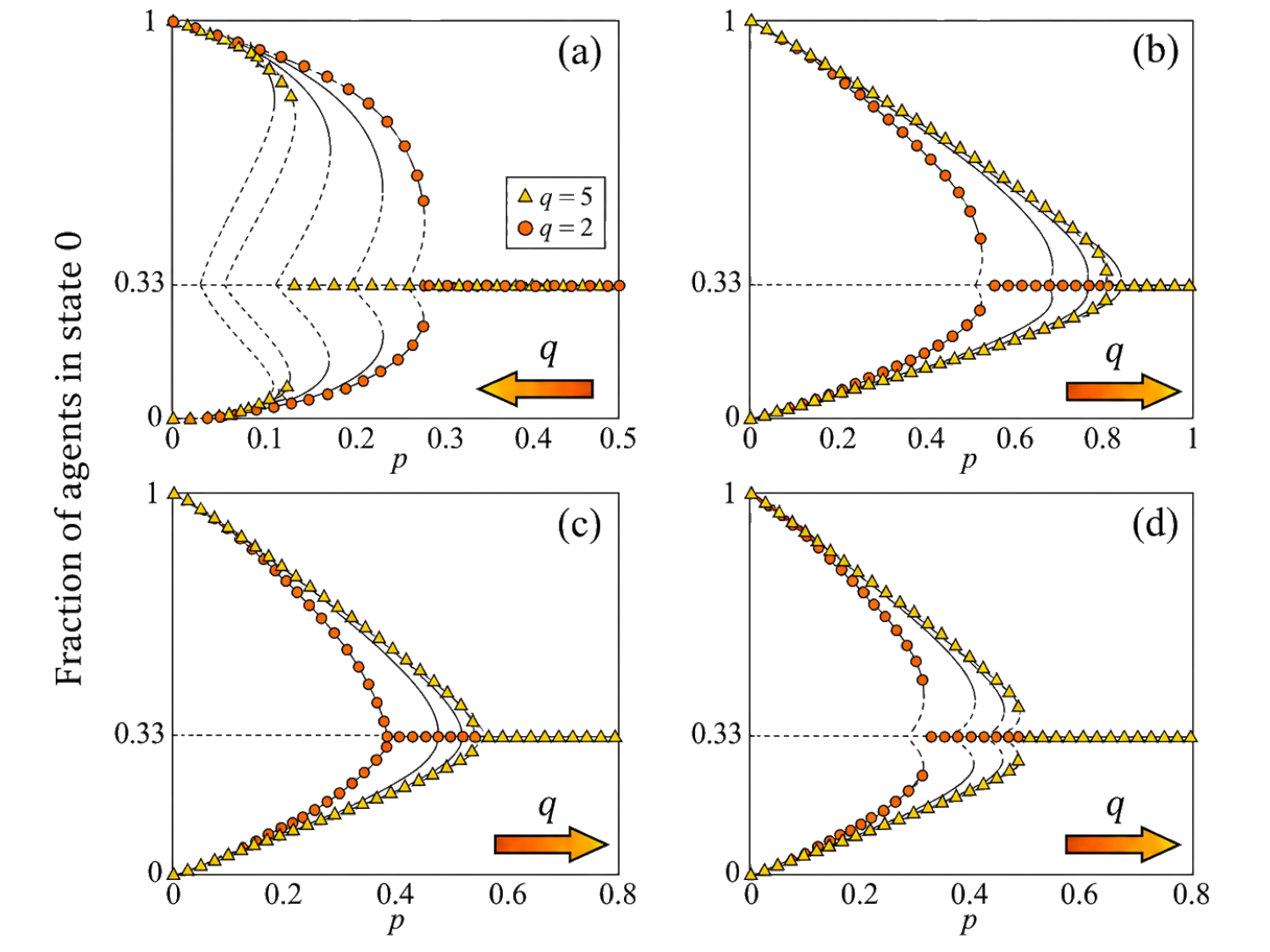}
\caption{Stationary fraction of agents in state 0 for the 3-state $q$-voter model on a complete graph as a function of the probability $p$, representing (a) annealed independence, (b) quenched independence, (c) annealed anticonformity, and (d) quenched anticonformity. The arrows in the bottom right corners of the panels indicate the direction in which the size of group q increases. This information is also included in the legend, which applies to all four panels. The figure is based on results reported in \cite{nowak_discontinuous_2021} for the model with independence (a,b) and \cite{nowak_switching_2022} for the model with anticonformity (c,d). Lines represent mean-field calculations, while symbols correspond to Monte Carlo simulation results.}
\label{fig:quenched_annealed}
\end{figure}

The puzzle  arises in the multistate extension of the model \cite{nowak_switching_2022}, where each agent holds one of $G$ unordered opinions. In contrast to ordered opinions, such as those measured on a Likert scale from strongly disagree to strongly agree, unordered opinions correspond to categories with no natural ranking. For example, choosing a preferred composer from Beethoven, Grieg, or Dvořák involves distinct options that cannot be meaningfully ordered. In such models, agents are not restricted to interacting only with others holding similar opinions, nor are they limited to changing opinions incrementally, as is typically the case in discrete opinion models with bounded confidence \cite{lipiecki_polarization_2022,lipiecki_depolarizing_2025}. In each update: 
(1) a random agent is selected, 
(2) a group of $q$ random neighbors is chosen, 
(3) if the group is unanimous, with probability $p$, the focal agent adopts a randomly chosen opinion different from the group’s, while with probability $(1 - p)$, it adopts the same opinion. 
Both pair approximation, as well as the mean-field calculations and Monte Carlo simulations show the existence of an order-disorder phase transition. 
In the ordered phase, one opinion dominates, and in the disordered phase, all $G$ opinions coexist and the order parameter vanishes. 
Under the annealed approach, this transition is continuous regardless of the number of states $G$ or the size of the group $q$.
However, in the quenched approach, a discontinuous phase transition appears for $G \ge 3$, even for small values of $q > 1$. Results for the $3$-state model are shown in Fig. \ref{fig:quenched_annealed}.
This is surprising, since quenched randomness typically rounds or completely eliminates discontinuous transitions, as briefly reviewed in \cite{nowak_switching_2022}. 

\paragraph{Contrarians.}
In the field of opinion dynamics, the concept of contrarian choices was introduced probably for the first time by Galam in his majority model \cite{galam_contrarian_2004}. 
Contrarian agents always adopt the opinion opposite to the prevailing choice in the group. 
Originally, Galam formulated this dynamics within the annealed framework \cite{galam_contrarian_2004}, i.e., each agent has a certain probability to behave like a contrarian \cite{borghesi_chaotic_2006}.
 In the original model, such a probability of a contrarian choice was a fixed parameter, leading to a continuous order-disorder phase transition, as in the majority-vote model. 
When the probability of contrarian behavior depends on the current level of support for a given opinion, the system exhibits chaotic dynamics \cite{borghesi_chaotic_2006}.

Contrarians have also been studied within the kinetic-exchange model family [see \sref{subsubsec:opinion_assimilation} and \textcite{biswas_social_2023} for more details]. In particular, the Biswas-Chatterjee-Sen (BCHS) model introduced probabilistic contrarians that interact negatively with probability $p$
\cite{biswas_disorder_2012}.
This modification leads to a behavior different from the LCCC model  \cite{lallouache_opinion_2010}: below the critical point, the BCHS model enters a sociologically relevant fragmented state where opposite opinion polarities ($\pm 1$) coexist. 
The BCHS model demonstrates an order-disorder phase transition, and its critical exponents are close to those of the equilibrium Ising universality class \cite{mukherjee_ising_2021}.

A variant of the BCHS model introduces a ``social temperature'' $T$ \cite{anteneodo_symmetry_2017}, which modulates randomness via a hyperbolic tangent function. 
High ``temperature'' can destroy order, leading to fragmentation, or induce phase transitions between symmetric (fragmented) and asymmetric (consensus) states, with critical behavior belonging to the directed percolation universality class. 
The BCHS model has also been studied on three-dimensional lattices, where each dimension may reflect a different environment (for instance, home vs workplace), showing a second-order phase transition \cite{oliveira_biswaschatterjeesen_2024}. 
Analysis of the critical exponents showed that the 3D BCHS model belongs to a different universality class than its respective 1D and 2D network models, as well as the standard Ising model on the same network.
The BCHS model has also been studied on a modular interaction structure (two groups with high intra-group and low inter-group interaction), revealing a new antisymmetric ordered state where each group maintains an opposite dominant opinion \cite{suchecki_biswas-chatterjee-sen_2025}.
This state is stable within certain limits of interaction strength and noise, exhibiting a discontinuous transition when those thresholds are exceeded.

The influence of contrarians within the quenched approach was later explored in various other models, as briefly reviewed in \cite{gambaro_influence_2017, khalil_noisy_2019}. A particularly interesting contribution was presented in \cite{masuda_voter_2013}, where three extensions of the linear voter model incorporating contrarians in the quenched framework were analyzed. These extensions differ in the behavioral rules assigned to contrarians. In the first version, contrarians oppose conformists (referred to as congregators in the paper) but align with other contrarians. For instance, a contrarian in state $-1$ flips to state $1$ at a rate determined by the number of conformists in state $-1$ and contrarians in state $1$. In the second version, contrarians align with conformists but oppose other contrarians. In the third version, contrarians oppose both conformists and other contrarians. This paper is a very good example of how different assumptions can exist under the same label, in this case, 'contrarian'

Another intriguing concept is that of reasonable contrarians, agents who usually oppose the average opinion of their neighbors but conform when confronted with an overwhelming majority \cite{bagnoli_bifurcations_2015}. Analyzed using mean-field approximation as well as on small-world and scale-free networks, the model demonstrates that a high proportion of conformists results in opinion fragmentation, whereas a majority of reasonable contrarians leads to coherent oscillations and complex behaviors such as period-doubling and chaos.

\subsubsection{Zealotry as independent behavior}
\label{sec:zealotry}

\paragraph{Flexible and inflexible zealots.}
The notion of zealotry in opinion dynamics dates back to \cite{galam_towards_1991,galam_rational_1997} and \cite{mobilia_does_2003}, although Galam used the term \textit{biased individual} rather than \textit{zealot}. In the context of the voter model, zealots were initially defined as biased individuals, that is, agents favoring one opinion, but not necessarily fully committed to a single state. What distinguished zealots from other agents in \cite{mobilia_does_2003} was that they could return to their favored opinion at a given rate, independently of interactions with other agents. Later, the general understanding appears to have shifted, and zealots are now more commonly regarded as agents who never change their opinion state \cite{mobilia_role_2007}. We also note the work of \cite{chinellato_dynamical_2015}, available as a preprint already in 2007, which considers a very similar model, as acknowledged in \cite{mobilia_role_2007}. Such agents are also known as \textit{inflexibles} or stubborn individuals, particularly in the context of the Galam majority-rule model \cite{galam_role_2007}. Voter models with zealotry have been used to describe phenomena such as market crises and US presidential elections; see, for example, \cite{harmon_anticipating_2015,braha_voting_2017}.

In this section, we briefly review the effects of zealotry on opinion dynamics models, by mostly focusing on fully inflexible zealots \cite{galam_rational_1997, galam_role_2007, mobilia_role_2007,martins_building_2013, mobilia_nonlinear_2015,galam_stubbornness_2016}.
 We distinguish between models with discrete and continuous opinion states. 
 We highlight that zealotry can be `one-sided' (all zealots in the population have the same opinion state), or mixed (multiple groups of zealots, each holding a different opinion).

\paragraph{Discrete opinions.}

\textcite{mobilia_role_2007} and \textcite{chinellato_dynamical_2015} studied the role of zealots in the conventional two-state voter model. 
For equal numbers of zealots supporting each of the two opinions, the
distribution of the number of agents in either state is approximately Gaussian in populations with all-to-all interaction, and in lattices with one and two dimensions. 
The width of the opinion distribution is proportional to the inverse square root of the number of zealots, independent of the total number of voters. 
As \textcite{mobilia_role_2007} conclude, a small number of zealots can thus prevent the formation of a robust majority, and consensus becomes impossible. The paper by \textcite{chinellato_dynamical_2015} focuses on the transition between unimodal and bimodat stationary states, and also addresses the effects of zealotry on networks.

\textcite{mobilia_nonlinear_2015} studied the effects of (mixed) zealotry in the nonlinear two-state voter model, finding a continuous phase transition. 
Below a critical zealotry density, the system is in a bistable phase, where the opinion distribution is bimodal, and the higher peak corresponds to the opinion supported by the larger number of zealots. 
For equal numbers of zealots in either state, the distribution is symmetric. 
Above a critical level of zealotry, the opinion distribution becomes unimodal, and most susceptible agents (who are not zealots and can therefore change opinion) support the opinion represented by the larger number of zealots. 
Transitions between bimodal and unimodal states at low and high levels of zealotry were also demonstrated in the linear voter model \cite{chinellato_dynamical_2015}. 
The authors provide closed-form expressions for the stationary state and introduce the idea of fractional zealots. 

 In two-state noisy nonlinear voter models, zealots can change the nature of the unimodal and bimodal states \cite{khalil_zealots_2018}. 
 For a given number of zealots, a symmetry-breaking transition can be found in the stationary distribution for the number of agents supporting a certain opinion, but only if there are unequal numbers of zealots supporting the two different opinion states.
 In noisy voter models with more than two opinion states,
 up to six possible qualitatively different types of stationary distributions are possible \cite{khalil_zealots_2021}. 
 Situations in which zealots only affect a fraction of agents can further enrich the phase diagram. 
 Such models can be mapped to a single-community model with a fractional number of zealots, which can be studied via a reduction to an effective birth-death model for the number of agents holding a selected opinion. The effects of zealotry in voter models has also been studied on static \cite{tunstall_how_2025,masuda_opinion_2015} and adaptive networks \cite{klamser_zealotry_2017}.

\textcite{fudolig_analytic_2014} focused on a voter model with zealots of one single opinion.
They determined how the time to reach consensus depends on the number of zealots and the number of susceptible (non-zealot) agents in populations with all-to-all interaction, and on networks. \textcite{brede_resisting_2018} studied the influence of optimal opinion control in a modified two-state voter dynamics. 
More specifically, they introduced a controller node with the ability to exert influence on a fixed number of other nodes. 
The question they asked is then what set of controlled nodes to choose to maximize the effect of the controller on the network as whole.
This was then extended to model networks and with different time horizons  \cite{brede_effects_2019}.

A nonlinear voter model with all-to-all interaction between two groups of susceptible agents (with different types of nonlinearity $q_1$ and $q_2$) and zealotry was investigated in \cite{mellor_characterization_2016, mellor_heterogeneous_2017}. Detailed balance can be broken in this model (despite the all-to-all interaction). 
For low zealotry, the model shows a bimodal opinion distribution; when zealotry is above a critical level, the distribution becomes single-peaked.

\textcite{mobilia_commitment_2013} provides a rare example of an analytically tractable three-state voter model with zealotry. \textcite{mobilia_polarization_2023} studied a three-state voter model with external influence that can fluctuate between two states. %
The population eventually either reaches consensus at one of the three opinions, or a polarization state consisting of a frozen mixture of leftists and rightists. 
The rate at which the external influences change is found to affect the final state and the mean time to reach this state. 
Time-dependent zealotry was also studied by \cite{caligiuri_noisy_2023} with methods from the theory of piecewise-deterministic Markov processes. 
It was shown that fragmented multi-peak opinion distributions can arise when there are multiple groups of influencers.

The ordering process in multistate voter models (without zealotry) proceeds through successive elimination of individual opinions. Between these extinction events the system remains in long-lived fragmented states. 
\textcite{ramirez_local_2022} showed that these partially ordered states can be stabilised by adding precisely one zealot representing each of the different opinions. 
It it thus possible to engineer fragmented states as genuinely stationary outcomes. 
\textcite{kitching_breaking_2025} studied how zealotry and nonlinear social impact affect the breaking of consensus in voter models, evolutionary games, and the partisan voter model (a two-state voter model in which each agent has a preference for one opinion state \cite{masuda_heterogeneous_2010, masuda_can_2011, llabres_partisan_2023}). As shown in \cite{kitching_breaking_2025} nonlinear social impact can override the effects of zealotry, when social impact is sublinear, no amount of one-sided zealotry can enforce consensus, and the population remains fragmented. 

In the Galam majority model, a concept similar to zealotry was introduced as ``inflexible" or ``stubborn" agents \cite{galam_role_2007,galam_stubbornness_2016}. In the majority model without inflexible agents, and dividing the population into groups containing an odd number of agents, the opinion held by the initial majority determines the final consensus. When inflexibles supporting only one of the two opinions are introduced, the basin of attraction for that opinion increases. Moreover, consensus on the opposing opinion can no longer be reached, even if the initial majority supports it.

\paragraph{Continuous opinions.}
\label{subsubsec:stubbornness}

Zealotry or stubbornness can also sustain fragmentation in models with continuous opinion states, even on connected networks and under otherwise purely assimilative dynamics.

The Friedkin–Johnsen (FJ) model~\cite{friedkin_social_1990,friedkin_choice_1999} extends the classical assimilation frameworks of French~\cite{french_formal_1956} and DeGroot~\cite{degroot_reaching_1974} by assigning every agent $i$ a fixed opinion (or \emph{prejudice}) $u_i$, which encodes the agent’s \emph{stubbornness}.
In the FJ model opinion evolve as 
\begin{equation}
    \mathbf{x}(t+1) \;=\; \Lambda\,\mathbf{A}\,\mathbf{x}(t) \;+\; (\mathbf{I}-\Lambda)\,\mathbf{u},
\end{equation}
where $\mathbf{A}$ is a row‑stochastic influence matrix,  
$\Lambda=\operatorname{diag}(\lambda_1,\dots,\lambda_N)$ contains each agent’s susceptibility $\lambda_i\in[0,1]$ to social influence, and $\mathbf{u}$ is the vector of prejudices.
The term $(\mathbf{I}-\Lambda)\mathbf{u}$ weights these predispositions by $(1-\lambda_i)$:
the larger the weight placed on $u_i$, the less the agent is willing to shift toward the opinions of others and adheres to their opinion $u_i$.
For $\Lambda=\mathbf{I}$, the inhomogeneity of vanishes and the FJ model turns into the DeGroot model (see \sref{subsubsec:opinion_assimilation}).
Importantly, in the FJ model persistent disagreement is possible, even in connected networks, and it can be shown that this dynamics converges to a fixed point if the matrix $\lambda_i A_{ij}$ is regular~\cite{grabisch_survey_2020}. 

Extensions to the FJ model have been used to investigate time-dependent levels of agent stubornness \cite{zhou_multidimensional_2022}, opinion dynamics in multi-dimensional opinion spaces \cite{parsegov_novel_2017}, homophily in interactions \cite{disaro_extension_2023}, social influence on signed graphs \cite{zhou_friedkin-johnsen_2024}. Furthermore, the FJ model has been extensively used to investigate opinion dynamics in terms of an optimization problem. These works are discussed in \sref{subsec:algo-appr}.

{\em Taylor model.} The Taylor model extends Abelson's model (\sref{subsubsec:opinion_assimilation}) and includes the influence of $m$ external communication sources that provide the opinions $s_1,\dots, s_m\in\mathbb{R}$ \cite{taylor_towards_1968}, which can also be interpreted as the opinions of stubborn individuals \cite{baumann_laplacian_2020}. The Taylor model is formulated as
\begin{equation}
    \dot{x}_i(t) = \sum\limits_{j=1}^{N} A_{ij} (x_j - x_i)+\sum\limits_{k=1}^{m} B_{ik} (s_k - x_i)\,,
\end{equation}
where $B_{ij}$ is the nonsquare $N\times m$ matrix that defines the persuability constants. In particular, the value of $B_{ij}$ captures to what extent agent $i$ is influenced by communication source $s_j$. If $b_{ik}=0\;\forall\;k$, agent $i$ is not influenced by any communication source. 
Taylor~\cite{taylor_towards_1968} demonstrated that persistent external influences drive the system toward a fragmented configuration that is asymptotically stable, converging to a unique equilibrium set by the communication source $s_k$. Second-order extensions of the Taylor model---incorporating momentum through a second derivative---have likewise been employed to analyse consensus formation in robotic multi-agent systems \cite{wenwu_yu_second-order_2010,yu_second-order_2011,baumann_periodic_2020}.

Inflexible zealots have also been introduced in the kinetic-exchange models [see \sref{subsubsec:opinion_assimilation} and \textcite{biswas_social_2023}].
The fraction of inflexible agents lowers the critical point for the onset of order \cite{crokidakis_impact_2014}.
These inflexible agents can be randomly distributed or restricted to extreme opinions ($\pm 1$), which affects the resulting phase boundary but leaves the universality class unchanged. 
Related models include those where the conviction parameter $\lambda$ is a discrete random variable (e.g., $0, \pm 1$), which, while not altering the universality class, lowers the critical point if $\lambda = 0, -1$ \cite{crokidakis_role_2012}.
Models allowing agents to independently select opinions, irrespective of interaction, similarly result in critical behavior consistent with the mean-field Ising model \cite{vieira_consequences_2016}.

\subsubsection{Bounded confidence}
\label{subsubsec:bounded_confidence}

Bounded confidence models (BCMs) extend opinion assimilation models (\sref{subsubsec:opinion_assimilation}) by introducing a similarity bias where individuals influence each other only if their opinion distance is below a fixed threshold, i.e., the \emph{confidence bound}.
We first introduce two canonical BCMs, the Hegselmann--Krause model (\hk)~\cite{hegselmann_opinion_2002} and the Deffuant–Weisbuch model (\dw) \cite{deffuant_mixing_2000}, and then review their most important extensions.

\paragraph{Hegselmann--Krause model.} The \hk is formulated in discrete time and can be written as $\mathbf{x}(t+1) = \mathbf{A}(\mathbf{x}(t))\mathbf{x}(t)$, where $\mathbf{A}(\mathbf{x}(t))$ denotes the influence matrix whose element $A_{ij}$ specifies how strongly the opinion of agent $i$ is influenced by the opinion of agent $j$. The key feature of the \hk is that the influence matrix $\mathbf{A}(\mathbf{x}(t))$ depends on the opinions.
Specifically, an agent $i$ considers \textit{only} those agents $j$ whose opinions differ from their own by no more than $\epsilon$---the confidence bound. 
For a given opinion distribution $\mathbf{x}$, at each time step, the set of agents $j$ influencing $i$ can be formally defined as $I(i, \mathbf{x})=\left\{1\leq j\leq N | \abs{x_i-x_j} \leq \epsilon \right\}$. 
Generally, the weight associated to each neighbor $j\in I(i, \mathbf{x})$ is equal, i.e. we get $A_{ij}(\mathbf{x})=0$ for $j\notin I(i, \mathbf{x})$ and $A_{ij}(\mathbf{x})=|I(i,\mathbf{x})|^{-1}$ for $j\in I(i, \mathbf{x})$, where $|I(i,\mathbf{x})|$ denotes the number of agents in the set $I$. 
The opinion dynamics evolves in discrete time steps according to
\begin{equation}\label{eq:HK_model}
x_i(t+1)=|I(i,\mathbf{x})|^{-1}\sum\limits_{j\in I(i, \mathbf{x}(t))}x_j(t)\,.
\end{equation}

Note that \eref{eq:HK_model} must be modified if the \hk is considered on a social network $B_{ij}$. In particular, the set $I(i, \mathbf{x})$ will be limited to individuals that are connected to agent $i$ and whose opinions are within the confidence bound, i.e. we have $I(i, \mathbf{x})=\left\{1\leq j\leq \mathcal{N}(i) | \abs{x_i-x_j} \leq \epsilon \right\}$, where $\mathcal{N}(i)$ denotes set of agent $i$'s neighbors in $B_{ij}$.

The \hk was mainly investigated through simulations for both symmetric and asymmetric confidence intervals \cite{hegselmann_opinion_2002}. For confidence bounds below a critical value $\epsilon=\epsilon_c$, consensus becomes unstable and the final distribution of opinions is fragmented.  The transition from fragmentation to consensus is non-monotonic, not smooth and irregular \cite{lorenz_consensus_2006, hegselmann_bounded_2023}.
In the limit of large populations ($N\rightarrow\infty$), $\epsilon_c$ assumes different values depending on the average degree of the underlying network $\langle k\rangle$. If $\langle k\rangle$ is constant, $\epsilon_c=1/2$, whereas if $\langle k\rangle \rightarrow\infty$ is constant, $\epsilon_c\approx 0.2$ \cite{fortunato_consensus_2005}.
Using a formulation of the \hk in the form of a rate equation, it was later also shown that, starting from a uniform distribution on the interval $x\in[0,1]$, the final opinion distribution is symmetrical with respect to the average opinion $1/2$ \cite{fortunato_vector_2005}.

\paragraph{Deffuant-Weisbuch model.} 
The \dw, introduced by \textcite{deffuant_mixing_2000}, is similar to the \hk, with one crucial difference: the opinion updating scheme.
In the \hk (Eq.~\ref{eq:HK_model}), \textit{all} opinions are updated synchronously. 
By contrast, the \dw assumes that, in each time step, two randomly selected individuals $i$ and $j$ update their opinions according to 
\begin{align}
    x_{i}(t+1) &= x_{i}(t) + \mu \left[ x_{j}(t) - x_{i}(t) \right] \\
    x_{j}(t+1) &= x_{j}(t) + \mu \left[ x_{i}(t) - x_{j}(t) \right]  
\end{align}
provided that their opinion difference (prior to the interaction) satisfies $|x_i(t) - x_j(t)| \leq \epsilon$, otherwise opinions stay constant. The parameter $\mu \in [0, 1/2]$ is the convergence parameter that determines the extent to which opinions adjust. 

The dynamics of the \dw preserves the mean opinion of the interacting pair for all values of $\mu$.  
In the original formulation, opinions are typically initialized from a uniform distribution over the interval $[0,1]$. 
Under these conditions, the \dw dynamics leads to the gradual emergence of opinion clusters, which either merge into a (complete) consensus state---for sufficiently large $\epsilon$---or remain separated when the opinion differences between clusters exceed the confidence bound. In the latter case, agents continue interacting only within their respective clusters, ultimately yielding a asymptotic configuration characterized by a series of Dirac $\delta$-distributions.  
For a broad range of $\mu$ values, the number and distribution of final opinion clusters are primarily determined by $\epsilon$. 
Most importantly, numerical simulations \cite{fortunato_universality_2004}, later confirmed by rigorous mathematical analysis \cite{lorenz_about_2007}, showed that the \dw leads to complete consensus at $x_i \equiv 1/2$ provided that the confidence bound satisfies, $\epsilon > \epsilon_c = 1/2$, under the assumption of an infinitely large population with opinions initially drawn from a uniform distribution. This result also holds for various underlying network structures including regular lattices and scale-free networks.

Interestingly, however, for small values of $\mu$, the convergence parameter itself also influences the resulting opinion distribution 
\cite{porfiri_decline_2007}. As argued in \cite{laguna_minorities_2004}, the reason for this dependence on $\mu$ is that when the dynamics proceed slowly (small $\mu$), many agents whose opinions are initially far from the future consensus value (further than $\epsilon$), can nevertheless, through interaction with intermediate agents,
be enabled to join more centrist clusters. 
Conversely, if the dynamics evolves faster, those intermediate opinions do not exist long enough and a large number of agents will be therefore stuck in extremist opinion clusters.

An alternative approach to studying BCMs, both the \hk \cite{lorenz_continuous_2007} and the \dw \cite{ben-naim_bifurcation_2003,chu_density_2023}, is to convert the models into an opinion density formulation. Following the formulation by Ben-Naim et al. \cite{ben-naim_bifurcation_2003}, the \dw can be described in terms of the following integro-differential equation, which reads

\begin{eqnarray}
\label{eq:deffuant_rate_equation}
    \frac{\partial}{\partial t} P(x,t) &=& \iint\limits_{|x_1 - x_2| < 1} \, dx_1 \, dx_2\,P(x_1,t) P(x_2,t) \nonumber \\
    &&\times\left[ \delta\left(x - \frac{x_1 + x_2}{2}\right) - \delta(x - x_1) \right] ,
\end{eqnarray}
where $\epsilon=1$ and $\mu=1/2$, and the opinion interval is $x=[-\Delta, \Delta]$ \cite{castellano_statistical_2009}. The term $P(x,t)\,dx$ denotes the fraction of agents with opinions in the interval $(x, x + dx)$ at time $t$. When the opinion difference between two agents is less than $\epsilon=1$, 
they interact and update their opinions to the mean value of their current opinions, which is represented by the gain and loss terms at $(x_1 + x_2)/2$, and $x_1$ and $x_2$, respectively. 

The formulation of the \dw in terms of the rate equation in Eq.~\eqref{eq:deffuant_rate_equation} allows us to draw some general conclusions. 
First, Eq.~\ref{eq:deffuant_rate_equation} conserves the average opinion $\int_{-\Delta}^{+\Delta}xP(x,t)$. 
Second, starting from a uniform distribution $P(x, t=0)=1/(2\Delta)$, on the interval $x=[-\Delta, +\Delta]$, and for $\Delta<1/2$, Eq.~\eqref{eq:deffuant_rate_equation} is integrable and the population converges to $P(x, t\rightarrow\infty)=\delta(x)$.

\paragraph{Extensions of bounded confidence models.}
There is a vast body of literature investigating extensions to classic BCMs. 
One focus has been to capture various forms of heterogeneity, including heterogeneous and adaptive confidence bounds~\cite{lorenz_heterogeneous_2010,li_bounded-confidence_2024,torok_opinions_2013,iniguez_modeling_2014}, behavioral diversity induced by extremist and stubborn agents, or zealots \cite{deffuant_how_2002,weisbuch_persuasion_2005,deffuant_comparing_2006,franks_extremism_2008,mathias_bounded_2016}, variations in node activity in temporal networks~\cite{li_bounded-confidence_2023}, interactions with arbitrary waiting-time distributions~\cite{chu_bounded-confidence_2024}, as well as structural heterogeneity induced by underlying social network topologies~\cite{meng_opinion_2018,schawe_when_2021,amblard_role_2004} and hypergraphs \cite{schawe_higher_2022,hickok_bounded-confidence_2022}.

Anther line of research has extended the interaction mechanism of classical BCMs.
For instance, \textcite{brooks_emergence_2023} replaced the sharp confidence threshold by a smooth sigmoidal-shaped influence function~\cite{brooks_emergence_2023}. Furthermore, pairwise interactions have been replaced by higher-order interactions, where an individual is influenced by a neighbor only if the opinions of that neighbor's neighbors are sufficiently close \cite{krishnagopal_bounded-confidence_2024}. Additionally, BCMs were considered that include repulsive interactions to capture disagreement dynamics~\cite{martins_mass_2010}; these repulsion‑based models are examined in greater detail in Sec.~\ref{subsec:opinion_repulsion}. Another line of research investigated the role of noise in BCMs, where the nature of noise differed across papers \cite{schweitzer_modelling_2000,carro_role_2013,pineda_diffusing_2011,pineda_noisy_2009}. Some models introduced noise that allowed agents to adopt random opinions \cite{pineda_noisy_2009}, reset individuals to a previously held opinion \cite{carro_role_2013}, or noise that was applied to the confidence bound itself \cite{steiglechner_noise_2024}. 

The bounded confidence mechanism has also been studied within kinetic-exchange models \cite{biswas_social_2023}, by the introduction of a bound $\delta$, while $\lambda$ is the conviction parameter. 
The model's behavior drastically changes with the introduction of $\delta$, showing three distinct regions in the $\delta-\lambda$ phase diagram, with evidence of a first-order phase transition present for $\delta \geq 0.3$ \cite{sen_nonconservative_2012}.

Relatedly, a stream of work explored the \emph{co‑evolution} of opinions in BCMs and other quantities of interest. 
\textcite{bagnoli_dynamical_2007} examine a DW‑style model in which agents’ opinions co‑evolve with pairwise affinities that gate whether two individuals interact and update their views.  
Through numerical simulations they uncover a continuous phase transition from consensus to fragmentation, characterized by the rescaled control parameter $1/\sqrt{\sigma\alpha_c}.$  
Here, \(\alpha_c\) is the critical affinity that sets the interaction threshold, whereas the social temperature \(\sigma\) regulates the likelihood of encounters between dissimilar agents; sufficiently large \(\sigma\) enables cross‑cutting interactions and thus influences the onset of fragmentation.
Other BCMs that consider the co-evolution of opinions and the network structure \cite{kan_adaptive_2022} are discussed in detail in \sref{subsubsec:co-evo-cont-op}.

The idea of bounded confidence has also been applied to models in which opinions take discrete integer values \cite{fortunato_krause-hegselmann_2004}. In particular, in three-state models \cite{vazquez_constrained_2003,mobilia_fixation_2011,lipiecki_polarization_2022}, although usually without explicitly referring to the confidence bound. 
For example, in the three-state constrained voter model, opinions $+1$ and $-1$ represent radical positions (e.g., leftist and rightist), while opinion $0$ represents a centrist or intermediate state \cite{vazquez_constrained_2003,mobilia_fixation_2011} (see also Section \ref{sec:MSVM}). 
Interactions are constrained such that agents holding $+1$ and $-1$ do not interact directly with each other, but only through the intermediate state $0$. 
This restriction effectively mimics bounded confidence: individuals with opposing radical views are too far apart in opinion space to influence each other directly. 
In this model, regardless of spatial dimension, the system eventually reaches either a consensus on one of the three possible opinions or a frozen state composed solely of leftists and rightists (polarized state). 
A similar idea was later used in the three-state $q$-voter model with anticonformity, where the authors explicitly referred to bounded confidence \cite{lipiecki_polarization_2022,lipiecki_depolarizing_2025}. 
The presence of bounded confidence supports polarization, but a surprising result from this model was that anticonformity can promote consensus \cite{lipiecki_polarization_2022} and even depolarize initially polarized groups \cite{lipiecki_depolarizing_2025}.

\subsection{Polarization}
\label{subsec:polarization}

This section reviews models explaining opinion polarization, defined here as a specific form of opinion fragmentation. 
A polarized state is characterized by a bimodal opinion distribution, where most individuals cluster at extreme positions, and only a small minority holds moderate views 
(see Fig.~\ref{fig:polarization1} in Sec.~\ref{sec:macro:polarization}). 
We stress that this is not to be confused with a bimodal distribution for the number of agents holding a particular opinion, see the detailed discussion in Sec.~\ref{para:caution}. 
We focus on \emph{ideological polarization}---the divergence of individual opinions---and do not treat \emph{elite polarization}\cite{leonard_nonlinear_2021}, which corresponds to diverging positions among political leaders, or \emph{affective polarization}, which refers to the increasing hostility that individuals feel towards members of an opposed group
\cite{tornberg_modeling_2021}.

Most models of opinion polarization reduce opinions to a single continuous axis, encoding each agent’s opinion as a real-valued scalar;
yet, because these models employ different micro-mechanisms, they arrive at varying (and sometimes conflicting~\cite{mark_culture_2003}) conclusions about the origins of polarization.
We first survey one-dimensional models and group them by the primary mechanism driving polarization: (i) Opinion distancing (or repulsion), (ii) directional updating, (iii) biased assimilation, and also consider (iv) the impact of algorithmic mediation on opinion polarization.
Finally, we cover aspects of polarization, such as issue alignment and hyper-polarization, that can only be captured when multiple opinion dimensions are considered (see \sref{sec:macro:alignment}).

\subsubsection{Opinion distancing}
\label{subsec:opinion_repulsion}
One of the simplest mechanisms leading to stable polarization is opinion distancing (or repulsion), where exposure to strongly opposing views reinforces opinion differences.
Distancing models typically extend bounded confidence models (Section~\ref{subsubsec:bounded_confidence}), by adding a repulsive mechanism \cite{jager_uniformity_2005,salzarulo_continuous_2006}. 
Opinion repulsion can be state-dependent, activating when opinion distance exceeds~\cite{jager_uniformity_2005} or opinion overlap falls below a repulsion threshold~\cite{sirbu_opinion_2013}.
Alternatively, repulsion may be structural, hard-wired via fixed negative ties~\cite{martins_mass_2010}, whose weights can vary non-linearly with opinion dissimilarity~\cite{dignum_cultural_2014}. We note the work by \cite{macy_polarization_2003} on polarization in Hopfield-like models of opinion dynamics; see also \cite{mark_culture_2003}.

Opinion repulsion can drive an agent’s opinion $x_i(t)$ beyond the initial population range $[x_{\min},x_{\max}]$, allowing it to migrate toward more extreme positions, i.e., $x_i(t)<x_\mathrm{min}$ or $x_i(t)>x_\mathrm{max}$.  
To prevent opinions from diverging without bound, models typically define boundary conditions that reset opinions to a fixed interval \cite{martins_mass_2010,flache_small_2011}, assume smoothing functions \cite{feliciani_how_2017,flache_why_2008}, or employ interaction dynamics that render the opinion space self‑contained \cite{huet_openness_2010, flache_models_2017}.

A classical repulsion model was introduced by \textcite{jager_uniformity_2005}. 
The model is an extension of the \dw \cite{deffuant_mixing_2000} and defines a \emph{repulsion threshold} $r$, with $r>c$, where $c$ denotes the confidence bound (see Sec.~\ref{subsubsec:bounded_confidence}). For $|x_i-x_j|>r$, the opinions of agents $i$ and $j$ are updated repulsively, $x_i(t+1) = x_i(t)+\mu(x_i(t)-x_j(t))$, 
where $\mu\in(0,1/2]$ is the influence strength. 
For $|x_i-x_j|\le c$, the dynamics corresponds to the one of the original \dw (Sec.~\ref{subsubsec:bounded_confidence}). The gap, $d=r-c>0$, is critical for the dynamics and polarization is most pronounced when both $c$ and $d$ are small. 
In one‑dimensional distancing models with $c=r$, \textcite{cornacchia_polarization_2020} recently showed that the dynamics allows only two absorbing states: consensus, or a polarized state with two homogeneous clusters. Instead, fragmentation into three or more opinion clusters is impossible. 

Distancing was also investigated in the context of discrete opinions \cite{radillo-diaz_axelrod_2009,macy_polarization_2003,krause_repulsion_2019}. \textcite{radillo-diaz_axelrod_2009} incorporated repulsion into the Axelrod model through a similarity threshold $\gamma$: if the proportion of shared traits between two agents exceeds $\gamma$, interaction is assimilative; otherwise one agent flips a randomly chosen trait to a value different from its partner’s, increasing their dissimilarity. Increasing values of $\gamma$ promote fragmentation and polarized states.
\textcite{macy_polarization_2003} have investigated the impact of distancing on polarization within a Hopfield model \cite{hopfield_neural_1982}. Here, binary opinion states ($x=\pm 1$) coevolve with the weights of the social network $A_{ij}\in[-1,+1]$ according to structural learning: $A_{ij} \gets A_{ij}(1-\lambda)+\frac{\lambda}{K}\sum_k x_{jk}x_{ik}$, where $\lambda$ is the learning rate and $K$ is the number of distinct opinions.

Repulsive interactions not only foster polarization but can also increase a population's susceptibility to external influences. For instance, \textcite{martins_mass_2010} examined a bounded confidence model, where a fraction of links $p$ are repulsive. Even for small values of the confidence bound $c$, the entire population eventually aligns with the value of an external signal $S$, in contrast to what would happen in the classical \dw \cite{carletti_how_2006}. 

\subsubsection{Directional updating}\label{subsubsec:opinion_reinforcement} 

Models of directional updating are typically based on a symmetric opinion space with a well-defined neutral point, allowing one to separate an opinion’s stance from its strength.  In the standard one-dimensional, continuous formulation the opinion axis is the real line, $x\in(-\infty,\infty)$: the stance is given by $\operatorname{sign}(x)$, while the conviction (strength) is the magnitude $|x|$. When an agent encounters a message or peer whose stance is positive $\left(\operatorname{sign}(x)>0\right)$, the agent shifts its own opinion in the positive direction; a negative stance triggers motion in the opposite direction. A schematic illustration of this mechanism is provided in Fig.~\ref{fig:micromechanism}. 

Directional updating comes in many forms, including the exchange of arguments~\cite{mas_differentiation_2013,fu_opinion_2016}, reinforcement‑learning paradigms~\cite{banisch_opinion_2019}, Bayesian updating~\cite{martins_continuous_2008,martins_mobility_2008,arehart_biased_2025}, as well as more classical continuous~\cite{vazquez_role_2020,saintier_model_2020,baumann_modeling_2020,shin_tipping_2010} and discrete~\cite{baumgaertner_opinion_2016} models.
In some models of directional updating, opinions and the interaction network co‑evolve, i.e., agents rewire their ties based on homophily, preferentially forming new links with like‑minded others.  
This feedback tends to drive opinions toward the extremes and can generate polarization.  
In static networks, directional updating typically promotes polarization in clustered networks. 

\paragraph{Argument exchange and information accumulation.} 

\begin{figure}[tbp]
\centering
\includegraphics[width=1.\columnwidth]{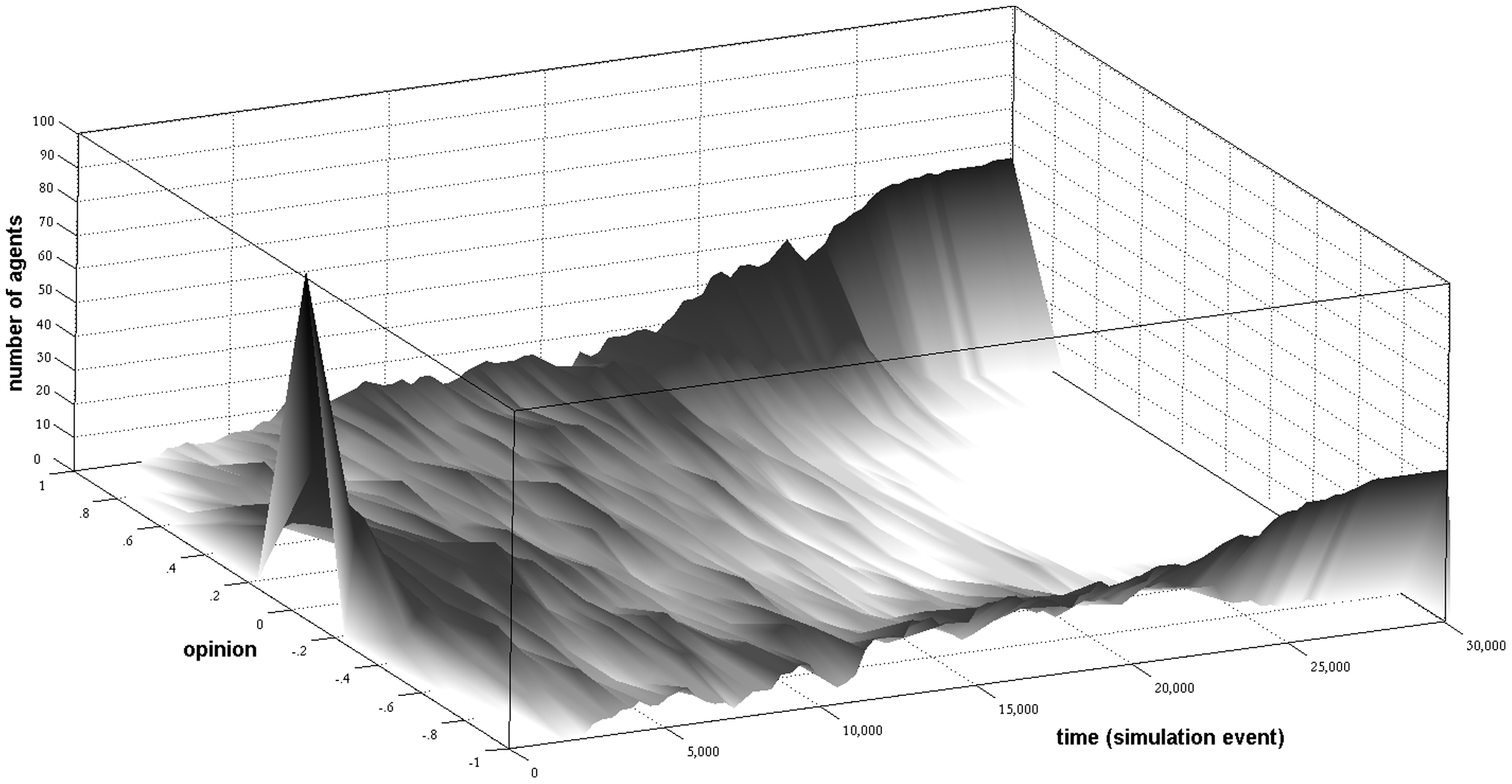}
\caption{In the ACTB model, strong homophily drives the population from initial consensus to a bi-polarized state with two opposing groups. Figure from \textcite{mas_differentiation_2013}.}\label{fig:mas_polarization}
\end{figure}

\textcite{mas_differentiation_2013} introduce a model of polarization derived from their \textit{argument-communication theory of bi-polarization} (ACTB), which is inspired by earlier work on persuasive argument theory \cite{isenberg_group_1986} and group polarization \cite{sunstein_law_2002, schkade_when_2010} (see Sec.~\ref{subsec:micro_mech}).

According to ACTB, individuals hold $K$ binary arguments $a_j\in\{-1,+1\}$, either in favor ($+1$) or against ($-1$) an issue, and exchange those arguments upon interaction. 
In a simplified version of the model, the opinion of agent $i$ is given by their average argument, i.e., $x_i=K^{-1}\sum_{j=1}^K a_j\in[-1,1]$.
For example, if $K=4$ and an agent has four arguments in favor of the issue, their opinion is one ($x_i=1$). 
Instead, if they hold two pro- and two con-arguments, their opinion is zero ($x_i=0$). 
Interactions are ruled by homophily, i.e., two agents $i$ and $j$ interact with probability $p_{ij}= (\Delta x_{ij})^h/\sum_j (\Delta x_{ij})^h$,
where $\Delta x_{ij}= (2-|x_i-x_j|)/2$ denotes their opinion-similarity and $h$ is the homophily parameter.
Social influence is mediated by the exchange of arguments, where agent $i$ adopts a randomly selected argument from agent $j$.

Through simulations, it was demonstrated that ACTB can give rise to opinion extremization, where opinions leave the initial interval, and opinion bi-polarization for sufficient levels of homophily, as illustrated in Fig.~\ref{fig:mas_polarization}.
The finding was later confirmed by \textcite{fu_opinion_2016}, 
who demonstrated the strong influence of the network structure on ACTB, where certain networks, e.g., small-world networks, promote the emergence of polarization. Additionally, 
analytical arguments were provided that the ACTB model can only give rise to two stable equilibria: a one-sided consensus at one of the extreme opinions ($+1$ or $-1$) or a bi-polarized state. 

Directional updating, as implemented by ACTB, can generate polarization even without homophily.  
\textcite{banisch_biased_2023} extend the original ACTB framework by introducing biased argument processing, where the probability that an agent accepts an argument decreases with the distance between that argument and the agent’s current opinion. 
When this bias is sufficiently strong, the model gives rise to bi-polarization despite the absence of homophilic interactions.

\begin{figure}[tbp]
\centering
\includegraphics[width=1.\columnwidth]{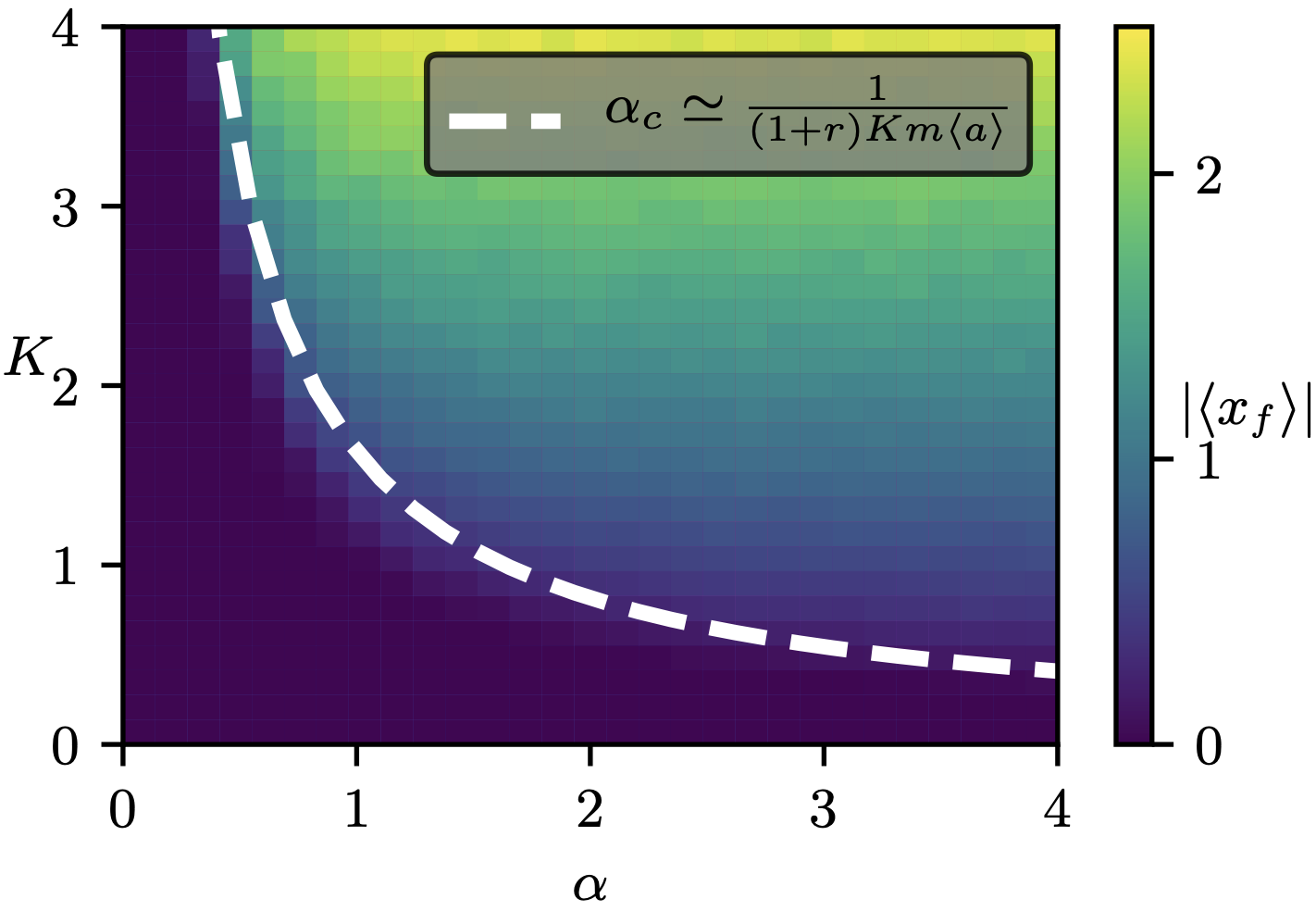}
\caption{Transition from consensus to radicalized states in the model proposed by \textcite{baumann_modeling_2020}.
Absolute values of the average final opinions $|\langle x\rangle_f|$ obtained from simulations are color coded.
In the dark region, the system approaches a neutral consensus, while in
the lighter areas, the population undergoes radicalization dynamics, which lead to polarization for high homophily.
The dashed line indicates a mean-field prediction separating the two regimes.
Figure from \textcite{baumann_modeling_2020}. 
}
\label{fig:baumann_transition}
\end{figure}

A more abstract type of directional updating is implemented in \cite{baumann_modeling_2020} and its extensions \cite{baumann_emergence_2021,gajewski_transitions_2022, santos_link_2021,pal_depolarization_2023,currin_depolarization_2022}. 
Instead of relying on an explicit exchange of arguments, opinions evolve according to a coupled system of differential equations $\dot{x}_i=-x_i+K\sum_i A_{ij}(t)\tanh(\alpha x_j)$\, that models three key mechanisms: (i) an agent's finite memory, the (ii) directional updating of opinions through a tunable sigmoidal nonlinearity, $\tanh(\alpha x_j)$, where $\alpha$ models the controversialness of the debate, and (iii) a temporal network $A_{ij}(t)$ based on the activity-driven model (Sec.~\ref{subsubsec:consensus-temp}) that co-evolves with the opinions, based on a homophily mechanism similar to \cite{mas_differentiation_2013}. 
The transition between consensus and non-consensus states is captured by $\alpha\simeq 1/(2Km\langle a\rangle)$, relating the controversialness $\alpha$ to the strength of social influence $K$, and the parameters of the temporal network $\langle a\rangle$ and $m$ that stand for the average activity and number of contacts per agent, respectively (see \fref{fig:baumann_transition}). For high homophily, non-consensus states evolve into opinion bi-polarization.
A similar model, based on static and random coupling matrices $A_{ij}$, was earlier used to investigate the transition to chaos in neural networks using dynamic mean-field theory \cite{sompolinsky_chaos_1988,kadmon_transition_2015}.

\textcite{evans_opinion_2018} investigated opinion evolution driven by a fitness-maximization process; an approach inspired by co-evolutionary games \cite{perc_coevolutionary_2010}. The fitness of agent $i$ is given as $f_i=\alpha |x_i|+\beta \sum_jA_{ij}x_ix_j+\gamma \sum_j A_{ij}$, where $\alpha$, $\beta$, and $\gamma$ define the weights of three contributions: strong conviction ($\alpha$), being connected to like minded-peers, i.e. homophily ($\beta$), and strong connections ($\gamma$). The model's connection to frameworks of directional updating is revealed by $\partial f_i/\partial x_i=\alpha\,\mathrm{sign}(x_i)+\beta\sum A_{ij}x_j$: to maximize fitness (i.e., generate positive values of $\partial f_i/\partial x_i$) agent $i$ increases their conviction $|x_i|$ in the direction (positive or negative) of their neighbors. Similarly, network rewiring unfolds by strengthening connections between like-minded individuals as $\partial f_i/\partial A_{ij}=\beta x_ix_j+\gamma$. Simulations reveal that the parameter choices of $\gamma$ and $\beta$ mainly determine the dynamics of the model, leading to polarized states for high values of homophily ($\beta$) and to a one-sided consensus for low $\beta$.

\textcite{shin_tipping_2010} implement directional updating as a process of information accumulation, in which the opinion of agent $i$ evolves according to $x_i^{t+1} = (1-\gamma)x_i^{t} + (1-|x_i^{t}|)\sum_j A_{ij}x_j^{t}$, where $\gamma$ captures finite memory of agents, $A_{ij}$ is the influence weight from $j$ to $i$, and $(1-|x_i^{t}|)$ is a saturation factor. 
A mean-field analysis of two interacting communities (each represented by a single logistic map) shows that sustained opinion diversity across communities (i.e., polarization) persists only when within-community coupling is sufficiently strong and between-community coupling sufficiently weak.

\paragraph{Internal and expressed opinions.} 
Some models of directional updating relax the classical assumption that a single variable captures both agents’ internal states (i.e., opinions) and observable actions (e.g., expressed opinions)~\cite{banisch_opinion_2019, martins_continuous_2008, martins_mobility_2008,zino_two-layer_2020,kaminska_impact_2025}. 
\textcite{banisch_opinion_2019} introduced a polarization model in which agents express binary opinions $s_i\in\{-1,+1\}$, but hold internal opinions, $x_i(s_i)$, which correspond to their convictions of each of the two expressed opinions. If $x_i(s_i=+1)>x_i(s_i=-1)$, agent $i$ expresses opinion $s_i=+1$, and vice versa. 
At each time step, agents interact with randomly selected neighbors in a static network and update their convictions $x_i$ according to a $Q$-learning algorithm, 
$x_i(s_i)\gets(1-\alpha)x_i(s_i)+\alpha s_is_j$ 
, where $\alpha$ corresponds to the learning rate \cite{clifton_q-learning_2020}. The term $\alpha s_is_j$ leads to the reinforcement of similar opinions and the weakening of dissimilar opinions. 
For example, if agent $i$ (with $s_i=+1$) gets feedback from agent $j$ expressing opinion $s_j=+1$, $x_i(+1)$ will increase and $x_i(-1)$ will decrease.
In segregated networks with pronounced community structure, the model yields local group polarization within communities and global bi-polarization between increasingly opposed groups.

Similarly, the Continuous Opinions and Discrete Actions (CODA) model is based on a combination of external and internal opinions \cite{martins_continuous_2008, martins_mobility_2008,martins_mobility_2008}. Agents hold internal continuous opinions $x_i \in [0,1]$, which encode their support for a binary issue, and are not directly influenced by the opinions of others. Instead, opinion updates of focal agents are driven by the discrete and observable actions of their peers, $s_i = \mathrm{sign}(x_i - 0.5)$, which are associated to expressing support ($s_i = +1$) or opposition ($s_i = -1$) to the issue. More specifically, when an agent observes a peer choosing $s_i = -1$, a Bayesian update rule adjusts the agent’s log-odds, $v_i = \ln[x_i / (1 - x_i)]$, by decreasing it by an amount $a$, which depends on the perceived likelihood that the peer favors the "correct" choice. Applied to the voter model \cite{holley_ergodic_1975} and the Sznajd model \cite{sznajd-weron_opinion_2000} on regular lattices, the CODA update rule can lead to qualitatively similar polarization patterns, suggesting that this behavior is a robust feature of the Bayesian update mechanism rather than a consequence of the specific interaction topology.

\subsubsection{Biased assimilation} 
Homophily can lead to polarization not only in directional-updating models (see previous section) but also in modified assimilation frameworks.  
A key example is \emph{biased assimilation}, a form of similarity bias (\sref{subsec:micro_mech}), in which evidence consistent with an agent's opinion is weighted more heavily than contradictory information \cite{lord_biased_1979}.

\textcite{dandekar_biased_2013} model biased assimilation building on the DeGroot model (\sref{subsubsec:opinion_assimilation}). In contrast to DeGroot's original work,
it is assumed that agents process information in a biased way, i.e., individuals weigh confirming evidence more heavily. In particular, individual $i$ weighs their peer's support for 1, i.e., $x_i(t)$, by an additional factor $x_i(t)^{b_i}$, and their peer's support for 0, i.e., $1-x_i(t)$, by $(1-x_i(t))^{b_i}$, where $b_i$ is the bias parameter. 
Individuals update their opinion according to 
\begin{equation}
    x_i(t+1)=\frac{A_{ii}x_i(t)+(x_i(t))^{b_i}s_i(t)}{A_{ii}+(x_i(t))^{b_i}s_i(t)+(1-x_i(t))^{b_i}(d_i-s_i(t))}\,,
\end{equation}
where $s=\sum_jA_{ij}x_j(t)$ is the weighted sum of opinions of individual $i$'s neighbors and $d_i=\sum A_{ij}$ is the weighted degree of $i$. 
The mechanism of biased assimilation is different to the idea of bounded confidence models (cf. Section~\ref{subsubsec:opinion_assimilation}), where agents do not interact at all with peers that have an opinion sufficiently different from them. 
Biased assimilation can promote polarization, in particular, when the underlying network exhibits sufficient levels of homophily.
Several subsequent works building on biased assimilation combined it with a backfire mechanism \cite{chen_opinion_2021}, and investigated it on signed networks \cite{wang_biased_2020,wang_signed_2022}.

In modeling frameworks where opinions and influence networks co-evolve (see Sec.~\ref{subsec:coevolution} for an in-depth discussion), even unbiased forms of assimilation, which usually promote consensus (Sec.~\ref{subsubsec:opinion_assimilation}), can lead to persistent polarization. \textcite{liu_emergence_2023} couple opinion assimilation with homophily in a system of stochastic differential equations  
$d x_i =\mu_i(\mathbf{x},\mathbf{A})\,dt + \sigma(x_i)\,dW_t$, where $W_t$ is a standard Wiener process and the drift term is $\mu_i = \sum_j A_{ij}F(x_i,x_j)$. Homophily operates through link creation and deletion at rates $\gamma_{\pm}(x_i,x_j)/N$.  
A minimal model is then investigated with $F(x_i,x_j)=\lambda(x_j-x_i)$, capturing opinion assimilation, and $\gamma_{\pm}(x_i,x_j)=r_{\pm}\bigl(1+J_{\pm}\bigr)$, where $|J_{\pm}|\le 1$ quantifies the homophilic bias of agents. Analytical steady-state results show that polarization emerges when the opinion-homogenization rate $\lambda$ is small and the birth-to-death ratio $r_{+}/r_{-}$ is large: slow assimilation hinders the emergence of a global consensus, while high $r_{+}/r_{-}$ limits exposure to dissimilar views, effectively partitioning the network into two almost-isolated components in which local assimilation prevails. The proposed framework gives new analytical insights by predicting universal scaling laws of stationary opinion distributions, $p(x)\sim(1\mp x)^{\delta_{\pm}}$, with exponents $\delta_\pm$, that are in line with the scaling law discovered in real-world networks.
However, the model does not account for opinions becoming more extreme than the initial range, similar to bounded confidence models (Sec.~\ref{subsubsec:bounded_confidence}).

\subsubsection{Algorithmic mediation}
\label{subsec:algo-mediation-polarization}

A growing body of research shifts attention from purely human interactions to socio-technological systems in which intelligent algorithms mediate information flows between users \cite{brinkmann_machine_2023,wagner_measuring_2021}.  
Here we review models that examine how algorithmic mediation fosters or sustains opinion polarization. Most of these works extend well-established opinion-dynamics frameworks, including models based on confidence bounds \cite{sirbu_algorithmic_2019,cinus_effect_2022,pansanella_modeling_2022,pansanella_mass_2023}, (biased) assimilation \cite{ramaciotti_morales_auditing_2021,lanzetti_impact_2023,baumann_optimal_2024,zhou_modeling_2024}, attraction-repulsion mechanisms \cite{ferraz_de_arruda_modelling_2022}, and directional updating \cite{santos_link_2021,currin_depolarization_2022}.

A central theme is algorithmic filtering, which mediates content exposure. Different filtering strategies, such as random, recency-based, or belief-aligned, have been shown to influence the degree of polarization \cite{perra_modelling_2019}. Reinforcing existing views promotes polarization, especially in structured networks, while heterogeneity in connectivity can counteract this effect. Algorithmic curation that provides agents with personalized information has also been found to increase polarization in a voter model \cite{de_marzo_emergence_2020}. Similarly,  \textcite{peralta_opinion_2021} model biased exposure by deriving evolution equations that incorporate network topology, noise, and filtering asymmetry. Their analysis reveals that even mild algorithmic biases can steer the system toward consensus, coexistence, or polarization. Extending this perspective, a more general framework was introduced to link algorithmic bias with interaction structure \cite{peralta_effect_2021}. While pairwise interactions tend to amplify polarization, group interactions promote opinion coexistence, offering insight into which dynamics are more robust to algorithmic influence.
Relatedly, \textcite{bellina_effect_2023} investigate the co-evolution of a recommender system based on collaborative-filtering and the users’ own ratings (or opinions).
The analysis uncovers phase transitions from a disordered to either a consensus or polarized state, with higher levels of similarity bias promoting polarization.
While algorithmic personalization tends to increase polarization, definitive conclusions are difficult, as outcomes depend on the interplay between the type of algorithm and the micro-mechanism of opinion change \cite{mas_will_2015}.

While much attention has been paid to content filtering, other models highlight the role of algorithmic link formation. \textcite{santos_link_2021} propose a mechanism in which users form connections based on structural similarity (shared neighbors) rather than opinion alignment. This process can foster polarization even without intentional avoidance of opposing views. Similarly,
\textcite{cinus_effect_2022} examine how different link recommendation algorithms affect the polarization dynamics in the \dw (\sref{subsubsec:bounded_confidence}). Their study considers structurally driven methods, such as personalized PageRank and Jaccard similarity, alongside opinion-based algorithms that promote ties between individuals with similar views. Agent-based simulations reveal that the influence of recommender systems on polarization varies significantly. Algorithms that fail to bridge existing divides, measured by community structure metrics, tend to exacerbate polarization. A similar conclusion is reached in \cite{sirbu_algorithmic_2019}, where the \dw is extended to mimic an algorithmic bias in the way individuals interact. In particular, a homophily mechanism is introduced that replaces the random sampling of pairs of interacting individuals. Algorithmic bias both dramatically slows convergence (making consensus practically unattainable) and increases opinion fragmentation, especially under stronger bias, initially fragmented conditions, and in small populations subject to finite-size effects.
 
Models have also focused on how algorithmic mediation can help mitigate radicalization and polarization, a topic of growing interest \cite{ovadya_bridging_2023}. 
In particular, \textcite{pansanella_mass_2023} find that an algorithmic bias (based on homophily in the interactions) \cite{sirbu_algorithmic_2019} can mitigate the extremization of opinions induced by media sources.
Two other studies introduce nudging strategies that could be implemented as algorithmic interventions. \textcite{currin_depolarization_2022} propose the Random Dynamical Nudge (RDN), an intervention in which users receive input from randomly selected peers  effectively reducing or even preventing polarization. Similarly,  \textcite{pal_depolarization_2023} introduced a network-based nudge that encourages occasional random ties. While effective in reducing echo chambers, overly strong nudging can lead to radicalization. Both studies underscore the potential of minimal, privacy-preserving mechanisms for depolarization. %

\subsubsection{Multidimensional polarization}
\label{sec:multidim_pol}

When individuals are assumed to hold views on multiple issues, polarization may appear not only as bi-modal distributions on such issues but can also reveal strong correlations across issues, known as \emph{issue alignment}~\cite{converse_nature_2006,kozlowski_issue_2021}; see \fref{fig:polarization2} in \sref{sec:macro:polarization}.
Multidimensional models characterize each agent $i$ by a $D$-dimensional opinion vector, $(x_{i,1}\dots,x_{i,D})$, 
where $x_{i,d}(t)$ denotes the opinion of agent $i$ on issue $d$ at time $t$; in practice most works focus on the two-dimensional case ($D=2$).  
Multidimensional models specify (i) how agents interact, and (ii) a micro-mechanism for opinion change, which is often similar to those previously considered in one-dimensional models ($D=1$) and include attraction-repulsion dynamics \cite{huet_openness_2010,flache_why_2008,flache_small_2011,schweighofer_agent-based_2020,li_agent-based_2017}, confidence bounds \cite{schweighofer_agent-based_2020}, directional updating \cite{baumann_emergence_2021}, and formulations rooted in Heider’s balance theory \cite{schweighofer_weighted_2020}. 
In addition, multidimensional models require assumptions on how opinions on different issues influence each other.

The remainder of this section focuses on models that aim to explain the emergence of both issue alignment \cite{dellaposta_why_2015} and hyper-polarization \cite{huet_openness_2010,schweighofer_weighted_2020,baumann_emergence_2021}, i.e., the combination of issue alignment and polarization.

\textcite{huet_openness_2010} proposed a two-dimensional model based on bounded confidence where the opinion dynamics on one issue is coupled to that on the other.
Specifically, agents’ opinions attract each other on both dimensions if they are within the confidence bound on the first, but repel each other on the second if they differ on the first yet are too close on the second.
This coupling can lead to a "hyper-polarized" state with three clusters and strong issue alignment. 
In a similar model, \textcite{schweighofer_agent-based_2020} compared proximity and directional voting, finding that only the latter yields alignment, especially in low dimensions. Adding repulsion between distant opinions enables robust global alignment even in high-dimensional spaces with directional voting

\textcite{flache_why_2008,flache_how_2008} not only consider agents' continuous opinions on $D$ issues ($x_{i,d}$), but also $K$ binary and fixed demographic attributes $(l_{i,k})$. 
Social influence is moderated by the sign and strength of the interpersonal relation between individuals $-1\leq A_{ij}\leq1$, which co-evolve with their opinions in discrete-time: the change of agent $i$'s opinion on issue $D$ is proportional to $\Delta x_{i,d}\sim \sum_j A_{ij}(x_{j, d}(t)-x_{i,d}(t))$ %
and interpersonal relations are updated according to $A_{ij}(t+1)=1-(\sum_k|l_{i,k}-l_{j,k}|+\sum_d|l_{i,d}-l_{j,d}|)/(D+K)$, i.e., in response to the average distance between agents in the combined space of demographics and opinions.
Issue alignment emerges for sufficiently strong correlations between demographic characteristics, which are assumed a priori.
Without demographic attributes and a fixed social network, the model can still give rise to hyperpolarization, however, only in low dimensions \cite{flache_small_2011}.

\begin{figure}[tbp]
\centering
\includegraphics[width=1.\columnwidth]{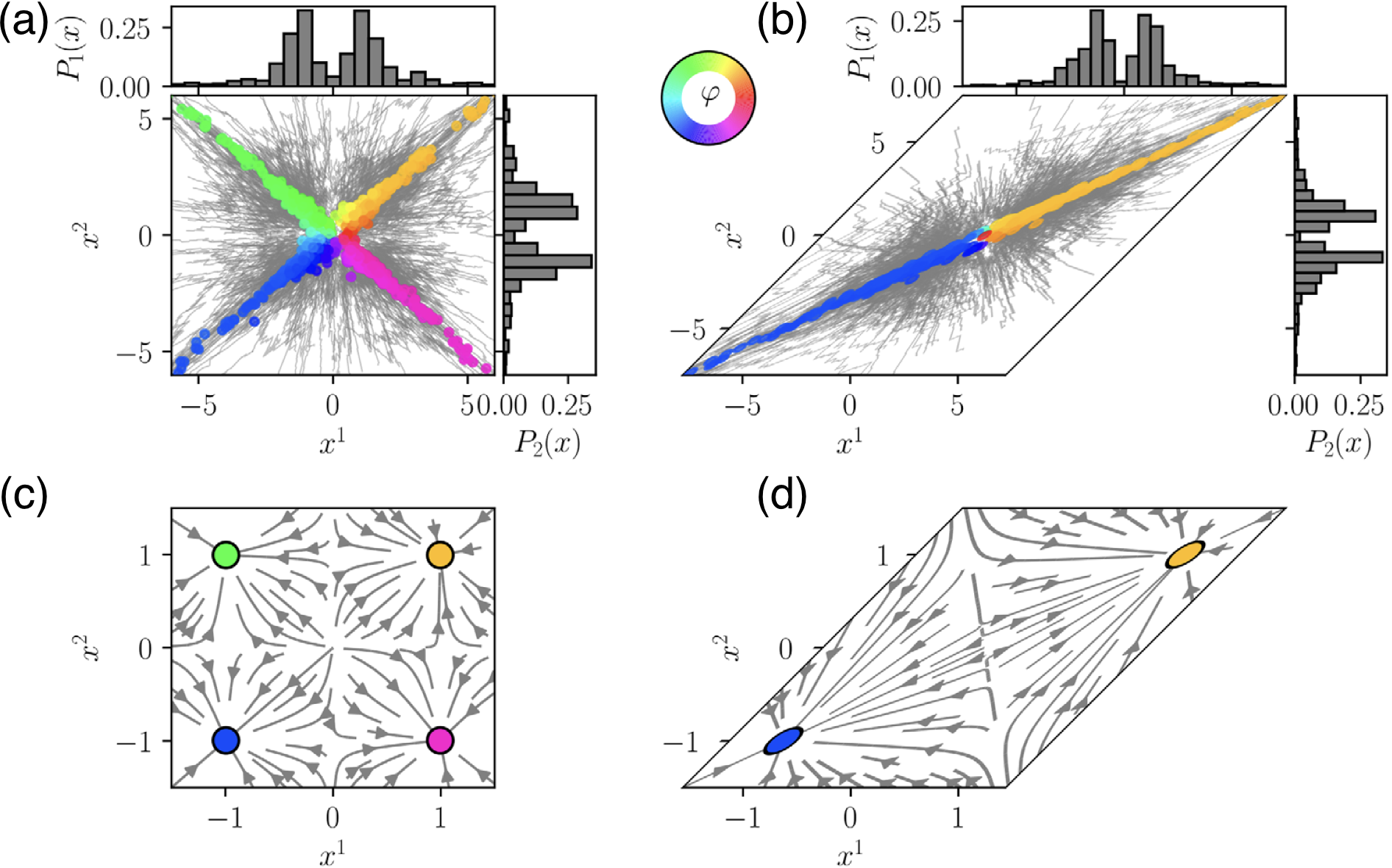}
\caption{Polarized states that emerge for high homophily, for two topics represented in two dimensions, in the model proposed by \textcite{baumann_emergence_2021}.
Color codes the opinions of agents with respect to the two topics, represented by the value $\varphi$.
Panels (a) and (b) show simulation results for small and high values of the topic overlap $\phi_{12}$, which leads to polarized states that are uncorrelated and correlated, respectively. 
Marginal opinion distributions for each topic, shown on the plot margins, are very similar. 
Panels (c) and (d) depict the corresponding deterministic dynamics and its attractors derived from a mean-field approximation. Figure adapted from~\cite{baumann_emergence_2021}.
}\label{fig:ideological_states_baumann}
\end{figure}

In another line of research, \textcite{schweighofer_weighted_2020} introduced a model inspired by Heider's balance theory \cite{heider_attitudes_1946}. 
In their "Weighted Balance Theory" (WBT), opinions on several issues, $x_{i,d}$, co-evolve with interpersonal attitudes, $A_{ij}$, such that opinions are updated in order to increase balance. 
If agent $i$ likes agent $j$ (i.e., $A_{ij}>0$) yet their views on issue $d$ differ, the triad is imbalanced; balance increases by $i$ shifting their opinion $x_{i,d}$ toward $x_{j,d}$. 
Conversely, if $i$ dislikes $j$ ($A_{ij}<0$), balance grows when $x_{i,d}$ moves away from $x_{j,d}$. 
Simulations of the WBT model show that it yields either consensus or hyperpolarized states. 
In particular, hyperpolarization only occurs if evaluative extremeness is considered, i.e., if interpersonal attitudes $A_{ij}$ are evaluated based on a sigmoidal non-linearity. 
Hyperpolarization arises when evaluative extremeness is included, i.e., interpersonal attitudes $A_{ij}$ are shaped by a sigmoidal non-linearity. As in earlier models, higher dimensional spaces destabilize hyperpolarization and promote consensus.

\textcite{baumann_emergence_2021} drop the assumption of orthogonal issue dimensions, allowing them to overlap.
Such topic overlaps shape the micro-dynamics of opinion change: when agents $i$ and $j$ interact, $i$’s opinion on issue $d$ is influenced not only by $j$’s opinion on $d$ but also by $j$’s stance on any other issue $\ell$, with the strength of that cross-issue influence captured by the symmetric overlap matrix $\phi_{d\ell}$. For high homophily the model displays three regimes---consensus, uncorrelated polarization, and hyper-polarization. The latter emerges when the topic overlap exceeds a critical threshold, causing the previously uncorrelated polarized state to lock into strongly aligned issue positions, as depicted in \fref{fig:ideological_states_baumann}.

When multiple topics are discussed at the same time, the correlations between opinions may play a role in both hyper-polarization and in the process of reducing it, also known as \emph{depolarization}. 
\textcite{ojer_modeling_2023} proposed the social compass model, where agents exert social influence across two interdependent topics represented in a polar space.
The model shows a phase transition from polarization to consensus, at a critical threshold of social influence, that can be computed analytically.
The transition nature changes depending on the correlations among initial opinions: A first-order phase transition emerges when opinions are uncorrelated, while the transition is continuous if opinions are correlated. 
\textcite{ojer_social_2025} extended the social compass model to any dimension, uncovering an upper critical dimension for uncorrelated initial opinions, distinguishing between discontinuous and continuous phase transitions.
Furthermore, when considering a heterogeneous network connectivity between agents, the threshold value approaches zero for an infinite network size, indicating that the presence of influential nodes or hubs can facilitate consensus by effectively lowering the social influence required for depolarization.

\subsection{Echo chambers}
\label{subsec:echo_chambers} 

Empirical research, particularly on social media platforms (\sref{sec:empirical_data}), has documented the existence of \textit{echo chambers} (ECs), i.e., settings in which individuals are predominantly exposed to views that reflect their own opinions (\sref{subsec:macro-phen}). 
Quantifying the extent to which social systems are characterized by echo chambers is not sufficient to answer the important questions about their origin and down-stream consequences.
Modeling studies have therefore begun to fill this gap, pursuing two main goals: first, identifying the micro-level processes that give rise to the emergence of ECs and, second, investigating the consequences.

\subsubsection{Emergence of echo chambers}

A key mechanism behind the formation of echo chambers is the \textit{co-evolution} of opinions and network structures.
Models that implement opinion-network co-evolution typically couple a micro-mechanism of opinion change with a link-rewiring rule, typically based on homophily \cite{baumann_modeling_2020,baumann_emergence_2021,wang_public_2020,liu_emergence_2023,sasahara_social_2021,evans_opinion_2018,piao_humanai_2023, vazquez_generic_2008, saeedian_absorbing-state_2020}.
In continuous opinion spaces, this feedback loop can either widen the gap between opposing individuals through directional updating [e.g., \cite{banisch_opinion_2019}, see \sref{subsubsec:opinion_reinforcement}] or deepen within-group similarity through assimilation [e.g., \cite{liu_emergence_2023}, see \sref{subsubsec:opinion_assimilation}].
Within co-evolving frameworks it was shown that echo chamber formation can be promoted by system-level properties such as controversial debates \cite{baumann_modeling_2020,baumann_emergence_2021}, political campaigns acting as external forces \cite{wang_public_2020}, individual cognitive factors such similarity bias \cite{wang_public_2020}, as well as combinations thereof \cite{evans_opinion_2018}, with synergies emerging among mechanisms: even minor individual biases, combined with infrequent social rewiring, can generate pronounced echo chambers \cite{sasahara_social_2021}.
Moreover, synergies between individual biases and algorithmic mediation (\sref{subsec:algo-mediation-polarization}) can further entrench ECs \cite{piao_humanai_2023,cinus_effect_2022}. In particular, \textcite{geschke_triple-filter_2019} introduce a modeling framework, incorporating cognitive, social, and technological filtering mechanisms. 
They find that, even in the absence of technological  and social filters, echo chambers can emerge purely from individual biases. Adding social and algorithmic filters further amplifies segregation, leading to more distinct and isolated opinion clusters.
Inspired by early models in sociology \cite{schelling_models_1969,schelling_dynamic_1971}, echo chambers have also been explored in \emph{physical} space by modeling individuals as random walkers on a two-dimensional plane \cite{starnini_emergence_2016}. Upon encounter, agents update their opinions through assimilation (see \sref{subsubsec:opinion_assimilation}), and homophily is implemented by letting like-minded pairs interact for longer periods, which can result in a metapopulation structure in which opinions are highly similar within spatial clusters yet different across clusters. Note that many earlier studies, in particular those focusing on the adaptive voter model, have analyzed similar co-evolutionary dynamics without explicitly invoking the term “echo chamber", yet their models often yield comparable macroscopic phenomena. For a detailed discussion of the models that give rise echo chamber-like phenomena in co-evolving settings, see Sec.~\ref{subsec:coevolution}.

Echo chambers can also form on \emph{static} social networks, provided the network is sufficiently modular \cite{banisch_opinion_2019,franken_cascades_2021,botte_clustering_2022,iannelli_filter_2022}. This is the case across models built on diverse micro-mechanisms of opinion change including directional updating in continuous opinion dynamics, e.g., \cite{banisch_opinion_2019,baumann_modeling_2020} which are discussed in \sref{subsubsec:opinion_reinforcement}. 
A complementary approach for discrete opinion dynamics is presented in \cite{botte_clustering_2022}, where opinions evolve according to a probabilistic majority model (\sref{sec:majority_rule}). Again, the model shows that clustered network structures (e.g., Watts-Strogatz networks) promote echo chambers, which are further amplified by algorithms that cater opinions to individuals that align with their current opinions. 
Interestingly, stubborn individuals (who resist opinion shifts) can reduce this effect. %
Beyond the classical paradigm of repeated interactions between individuals, information diffusion can lead to echo chambers on static networks. \cite{franken_cascades_2021}, for instance, demonstrate that information cascades combined with Bayesian opinion updating can be sufficient for echo chamber formation, even without explicit homophily or individual-level biases.

\subsubsection{Impact of echo chambers}

Echo chambers are often cited as a cause of opinion radicalization or polarization~\cite{mahmoudi_echo_2024}. However, the modeling results reviewed in \sref{subsec:polarization} suggest a more nuanced relationship, in which echo chambers tend to co-occur with polarization rather than directly cause it.
For example, in models of directional updating combined with homophily (\sref{subsubsec:opinion_reinforcement}), being connected with like-minded people, i.e., in an echo chamber, can indeed reinforce a person's opinion and even lead to polarization when two opposing camps are present. However, the reason for this dynamic is not the echo chambers, but a combination of simpler mechanisms that gives rise to all phenomena---radicalization, polarization, and ECs---at the same time.

Nevertheless, once echo chambers have formed, they can give rise to various phenomena, e.g. selective information filtering. Building on simple social–feedback principles, \textcite{banisch_modelling_2022} devised a $Q$-learning \emph{opinion-expression game} in which agents repeatedly choose to voice ($E$) or withhold ($S$) their stance; like-minded clusters buffer negative out-group feedback, encouraging minority members to speak up while simultaneously reinforcing the perceived dominance of each camp. These findings imply that, even if echo chambers do not directly cause polarization in some cases, they may reshape who speaks, what is heard, and how confidently different viewpoints are articulated.

This idea is also reflected in the main direction of research on the impact of ECs, namely their impact on information dissemination: Do ECs affect the way information diffuses through a population, and might this promote the spread of misinformation? A growing body of work has aimed to shed light on these questions by combining models of networks and social contagion \cite{tornberg_echo_2018,diaz-diaz_echo_2022,vendeville_modeling_2025,cota_quantifying_2019}.

\textcite{tornberg_echo_2018} models information diffusion as complex contagion on an Erdős–Rényi network with an embedded echo chamber, i.e., a cluster with stronger internal than external connectivity.
Inside the EC the threshold of individuals to accept information is lower than outside of it, a mechanism through which the EC significantly amplifies the spread of information, by acting as a bandwagon, facilitating the diffusion of information also outside of the EC.

Related work investigated information diffusion on scale-free networks with tunable levels of homophily between two node types A and B (associated to the a minority and majority, respectively) \cite{lee_homophily_2019}, and analyzed how different types of social contagion evolve on top of it \cite{diaz-diaz_echo_2022}.
For simple contagion, the diffusion shows little sensitivity to the homophily level, whereas for complex contagion the reach of information grows sharply with increasing homophily. The richest behavior appears in the Hybrid Contagion Model, introduced by the authors, where information either spreads as simple or complex contagion depending on the type of the receiving individual. 
Here, as in the case of complex contagion, information diffusion strongly depends on the level of homophily, but also on the source of the spreading process, i.e., an individual minority or majority. This asymmetry is an important insight, as it suggests that echo chambers can affect different communities unequally.

\textcite{cota_quantifying_2019} reported similar results based on empirical communication drawn from Twitter, where nodes represent users and a directed link from node $i$ to node  $j$ indicates that user $i$ has tweeted mentioning $j$.
The study finds that echo chambers manifest as well-separated communities with asymmetric information-spreading capacities, and users' effectiveness to disseminate content critically depends on the ECs they are residing in.

\subsection{Coevolution of network and opinions}
\label{subsec:coevolution}

\begin{figure}[tbp]
\centering
\includegraphics[width=\columnwidth]{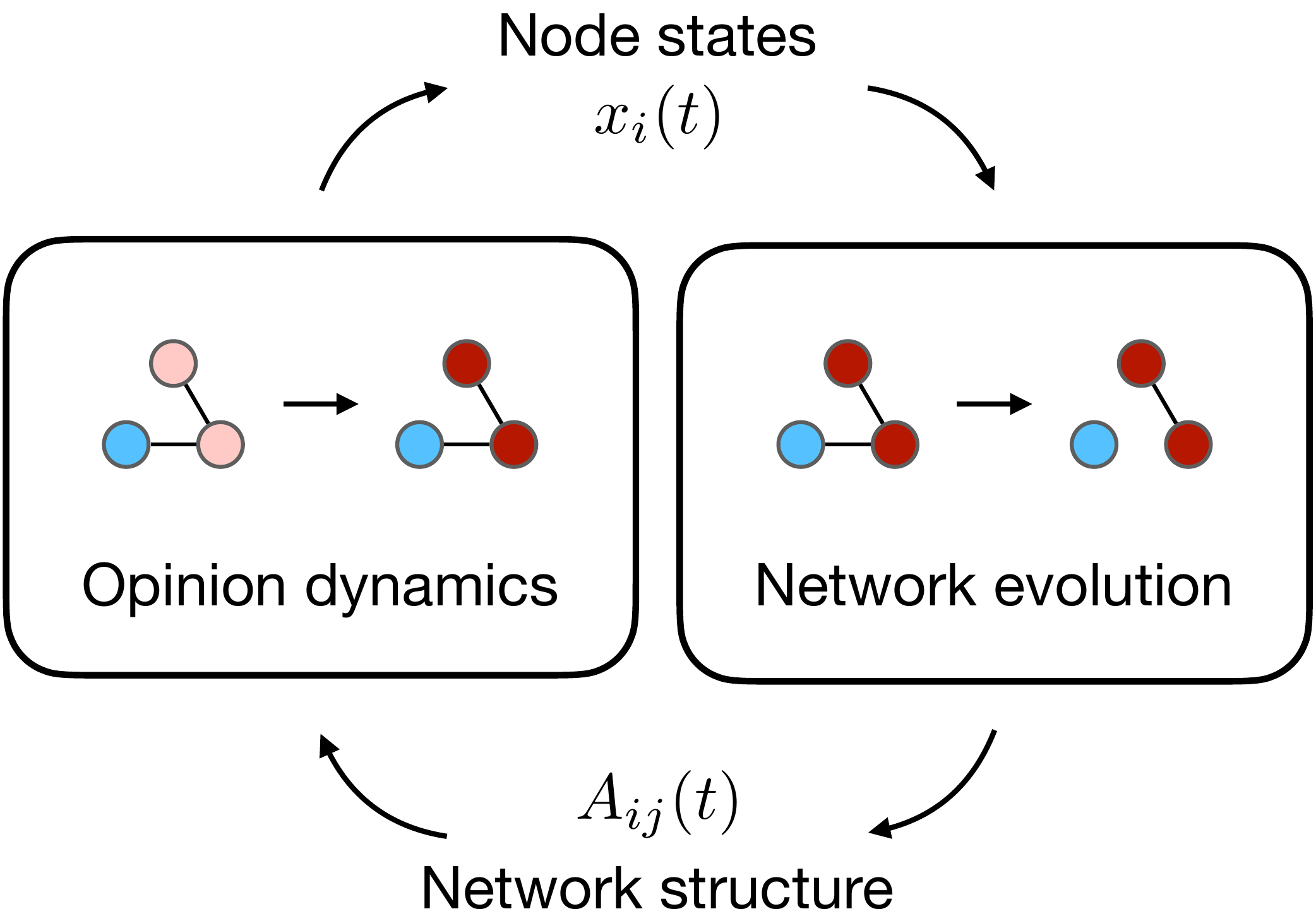}
\caption{In coevolutionary or adaptive networks, node states $x_i(t)$ (coded by different colors) affect the network structure $A_{ij}(t)$ and vice versa. Thus, opinion dynamics and network evolution engage in a feedback loop with a dynamical exchange of information. Figure adapted from \textcite{gross_adaptive_2008}.
}
\label{fig:adaptive figure}
\end{figure}

In coevolutionary or adaptive opinion dynamics, the time evolution of network edges and node opinions are coupled to each other (see \sref{sec:net_adapt} and \fref{fig:adaptive figure}). Coevolving dynamics have been extensively studied for discrete and continuous opinions \cite{peralta_opinion_2022}, as well as in a diversity of phenomena including boolean networks \cite{bornholdt_topological_2000}, synchronization \cite{ito_spontaneous_2001}, game-theoretical dilemmas \cite{ebel_coevolutionary_2002}, epidemic spreading \cite{gross_epidemic_2006}, cultural differentiation \cite{vazquez_time-scale_2007,centola_homophily_2007}, socio-economic dynamics \cite{schelling_models_1969,biely_socio-economical_2009}, resource allocation \cite{wiedermann_macroscopic_2015}, and language competition \cite{carro_coupled_2016, charalambous_language_2023}. We refer the reader to the reviews of \textcite{gross_adaptive_2008,gross_adaptive_2009,zschaler_adaptive-network_2012,sayama_modeling_2013,berner_adaptive_2023}.

Models of network and opinion coevolution typically start from one of the standard classes of opinion dynamics on static networks (voterlike, majority rule, bounded confidence, or any of the models in \sref{sec:models}). Additionally, there are rules on how nodes and/or links may be created, destroyed, or rewired based on the opinions of nodes. Edge dynamics might be motivated by social mechanisms such as local triadic closure (a connection with the neighbor of a neighbor) or global focal closure (a connection based on node attributes, regardless of network distance) \cite{kossinets_empirical_2006,klimek_dynamical_2016}. These mechanisms might be random or based on opinion similarity, and can be implemented in networks with either a constant or growing number of nodes \cite{toivonen_comparative_2009}. Since the evolution of network structure depends on the dynamics of the nodes and viceversa, coevolving models explore feedback loops with a dynamical exchange of information between networks and opinions (see \fref{fig:adaptive figure}) \cite{gross_adaptive_2008}. Given that opinions and links evolve on timescales with a ratio determined by a plasticity parameter $p$, and fragmentation typically occurs for some critical value $p_c$, these models tend to be harder to treat analytically than non-coevolving dynamics. Still, some mean-field or pair approximations and moment closure techniques do apply \cite{zschaler_adaptive-network_2012} (for details on the treatment of opinion dynamics as stochastic processes see \sref{sec:methods}).

\subsubsection{Adaptive voterlike dynamics}
\label{subsubsec:aVM}

\paragraph{Binary opinions.}

The \avm is a prominent example of coevolving dynamics for binary opinions. At each time step a random focal node is selected, along with one of its neighbors. If the opinions of these two agents are different, with probability $1 - p$ the focal node adopts the opinion of the neighbor (voterlike dynamics); otherwise, with probability $p$ the link is broken and the focal node connects to a random node with the same opinion (link rewiring). The model shows an absorbing phase transition in the thermodynamic limit at $p_c = (\langle k \rangle - 2) / (\langle k \rangle - 1)$, with $\langle k \rangle$ the average degree of the network and mean-field critical exponent 1 \cite{vazquez_generic_2008}. In finite systems, the transition separates an active phase ($p < p_c$) where nodes of different opinions are still connected for long times, before eventually reaching an absorbing state of consensus, from a frozen fragmented phase ($p > p_c$) with two disconnected components, each with nodes in one of the two states \cite{demirel_moment-closure_2014}.

Similar results hold even when rewiring is random regardless of opinion, but convergence times for low rewiring ($p \to 0$) depend on the details of the voter rule. If the focal node in each step adopts the opinion of the neighbor (direct voter), time to consensus is logarithmic on the system size $N$, but if the neighbor changes opinion instead (reverse voter), convergence time increases exponentially with $N$ and links between disagreeing nodes remain for long times \cite{nardini_whos_2008}. The active phase is characterized by a statistical conservation law between measures of magnetization of nodes and links, which is lost by crossing the transition into the fragmented phase \cite{toruniewska_coupling_2017}.

Details of the opinion update and rewiring rules in the \avm have effects on the convergence time to consensus  \cite{rogers_consensus_2013}. When link rewiring is replaced by link deletion, phases of consensus and fragmentation remain, separated by a critical point where very few links stay in the network \cite{zanette_opinion_2006,gil_coevolution_2006}. Edge rewiring based on preferential attachment rather than opinion similarity promotes consensus \cite{zhong_effects_2010}. A tunable amount of random rewiring vs. triadic closure also shifts the transition to consensus \cite{malik_transitivity_2016}. \textcite{li_how_2021, li_effects_2022} have studied a two-state voter model on a dynamic network with inflexible zealots. Instead of the commonly used link rewiring, agents in this model create and delete links (i.e. a fluctuating number of edges). In addition to fragmentation, the authors find a so-called ``overwhelming transition''.

Recently, the \avm has been extended to include aging, where either nodes or links of increasing age are less likely to change opinion or be broken \cite{min_aging_2025}. Link aging increases the critical plasticity $p_c$, promoting the active phase and eventual consensus, but node aging does the opposite, decreasing $p_c$ and thus enhancing frozen fragmentation.

An in-depth analysis of the role of noise and various rewiring schemes in the \avm has been done by \textcite{chodrow_local_2020}. Homogeneous noise in the \avm destroys the fragmented phase and introduces a new regime of dynamic fragmentation, while targeted noise moves the phase transition to larger $p_c$ \cite{diakonova_noise_2015}. The \avm has been extended to multilayer networks, characterized by a rewiring parameter for each layer and the degree of multiplexity (fraction of nodes common to both layers) \cite{diakonova_absorbing_2014}. Multiplexing prevents fragmentation, yet it also introduces a new ``shattered'' fragmentation phase in which large consensus groups of opposite opinion coexist with isolated nodes. A similar shattered fragmentation transition in the multiplex \avm with triadic closure has been used to describe the empirical distribution of cluster sizes in an online game \cite{klimek_dynamical_2016}. In simplicial complexes, the \avm retains its phase transition but has faster convergence times to consensus and fragmentation due to peer-pressure effects \cite{horstmeyer_adaptive_2020}.

\paragraph{More than two opinions.}

The multistate \avm allows for opinions to take one of $G \geq 2$ possible values, but otherwise follows the same opinion copying and link rewiring rules of the \avm. In the seminal model introduced by \textcite{holme_nonequilibrium_2006}, for example, $G$ increases with system size $N$ but is still small, such that $G/N$ is constant. When rewiring is dominant ($p \to 1$), final clusters coincide with the initial holders of each opinion, corresponding to a multinomial cluster size distribution for random initial conditions. For $p \to 0$ there is no rewiring and opinion groups correspond to the initial components of the network (e.g. a giant component and exponentially distributed small clusters for $\langle k \rangle > 1$). These two phases are separated by a continuous phase transition with a power-law distribution of cluster sizes and critical slowing down in convergence time. Finite size scaling suggests that the critical exponents and universality class of this transition are different from those of random graph percolation \cite{holme_nonequilibrium_2006}. The nature of this transition seems to depend on the particular rewiring scheme chosen \cite{durrett_graph_2012}.

The multistate \avm has additional fragmentation transitions to ``partially fragmented'' states where not every component is internally in opinion consensus \cite{bohme_fragmentation_2012}, which also holds for directed networks \cite{zschaler_early_2012}. When link rewiring is random and as $G \to \infty$, there are infinitely many of such transitions \cite{shi_multi-opinion_2013}. Edge rewiring in coevolving dynamics is usually a minimal implementation of the homophily principle in sociology, by which similarity breeds connection \cite{mcpherson_birds_2001,asikainen_cumulative_2020}. When the multistate \avm includes heterophily (via a parameter regulating rewiring between nodes with different opinions), the dynamics leads to bridges between groups that would otherwise be internally in consensus \cite{kimura_coevolutionary_2008}. A multistate \avm with variable homophily over spatially embedded networks has been used to describe individual attitude change and aggregated spatial polarization in panel survey data from Ukraine \cite{chu_microdynamics_2021}.

\paragraph{Role of nonlinearity.}

The binary \avm can be generalized to account for nonlinear voter interactions by updating the state or links of a node with probability $(m/k)^{\alpha}$, where $m$ is the number of neighbors with the opposite opinion, $k$ its degree, and $\alpha$ a parameter regulating nonlinearity  \cite{castellano_nonlinear_2009,schweitzer_nonlinear_2009} (see \tref{sec:transition_rates} for a longer discussion of transition rates in opinion dynamics). When $\alpha > 1$ ($\alpha < 1$), the node follows the majority (minority) opinion in its neighborhood, while $\alpha = 1$ recovers the standard \avm. The model shows phases of consensus, fragmentation, and active coexistence of opinions \cite{min_coevolutionary_2023}. Beyond the transition between active and fragmented phases in the \avm, nonlinearity leads to new discontinuous phase transitions and longer convergence times \cite{min_fragmentation_2017}. With noise, these phases show further substructure: consensus can keep the same state or alternate between opinions during the dynamics, while coexistence can be fully mixed or have community structure \cite{raducha_emergence_2020}. In multilayer networks with the same rewiring probability $p$ in each layer, the fragmentation transition moves to larger $p$ (i.e. multiplexity decreases fragmentation), while different values of plasticity $p$ and nonlinearity $\alpha$ across layers lead to new phases not seen in simple networks \cite{min_multilayer_2019}.

Substituting global rewiring by triadic closure, active coexistence becomes a ``shattering'' phase where a large fraction of nodes become isolated \cite{raducha_coevolving_2018}. When edge rewiring is random, the active phase shows spontaneous symmetry breaking in terms of the ratio of nodes holding each opinion \cite{jedrzejewski_spontaneous_2020}. A majority rule (similar to the nonlinear \avm with $\alpha > 1$), together with homophilic or heterophilic edge rewiring, leads to metastable states with a diversity of opinions, before ultimately reaching consensus \cite{benczik_lack_2008,fu_coevolutionary_2008}.

\paragraph{Node and link state dynamics.}

A complementary approach to adaptive voterlike dynamics considers time-dependent state variables for edges too, signaling the positivity or negativity of an interaction in the context of theories like structural balance \cite{heider_attitudes_1946}. The majority rule (analogous to a zero temperature Glauber dynamics) has been explored for link states rather than nodes \cite{carro_fragmentation_2014}. When coupled to rewiring of edges in the local minority, the dynamics leads to consensus and fragmented phases similar to those of the standard \avm. The model of \textcite{saeedian_absorbing_2019} describes the coupled evolution of binary states of nodes and links such that agents with similar opinions favor positive interactions and viceversa. Similarly to the \avm, there is an absorbing continuous phase transition from a frozen to a dynamically active phase as a function of the probability for link updates. This behavior seems to be robust even in the presence of link rewiring \cite{saeedian_absorbing-state_2020}. Indeed, the dynamics always tends towards balanced signed networks \cite{cartwright_structural_1956}, so for finite systems one needs to identify the transitions between topological configurations. The coevolution of node and link states has also been instrumentalized as a Hamiltonian including homophilic (ferromagnetic) opinion interactions and structural balance within triads \cite{minh_pham_effect_2020}. A minimization of the Hamiltonian via a Metropolis algorithm uncovers phases of opinion coexistence and network fragmentation. Another class of models related to adaptive opinion dynamics are Hopfield networks \cite{macy_polarization_2003}, where binary opinions coevolve with weights in a social network according to structural learning (see \sref{subsec:opinion_repulsion}).

\subsubsection{Role of adaptivity on opinion and network fragmentation}
\label{subsubsec:co-evo-cont-op}

\paragraph{Bounded confidence.}

Another opinion dynamics mechanism widely explored over adaptive networks is bounded confidence (see \sref{subsec:opinion_fragmentation} and \sref{subsec:polarization}). The adaptive bounded confidence model (\abcm) couples the opinion dynamics of the \dw on static networks \cite{deffuant_mixing_2000,fortunato_universality_2004,lorenz_about_2007} with network rewiring when opinions are too dissimilar \cite{stauffer_coevolution_2006}. At each time step with probability $1 - p$, a pair of randomly selected neighbors follows bounded-confidence dynamics with tolerance $d$; otherwise, with probability $p$ one of the nodes breaks the link and reconnects to another random node. The critical tolerance $d_c$ of the \abcm (leading to consensus when $d_c < d$) grows with $p$, making consensus more difficult to achieve in the presence of random rewiring. Conversely, the transition point $d_f$ separating the fragmented phase ($d < d_f$) from the polarized phase ($d_f < d < d_c$) goes to zero for $p>0$, since agents can more easily find agreeing nodes via rewiring, promoting polarization rather than fragmentation \cite{kozma_consensus_2008}. Random rewiring in the \abcm can be substituted by a notion of homophily by cutting links with large opinion differences and rewiring them to nodes with more similar opinions \cite{kan_adaptive_2022}. Homophily increases the consensus threshold $d_c$ even further, leading to ``pseudo-consensus'' states where two subgraphs in a component have opinions differing from each other by a small amount.

Sequential pair interactions in the \abcm can be replaced by simultaneous neighbor interactions by following the opinion update rules of the \hk instead \cite{hegselmann_opinion_2002}. Random rewiring leads to similar behavior as the baseline \abcm \cite{su_coevolution_2014}. When using homophilic rewiring, bounded confidence and random noise lead to a dynamics of influence maximization \cite{brede_how_2019}. In signed networks, homophilic rewiring and simultaneous opinion updates have been implemented as a preferential flip of edge sign between agreeing nodes that recovers the structural balance of triads \cite{parravano_bounded_2016}.

Variants of the \abcm are useful in describing opinion dynamics in online social media platforms. The model of \textcite{sasahara_social_2021} mixes bounded confidence with social sharing and selective rewiring by letting agents post/repost messages and unfollow disagreeing individuals. Even with minimal amounts of influence and unfriending, the dynamics rapidly leads to segregated and homogeneous communities, which the authors use to describe echo chambers in empirical Twitter data. The \abcm with homophilic rewiring has also been extended by adding a notion of algorithmic bias (promoting interactions between similarly minded individuals) and peer pressure to follow the majority opinion within triangles \cite{pansanella_modeling_2022}. While algorithmic bias promotes fragmentation [in agreement with other models over static networks \cite{peralta_opinion_2021,peralta_effect_2021,benito_mean-field_2022}], peer pressure enhances consensus. Relaxing the bounded confidence rule to allow interactions between highly disagreeing nodes, either with random rewiring or not, leads to the coexistence of two stable opinions in the network \cite{del_vicario_modeling_2017}. A model mixing bounded confidence and noise in adaptive networks has been explored via stochastic differential equations and applied to survey data on political identity and public opinion in the US  \cite{djurdjevac_conrad_co-evolving_2024}. Recently, a variant of the \abcm with a notion of node reputation has been used to describe the dynamics of political polarization surrounding the COVID-19 vaccination debate in empirical Facebook data \cite{lipatov_coevolution_2025}.

\begin{figure}[tbp]
\centering
\includegraphics[width=1.\columnwidth]{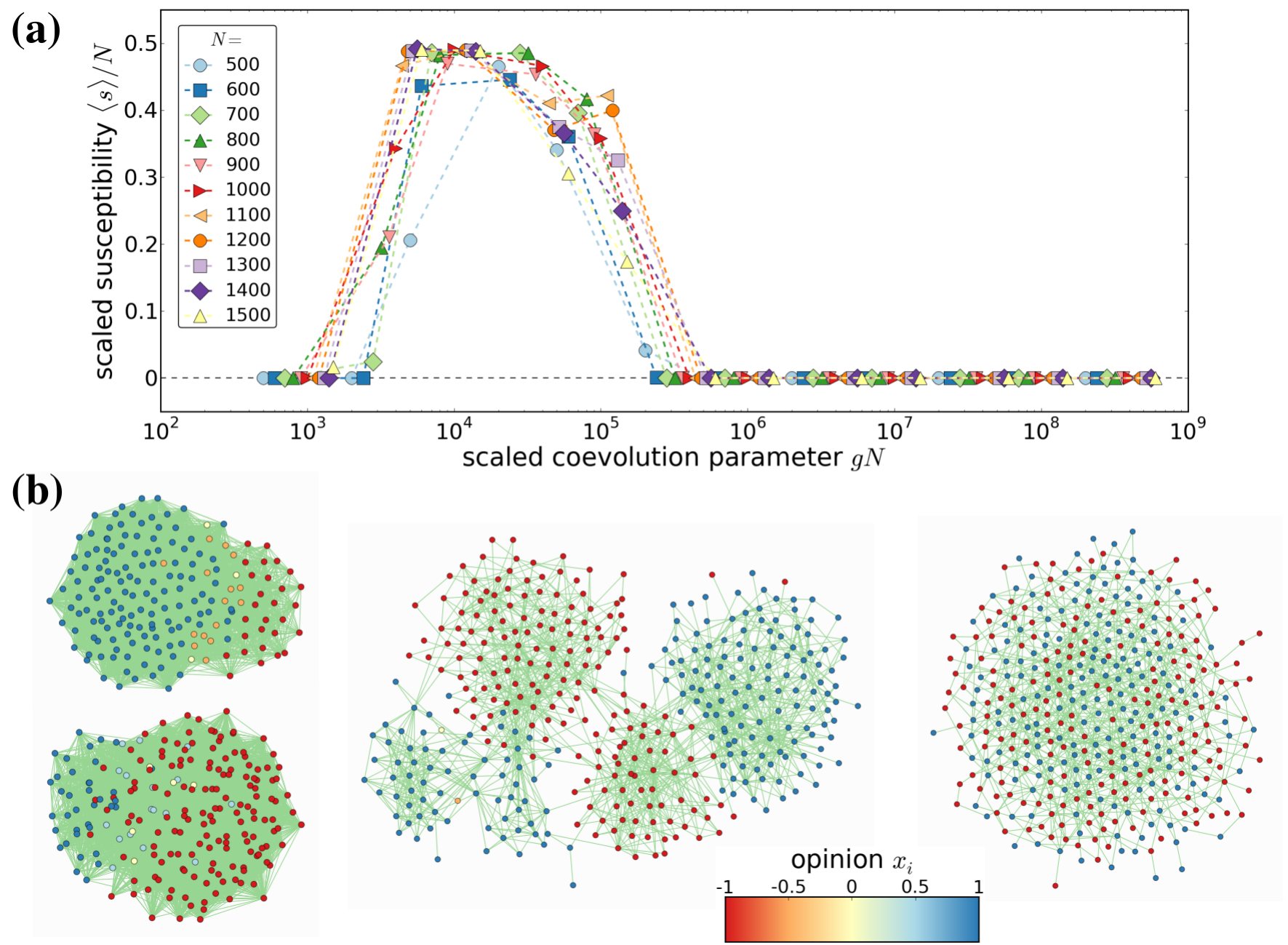}
\caption{(a) Scaled susceptibility $\langle s \rangle / N$ (average small component size) as a function of scaled coevolution parameter $gN$ in the model by \textcite{iniguez_opinion_2009}. The value $g$ is a ratio of time scales of network evolution and opinion dynamics (measuring the plasticity of the adaptive network), and large susceptibility indicates a fragmented network close to a percolation transition. (b) As $g$ increases, the stationary state transitions between a fragmented system with two components, a connected network with community structure, and a network without rewiring (shown for $N = 400$ and $g = 5, 10^3, 10^5$). 
Node color codes different opinions.
Figure adapted from \textcite{iniguez_opinion_2009}.
}
\label{fig:coevolution}
\end{figure}

\paragraph{Adaptivity and community structure.}

A signature feature of coevolving dynamics is the feedback loop between state change and link rewiring leading to a correspondence between opinion similarity and topological connectivity. In the case of voterlike dynamics, the \avm shows structured phases where almost disconnected components have mostly internal consensus \cite{timpanaro_emergence_2025}, as opposed to an active phase without group structure. For continuous opinions, the \abcm leads to a dynamic stationary state where groups have peaked opinion distributions within and nonzero connectivity between them \cite{yu_opinion_2017}. A separation of time scales between the dynamics of network and opinions is a potential factor in the emergence of these types of community structure, as suggested in the model by  \textcite{iniguez_opinion_2009}. On a fast time scale, continuous opinions evolve towards frozen binary states of total conviction ($x_i = \pm 1$) by following opinion averages weighted by topological distance. On a $g$-fold longer time scale parametrized by a coevolution parameter $g \geq 1$, links with disagreeing opinions are preferentially rewired, either locally via triadic closure, or globally to nodes with more similar opinions. For $g \to \infty$ there is no rewiring, while for $g \to 1$ the network fragments into two components. Intermediate values of the time scale ratio $g$ lead to networks with community structure, where groups have internal consensus and large clustering coefficient (see \fref{fig:coevolution}). A simplified dynamics (with node exchange over a ring instead of rewiring) retains the role of time scale separation in group formation \cite{iniguez_phase_2011}.

The model by \textcite{iniguez_opinion_2009} includes notions of quenched personal bias in its opinion dynamics. Contrarian agents who go against global average opinion increase the number of groups in the phase with community structure \cite{iniguez_modelling_2011}. Similarly, an assymetric external field coupled to opinion (simulating individual reactions to mass media) prevents consensus, which might explain heterogeneities in the public perception of scientific facts in survey data from Mexico and the EU \cite{iniguez_are_2012}. A variant of this model has been used to explore the nuanced role of deception in society by including both private and public opinions \cite{iniguez_effects_2014}. Agents can lie prosocially to feign opinion similarity, or antisocially to do the opposite. A certain amount of prosocial deception retains network cohesion, even when the edge weight dynamics is set to reward honesty. Antisocial liars, in turn, can end up bridging communities of different opinion \cite{barrio_dynamics_2015}. Some of these mechanisms have been coupled to variants of the ultimatum game \cite{snellman_modelling_2017} and the dictator game \cite{snellman_status_2019} to explore the interplay between opinion dynamics and status maximization in socio-economic networks. Similar models have also been recently introduced to emulate continuous opinion dynamics with homophily/heterophily and triadic closure \cite{wu_effects_2024}, as well as group emergence when nodes enter and leave the network \cite{bao_coevolutionary_2023}.

The adaptive coupling between topology and node states is arguably a key reinforcement mechanism in the radicalization dynamics leading to echo chambers and opinion polarization in social networks \cite{myers_group_1976} (see \sref{subsec:echo_chambers}). A minimal implementation of this feedback loop in terms of social influence, heterogeneous activity and homophily is the model by \textcite{baumann_modeling_2020}. Continuous opinions (with their sign and magnitude representing conviction in a binary issue) follow coupled differential equations where similar convictions between neighbors reinforce each other, and opposite convictions make opinions converge. Edge dynamics corresponds to an activity-driven temporal network \cite{perra_activity_2012} with link selection driven by opinion homophily, coupling the evolution of edge and node states. The model shows a transition between global consensus and a radicalized phase that emulates the statistics of polarized debates in Twitter. The model has been extended to opinions in multidimensional topic spaces, further separating the polarized phase into one with noncorrelated opinions and one with issue alignment (i.e. polarized ideology), which seems to capture empirical opinion polls in the US \cite{baumann_emergence_2021}. Alternative modelling approaches for the emergence of echo chambers as fitness maximization have also been proposed \cite{evans_opinion_2018}.

\subsubsection{General frameworks of adaptivity}

Coevolution shows how a time scale competition between dynamics of network and opinions generally leads to phase transitions between disordered or active phases (with no significant correlation between connectivity and node state) and non-trivial topologies with mesoscopic order (such as fragmentation or community structure) \cite{vazquez_time-scale_2007}. These are dynamical phase transitions driven by state-dependent link rewiring that lead to robust self-organization into emergent node classes, similar to the way plasticity promotes critical behavior in neuro-inspired adaptive dynamics \cite{droste_analytical_2013,sormunen_critical_2023}.

While most studies of adaptive networks focus on specific rules of node and link dynamics, a few have attempted to explore this link between adaptivity and collective self-organization within more general frameworks. The model of \textcite{herrera_general_2011} assumes that rewiring and node state change happen with probabilities $P_r$ and $P_c$, respectively, leading to a parameter space $(P_r, P_c)$ that characterizes a family of coevolution models. While node dynamics corresponds to the multistate \avm, link rewiring is implemented as a sequence of disconnection and reconnection actions, where each follows a mechanism of opinion \textit{dissimilarity}, \textit{randomness}, or \textit{similarity}. These lead, for example, to the rewiring rules of the standard \avm (combining dissimilarity and similarity) \cite{vazquez_generic_2008}, or the \abcm (combining dissimilarity and randomness) \cite{kozma_consensus_2008}. The authors find that only reconnections between nodes with similar states lead to network fragmentation, and that the fragmented phase in $(P_r, P_c)$ space decreases for denser networks. When voterlike opinion flips are noisy, the dynamics exhibits transient modular structure (groups with mostly internal consensus) that scale with system size and noise \cite{gonzalez-avella_emergence_2014}.

Another general framework of adaptive opinion dynamics has recently been introduced by \textcite{liu_emergence_2023}. Inspired by the polarization patterns seen in the model by  \textcite{baumann_modeling_2020,baumann_emergence_2021}, the authors focus on two coevolution mechanisms: \textit{opinion homogenization}, implemented as a stochastic differential equation driving the drift and diffusion of the opinions of connected agents, and \textit{homophily clustering}, where network links appear/disappear with rates dependent on the opinions of nodes at their endpoints. Opinion drift and diffusion are also rescaled to emulate a separation of timescales between opinion change and network evolution, similarly as in the model by \textcite{iniguez_opinion_2009}. In the minimal case of a voterlike linear dynamics for the drift kernel \cite{clifford_model_1973,holley_ergodic_1975}, the model shows phases of polarization, partial polarization and depolarization that recover some statistics of empirical data in Facebook and the blogosphere.

\section{Mathematical and computational methods}
\label{sec:methods}

The rules by which agents change state in models of opinion dynamics usually involve elements of randomness, placing the analysis of models of opinion dynamics in the area of stochastic processes.
The aim is to determine macroscopic quantities, such as the distribution of agents across opinions or the mean time to consensus.  A range of tools is available \cite{redner_guide_2001,van_kampen_stochastic_2007, gardiner_handbook_2009,toral_stochastic_2014,cox_theory_2017,kiss_mathematics_2017, jacobs_stochastic_2010}, including Markov chain methods, master equations, systems of coupled deterministic or stochastic differential equations, and generating functions.

Here we describe some of the most common analytical and computational tools used to study opinion models. 
\sref{sec:theory} and \sref{sec:heterogeneous_networks} contain a summary of principal exact and approximate analytical methods for complete and heterogeneous networks. 
Simulation techniques are described in \sref{sec:simulations}. 
Finally, \sref{subsec:algo-appr} contains an overview of optimization approaches to opinion dynamics.

\subsection{Exact and approximate methods for complete networks}
\label{sec:theory}

The simplest scenario, all-to-all interactions, assumes all agents are equivalent and interact with one another, simplifying the analytical treatment. 
This is also known as homogeneous mixing, well-mixed populations, equivalent-neighbor interaction, or mean-field approximation.

\subsubsection{Master equations for discrete opinions}
\label{sec:MEs}

\paragraph{Definition and agent-centered view.}

Master equations (\mes) describe the time evolution of the probability to find a system in a particular state. The most general form of a master equation for models of opinion dynamics derives from a \textit{node or agent point of view}. We write the state of node $i$ as $s_i$, where $s_i$ takes values in a discrete opinion space (this can be generalized to continuous opinions). Assuming that the interaction network is fixed in time, the state of the system is given by $\mathbf{s}=(s_1,\dots,s_N)$, where $N$ is the number of agents. We write $p_\mathbf{s}(t)$ for the probability to find the population (or network) in state $\mathbf{s}$ at time $t$. We also denote by $w_{\mathbf{s} \to \mathbf{s'}}$ the rate (probability per unit time) with which the system transitions from state $\mathbf{s}$ to $\mathbf{s'}$. For example, states $\mathbf{s}$ and $\mathbf{s'}$ may differ in the opinion of a single agent who has changed views. The quantity $w_{\mathbf{s} \to \mathbf{s'}} dt$ is the probability that the system moves from state $\mathbf{s}$ to state $\mathbf{s'}$ in a time interval $dt$ (with $dt$ assumed to be infinitesimally small). For a homogeneous Markov process, the rates $w_{\mathbf{s} \to \mathbf{s'}}$ are constant in time. Assuming that the system is in state $\mathbf{s}$, the probability that the system leaves this state in an interval $dt$ is $\sum_{\mathbf{s'} \neq \mathbf{s}} w_{\mathbf{s} \to \mathbf{s'}} dt$, while the probability that there is a change from any other state $\mathbf{s'}\neq\mathbf{s}$ to $\mathbf{s}$ is $\sum_{\mathbf{s'} \neq \mathbf{s}} w_{\mathbf{s'} \to \mathbf{s}} dt$. This outflux and influx leads to a change in $p_\mathbf{s}$. Ignoring the probability of multiple events in the time interval $dt$ (a justified assumption for $dt\to 0$), we obtain the so-called \textit{master equation} \cite{van_kampen_stochastic_2007, gardiner_handbook_2009,jacobs_stochastic_2010, toral_stochastic_2014}, 
\begin{equation}
\label{eq:ME}
\frac{d p_\mathbf{s}}{dt} = \sum_\mathbf{s'} \left[ w_{\mathbf{s'} \to \mathbf{s}} p_\mathbf{s'} - w_{\mathbf{s} \to \mathbf{s'}} p_\mathbf{s} \right],
\end{equation}
where the sum can be chosen to run over all states $\mathbf{s'}$ or only over $\mathbf{s'}\neq\mathbf{s}$ (the term for $\mathbf{s'}=\mathbf{s}$ cancels out). %
The \me is sometimes written more succinctly in terms of the net probability currents $J_{\mathbf{s} \to \mathbf{s'}} = w_{\mathbf{s'} \to \mathbf{s}} p_\mathbf{s'} - w_{\mathbf{s} \to \mathbf{s'}} p_\mathbf{s}$ from state $\mathbf{s}$ to state $\mathbf{s'}$. The \me can be generalized to time-dependent rates $w_{\mathbf{s} \to \mathbf{s'}}(t)$, or to continuous states. In the latter case, $p_\mathbf{s}$ becomes a probability density, and the sum in \eref{eq:ME} is replaced by an integral.

When the number of opinion states each $s_i$ can take is finite (say there are $M$ states), we can introduce a transition matrix $\mathbf{W}$ with $W_{\mathbf{s}\mathbf{s'}} = w_{\mathbf{s'} \to \mathbf{s}}$ for $\mathbf{s} \neq \mathbf{s'}$ and diagonal $W_{\mathbf{s}\mathbf{s}} = - \sum_{\mathbf{s'}\neq \mathbf{s}} w_{\mathbf{s} \to \mathbf{s'}}$ (i.e. the negative of the escape rate from state $\mathbf{s}$ to any other state). The matrix is of size $M^N\times M^N$. \eref{eq:ME} then takes the matrix form
\begin{equation}
\label{eq:ME_mat}
\dot{\mathbf{p}} = \mathbf{W} \mathbf{p}
\end{equation}
for the probability vector $\mathbf{p}(t) = \{ p_\mathbf{s} \}$ of length $M^N$. We highlight the linearity of the \me in terms of the probability vector $\mathbf{p}$. \eref{eq:ME_mat} has the formal solution $\mathbf{p} =  e^{t \mathbf{W}} \mathbf{p}(0)$ for an initial distribution $\mathbf{p}(0)$, which allows for a spectral decomposition into a superposition of eigensolutions \cite{hale_ordinary_2009,moya-cessa_differential_2011}. The condition $\sum_\mathbf{s'} W_{\mathbf{s}\mathbf{s'}} = 0$ for all states $\mathbf{s}$ implies the existence of at least one normalized eigenvector $\mathbf{p}^*$ with $\mathbf{W} \mathbf{p}^* = 0$, which we identify as a stationary opinion distribution. For a finite number of states, \eref{eq:ME_mat} always evolves towards one of the stationary solutions $\mathbf{p}^*$ as $t \to \infty$ \cite{van_kampen_stochastic_2007}. 

The \me in \eref{eq:ME} is frequently the starting point for further analysis or approximations. Before discussing such methods, it is useful to briefly present a description in terms of {\em occupation numbers} rather than the opinion states of agents. This entails a significant reduction in complexity (e.g. number of equations), and applies exactly to the case of all-to-all interactions.

\paragraph{Occupation-number perspective.}

If interactions are all-to-all, we do not need the opinion of each agent to describe the state of the entire system. Instead it is sufficient to know how many agents hold each of the $M$ opinions. We introduce occupation numbers $n_s$ as the number of agents in state $s=0,\dots, M-1$, and $\mathbf{n}=(n_1,\dots, n_M)$. The dynamics is then described by transition rates of the form $w_{\mathbf{n} \to \mathbf{n'}}$ and a \me similar to \eref{eq:ME}, where the only change is the replacement of $\mathbf{s}$ by $\mathbf{n}$ and $\mathbf{s'}$ by $\mathbf{n'}$. We note that $w_{\mathbf{n} \to \mathbf{n'}}$ has a different functional form than $w_{\mathbf{s} \to \mathbf{s'}}$. Indeed, the space of vectors $\mathbf{n}$ has a much lower dimensionality than that of $\mathbf{s}$. While the occupation number description is exact for systems with all-to-all interactions, it is also an intuitive starting point for \mes in heterogeneous networks (see \sref{sec:heterogeneous_networks}).

As an illustration, we briefly discuss the minimal case of binary opinions ($M=2$, $s = 0, 1$). We denote by $p(n, t)$ the probability that $n$ nodes are in state $s = 1$ at time $t$ (and the remaining $N-n$ nodes are in state $s = 0$). Assuming that each event in the dynamics only changes the state of one agent (i.e. all events are of the type $n\to n+1$ or $n\to n-1$), the \me can be written as
\begin{equation}
\label{eq:ME_occ}
\frac{\partial p(n, t)}{\partial t} = \sum_{\ell=\pm 1} (\hat{E}^{-\ell} - 1) \left[ w_{n}^\ell\, p(n, t) \right],
\end{equation}
where $\hat{E} f(n) = f(n + 1)$ is the step operator and $\hat{E}^{-1}$ its inverse, i.e. $\hat{E}^{-1} f(n)=f(n-1)$. The quantity $w_n^\ell$ is the global rate at which the system transitions from $n$ to $n+\ell$ agents with opinion $s = 1$ ($\ell=\pm 1$). Dynamics of this type are known as `one-step processes', since the only possible transitions are opinion swaps of a single node \cite{redner_guide_2001}. \mes like \eref{eq:ME_occ} can be generalized beyond one-step processes by extending the sum to arbitrary step sizes $\ell$, and for systems with more than one integer degree of freedom.

\paragraph{Stationary solutions for one-step processes.}

In principle, \eref{eq:ME} fully determines the distribution $p_\mathbf{s}(t)$ at any time $t$, and thus contains all information about the statistics of the system. In certain cases, it is possible to obtain explicit solutions. For one-step processes, the stationary solution $p^*(n) = \lim_{t \to \infty} p(n, t)$ of \eref{eq:ME_occ} can be found as \cite{herrerias-azcue_consensus_2019,toral_stochastic_2014,van_kampen_stochastic_2007}
\begin{equation}
p^*(n)=\frac{\prod_{m=1}^n \frac{w_{m-1}^+}{w_m^-}}{1+\sum_{m=1}^N\prod_{j=1}^m \frac{w_{j-1}^+}{w_j^-}}.
\end{equation}

Another method to obtain solutions of the \me (at least in simple scenarios) is via generating functions. We give here only a brief illustration for all-to-all interactions and binary opinions ($s = 0, 1$), starting from \eref{eq:ME_occ}. The method uses the probability generating function $g(z, t) = \sum_n z^n p(n, t)$. Writing $z=e^{ik}$, this is the characteristic function of the probability distribution $p(n, t)$, or the discrete Fourier transform of $p(n,t)$ with respect to $n$. The strategy is then to formulate a partial differential equation for $g(z,t)$, and to solve it where possible.

We illustrate the generating-function method with an example. We assume that any agent with opinion $s = 1$ switches to state $s = 0$ with rate $\lambda$, and that the reverse change occurs with rate $\mu$. We then have the population-wide rates $w^-_n = n \lambda$ and $w^+_n = (N - n) \mu $. \eref{eq:ME_occ} leads to the first-order partial differential equation
\begin{equation}
\label{eq:ME_pde}
\frac{\partial g}{\partial t} = (1 - z) \left[ (\lambda + \mu z) \frac{\partial g}{\partial z} - N \mu g \right].
\end{equation}
This equation can be solved with the method of characteristics \cite{kiss_mathematics_2017}, from which we obtain $p(n,t)$ by iterative differentiation of the generating function. For initial condition $p(n,0)=\delta_{n,0}$, one finds the binomial distribution $p(n, t) = \binom{N}{n} q^n (1 - q)^{N - n}$ with $q(t) = \lambda (1 - e^{-(\lambda + \mu) t}) / (\lambda + \mu)$.

\subsubsection{Approximations for large populations}
\label{sec:approx_large_pops}

\paragraph{Kramers--Moyal expansion and Fokker--Planck equations.}

A common approximation technique involves replacing discrete opinions by a continuous state space. This is accurate for large but finite populations. In most models of opinion dynamics, only one agent changes opinion in any given microscopic event. The idea is  that the relative change of one opinion flip is of order $1/N$, leading to a quasi-continuous change for $N\gg 1$.

We use \eref{eq:ME_occ} to illustrate how such an approximation can be formalized mathematically. Consider the Taylor expansion $(\hat{E}^\ell-1) f(n)=f(n+\ell)-f(n)= \sum_{m=1}^\infty \frac{\ell^m}{m!} \frac{\partial^m}{\partial n^m} f(n) $. Implementing the expansion into the right-hand side of \eref{eq:ME_occ} is known as the Kramers--Moyal expansion \cite{kramers_brownian_1940,gardiner_handbook_2009, jacobs_stochastic_2010}. Keeping terms up to second order leads to the \textit{\fpe}
\begin{equation}
\label{eq:FPE}
\frac{\partial p(n, t)}{\partial t} = \frac{\partial}{\partial n} \left [ -a(n) p(n) + \frac{1}{2} \frac{\partial}{\partial n} \left[ b(n) p(n) \right] \right],
\end{equation}
where $a(n) = \sum_\ell \ell w^{\ell}_n$ and $b(n) = \sum_\ell \ell^2 w^{\ell}_n$ are drift and diffusion coefficients, respectively. Original derivations of similar equations are due to Planck and Kolmogorov \cite{van_kampen_stochastic_2007}. We note that Pawula's theorem implies that retaining only first and second order terms in the expansion guarantees the positivity of the probability $p(n, t)$ \cite{pawula_approximation_1967}. This is not necessarily the case if the series is truncated at any higher finite order. The Fokker--Planck equation can be solved explicitly for certain functional forms of the drift and diffusion coefficients. The equation can also be generalized to the case of multiple degrees of freedom (i.e. systems with more than two opinion states).

\paragraph{Stochastic differential equations.}

\eref{eq:FPE} is equivalent to the (It\={o}) stochastic differential equation
\begin{equation}
\label{eq:sde}
\dot n = a(n) + \sqrt{b(n)}\xi,
\end{equation}
also known as a Langevin equation, where $\xi$ is standard Gaussian white noise, i.e., $\langle \xi(t)\rangle=0$, $\langle \xi(t)\xi(t')\rangle=\delta(t-t')$. In the example above, the rates $w^{\pm \ell}_n$ scale linearly with system size $N$, which makes it convenient to introduce a scaled variable $x=n/N$. \eref{eq:sde} can then be rewritten as $\dot x = \alpha(x)+\sqrt{\beta(x)/N}\xi$, where $\alpha(x)=a(xN)/N$ and $\beta(x) = b(xN)/N$. The term $\alpha(x)$ describes deterministic drift, while the noise amplitude $\sqrt{\beta(x)/N}$ scales as $N^{-1/2}$.

\paragraph{Rate equations for infinite populations.} For all-to-all interactions in the thermodynamic $N\to\infty$, stochastic fluctuations become irrelevant and the system is described by deterministic equations. This corresponds to retaining only the first term in the Kramers--Moyal expansion [\eref{eq:FPE}], resulting in the so-called \textit{Liouville equation} $\frac{\partial p(n, t)}{\partial t} = -\frac{\partial}{\partial n} \left [a(n) p(n, t)\right]$ or, equivalently, a rate equation $\dot n=a(n)$ instead of \eref{eq:sde}. We point here to a common misunderstanding. Rate equations are not definitions of the reaction rates in a system (i.e. the quantities $w^{\pm}_n$ in the \me for occupation numbers). They describe instead the time evolution of the mean number of various types of agents in a deterministic limit. Rate equations can also be obtained heuristically from simple mass-action kinetics \cite{kuriyan_molecules_2013}.

\paragraph{An example: The noisy voter model.}\label{para:noisy_vm}

In the noisy voter model with all-to-all interactions, the opinion of an agent can change through imitation of another individual, or spontaneously with rate $\varepsilon$ \cite{granovsky_noisy_1995,biancalani_noise-induced_2014}. We have then $w^+_n=n(N-n)/N+\varepsilon (N-n)$ and $w^-_n=n(N-n)/N+\varepsilon n$, which leads to $a(n)=\varepsilon(N-2n)$ and $b(n)=2n(N-n)/N+\varepsilon N$. Equivalently, we find $\alpha(x)=\varepsilon(1-2x)$ and $\beta(x)=2x(1-x)+\varepsilon$. Imposing zero-flux boundary conditions at $x=0$ and $x=1$, the stationary solution of the Fokker-Planck equation is \cite{biancalani_noise-induced_2014} 
\begin{equation}
    p(x)= {\cal N} [2x(1-x)+\varepsilon]^{\varepsilon N-1},
\end{equation}
where ${\cal N}$ is a normalisation factor (not to be confused with the population size $N$). 
The stationary distribution exhibits a transition between a bimodal shape at $\varepsilon N<1$ and a unimodal shape for $\varepsilon N>1$. In the former case, the population is mostly found near the consensus states (no polarization). In the latter, both states coexist, indicating a polarized population. In the deterministic limit we have the rate equation
\begin{equation}
    \dot x= \varepsilon (1-2x).
\end{equation}
Thus, infinite populations are driven to coexistence of both opinions ($x=1/2$) whenever there is spontaneous opinion change ($\varepsilon>0$). For $\varepsilon=0$ we recover the standard voter model, i.e. $\dot x=0$, in line with the conservation of magnetization in an average across realizations \cite{liggett_interacting_2005}. In the limit of large but finite $N$, the standard voter model can be described by the stochastic differential equation $\dot x = \sqrt{2x(1-x)}\xi$, highlighting the absorbing states $x=0$ and $x=1$, where the amplitude of the multiplicative noise vanishes. 

\begin{figure}[tbp]
\centering
\includegraphics[width=0.9\columnwidth]{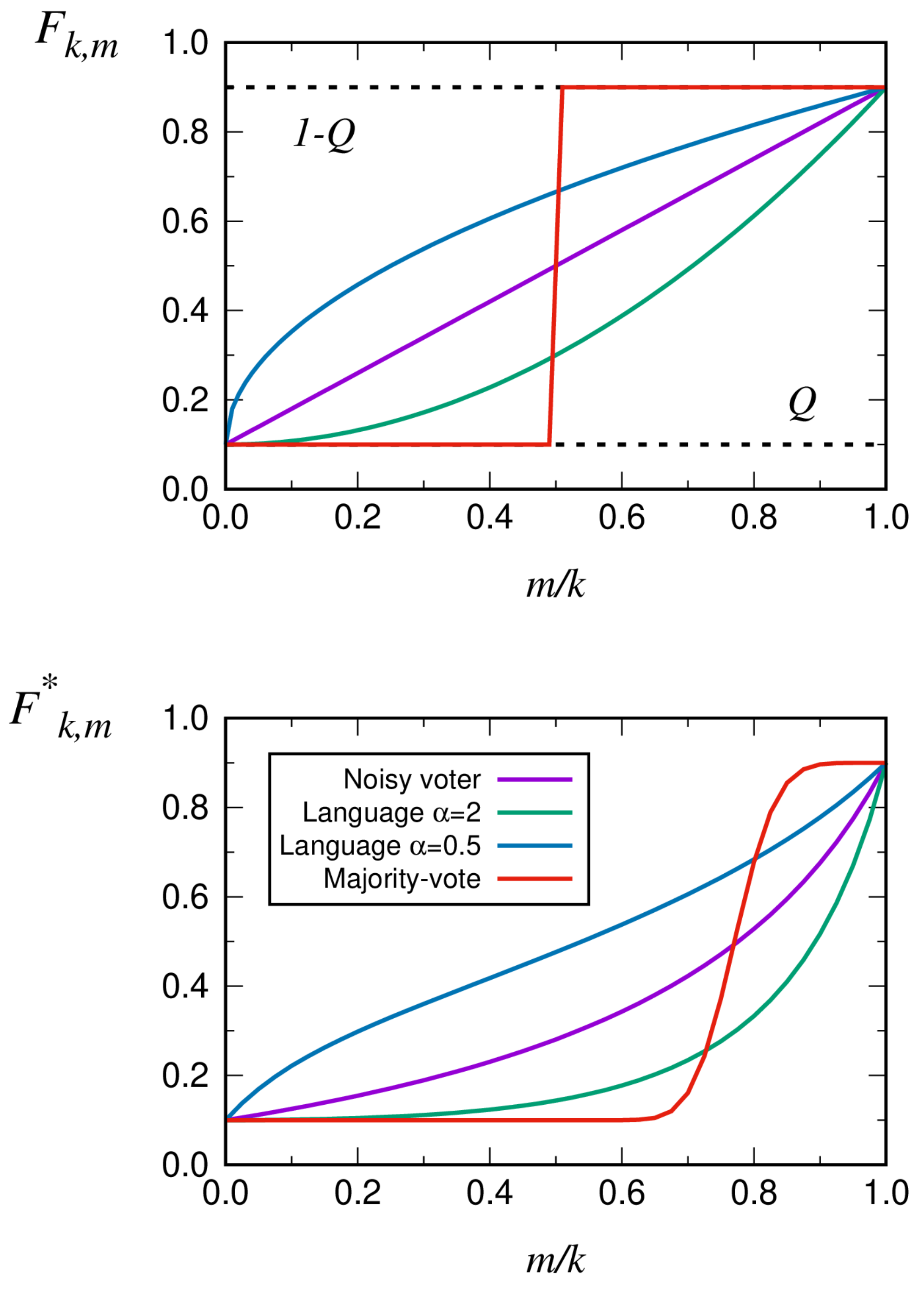}
\caption{(top) Infection rate $F_{k, m}$ as a function of fraction $m/k$ of infected neighbors for selected opinion models from \tref{tab:AMEmodels}: noisy voter with parameter $Q$, nonlinear noisy voter with parameter $\alpha$ (i.e. the language model of \textcite{abrams_modelling_2003}), and majority rule. (bottom) Under a notion of algorithmic bias (nodes only interact with a fraction $1 - b$ of nodes of opposite opinion, with $b$ a bias parameter), the effective infection rate $F^*_{k, m}$ decreases (see \sref{subsec:algo-mediation-polarization}). Figure adapted from \textcite{peralta_effect_2021}.}
\label{fig:transition_rates}
\end{figure}

\subsection{Approximations for heterogeneous networks}
\label{sec:heterogeneous_networks}

\subsubsection{Transition rates for binary opinion dynamics}
\label{sec:transition_rates}

To model opinion dynamics with binary states on heterogeneous networks, it is useful to exploit similarities with epidemic spreading and social contagion \cite{porter_dynamical_2016,gleeson_binary-state_2013,unicomb_dynamics_2021}. In this spirit, the opinion states $s = 0, 1$ are referred to as susceptible/inactive and infected/active, respectively. 
\begin{table*}[t]
	\centering
	\caption{\small Infection ($F_{k,m}$) and recovery ($R_{k,m}$) rates for selected standard opinion dynamics models with binary node states on simple networks, for nodes with degree $k$ and number of infected neighbors $m$ \cite{gleeson_high-accuracy_2011,peralta_opinion_2022}. The average degree of the network is denoted by $\langle k \rangle = \sum_k k p_k$. Models are listed by increasing number of parameters. Based on and extended from previous versions in \cite{gleeson_high-accuracy_2011,gleeson_binary-state_2013,unicomb_dynamics_2021,peralta_effect_2021,peralta_opinion_2021}. Transition rates can be generalized to networks of arbitrary complexity by vectorizing $k$ and $m$ \cite{unicomb_threshold_2018,unicomb_reentrant_2019,unicomb_dynamics_2021}.}
	\label{tab:AMEmodels}
	\begin{tabular}{p{3.7cm} c c c l}
		\toprule
		Model & Parameters & Infection rate $F_{k,m}$ & Recovery rate $R_{k,m}$ & Section \\
		\hline
        Voter & - & $\frac{m}{k}$ & $\frac{k - m}{k}$ & \sref{sec:voter_model} \\
        Link-update voter & - & $\frac{m}{\langle k \rangle}$ & $\frac{k - m}{\langle k \rangle}$ & \\
		Nonlinear-voter/q-voter & $\alpha$ & $(\frac{m}{k})^{\alpha}$ & $(\frac{k - m}{k})^{\alpha}$ & \sref{subsec:q_voter} \\
        Noisy-voter/Kirman & $Q$ & $Q + (1 - 2Q)\frac{m}{k}$ & $Q + (1 - 2Q)\frac{k - m}{k}$ & \sref{sec:voter_model} \\
        Majority rule & $Q$ & $\begin{cases} Q & \text{ if } m < k/2,\\ 1/2 & \text{ if } m = k/2,\\ 1 - Q & \text{ if } m > k/2 \end{cases}$ & $\begin{cases} 1 - Q & \text{ if } m < k/2,\\ 1/2 & \text{ if } m = k/2,\\ Q & \text{ if } m > k/2 \end{cases}$ & \sref{sec:majority_rule} \\
        Relative threshold & $\phi$, $p$ & $\begin{cases} 1 & \text{ if } m \geq k \phi,\\ p & \text{ otherwise }\\ \end{cases}$ & $0$ & \sref{sec:transition_rates} \\
        Absolute threshold & $M_{\phi}$, $p$ & $\begin{cases} 1 & \text{ if } m \geq M_{\phi},\\ p & \text{ otherwise }\\ \end{cases}$ & 0 & \sref{sec:transition_rates} \\
		\bottomrule
	\end{tabular}
\end{table*}

We assume that interactions occur only between nearest neighbors in the network. The probability of a node to flip is then determined by its current opinion, the number of neighbors $k$ (the node's degree), and by how many of those neighbors are in either of the two states. We denote by $m$ the number of neighbors in state $s = 1$ (the remaining $k-m$ neighbors are in state $s = 0$). The dynamics are then specified by the \textit{infection} and \textit{recovery} rates, $F_{k, m}$ and $R_{k, m}$ respectively, which are assumed to be constant in time. Thus, in a time interval $dt$, a node with opinion $s = 0$ with $k$ neighbors, $m$ of which have opinion $1$, flips to $s = 1$ with probability $F_{k,m} dt$, and a node in state $s = 1$ flips to $s = 0$ with probability $R_{k, m} dt$. These rates define the opinion dynamics model of interest (see \tref{tab:AMEmodels} and \fref{fig:transition_rates}). In the standard two-state voter model of \sref{sec:voter_model}, for example, a node is chosen uniformly at random and copies the opinion of one of its neighbors, selected at random too \cite{liggett_interacting_2005,sood_voter_2005}. Thus, a susceptible node in class $(k, m)$ becomes infected with probability $m/(Nk)$ in the next time step (where we have set $dt=1/N$), and an infected node becomes susceptible with probability $(k - m)/(Nk)$. This leads to the transition and infection rates of the voter model in \tref{tab:AMEmodels} [a linear dependence on $m$ characterizes most variants of the model, such as the link-update voter model \cite{suchecki_conservation_2004}].

The functional forms of the $F_{k, m}$ and $R_{k, m}$ allow us to identify mathematical equivalences between models that have otherwise been independently introduced or used to describe disconnected phenomena. For example, nonlinear-voter \cite{schweitzer_nonlinear_2009} and q-voter \cite{castellano_nonlinear_2009} models involve nonlinearity regulated by an exponent, just like some models of language dynamics \cite{abrams_modelling_2003} (see \fref{fig:transition_rates}). The transition rates of the noisy voter model \cite{peralta_analytical_2018} are the same as those of Kirman's herding model for ant colonies and financial markets \cite{kirman_ants_1993}. Both the majority rule \cite{de_oliveira_isotropic_1992} and threshold models of social contagion \cite{granovetter_threshold_1978,watts_simple_2002,centola_complex_2007,ruan_kinetics_2015} have stepwise transition rates, involving a threshold for the number of infected neighbors $m$.

\subsubsection{Approximate master equations, pair approximation and mean field}
\label{sec:AMEs}

\paragraph{Coarse-grained descriptions at different levels.}

The \me in \eref{eq:ME} drives the evolution of the system across a very large space of microstates (of size $2^N$ for binary opinions). There are several coarse-grained descriptions of such dynamics. In increasing order of reduction, these are \textit{approximate master equations}, \textit{pair approximations} and \textit{mean-field} theories. None of these approaches capture the full correlations of states beyond those of nearest neighbors, but their accuracy varies depending on the model and network topology. The approach has been introduced for epidemics \cite{petermann_cluster_2004,marceau_adaptive_2010,lindquist_effective_2011} and generalized for wider binary dynamics by \textcite{gleeson_high-accuracy_2011,gleeson_binary-state_2013}.

\paragraph{Approximate master equations.}

So-called ``approximate master equations'' (\ames) focus on the probabilities that a node of degree $k$ is infected or recovered, respectively, when $m$ of its neighbors are infected \cite{vazquez_systems_2008,ruan_kinetics_2015,gleeson_high-accuracy_2011,gleeson_binary-state_2013}. This approximates well the full opinion dynamics for large configuration-model networks \cite{newman_networks_2018}. We write $s_{k, m}(t)$ for the fraction of nodes with degree $k$ that are susceptible at time $t$ (i.e. holding opinion $s = 0$) and have $m$ infected neighbors. Similarly, $i_{k,m}(t)$ is the proportion of degree-$k$ nodes that are infected and have $m$ infected neighbors. Thus we have $\sum_{m=0}^k [s_{k,m}(t)+i_{k,m}(t)]=1$ for all $k$. The fraction of infected nodes among all degree-$k$ nodes is $\rho_k(t) = \sum_{m=0}^k i_{k, m} = 1 - \sum_{m=0}^k s_{k, m}$, while the fraction of infected nodes in the entire network is $\rho(t) = \sum_{k=0}^\infty P_k \rho_k$, with $P_k$ the degree distribution of the network. 

The \ames are coupled differential equations for $s_{k, m}$ and $i_{k, m}$. The rates at which a node of class $(k,m)$ changes from susceptible to infected or vice versa are given by $F_{k,m}$ and $R_{k,m}$. A node may also change, say, from $(k,m)$ to $(k,m+1)$ if one of its neighbors becomes infected. To formulate a closed set of equations for the $s_{k,m}$ and $i_{k,m}$, \textcite{gleeson_high-accuracy_2011,gleeson_binary-state_2013} proceeds by assuming there are global rates with which links of type $SS$ change to $SI$, $II$ to $SI$, or $SI$ to either $SS$ or $II$ respectively (where $S$ and $I$ denote the opinions of nodes at the endpoints of an edge). These rates can be expressed in terms of $s_{k,m}$ and $i_{k,m}$, by using $F_{k,m}$ and $R_{k,m}$ and disregarding higher-order correlations. The end result is a closed set of differential equations for $s_{k,m}$ and $i_{k,m}$. In a network of maximum degree $k_{\rm max}$, the \ame system has $(k_{\rm max} + 1) (k_{\rm max} + 2)$ degrees of freedom \cite{ruan_kinetics_2015,porter_dynamical_2016}.

\paragraph{Pair approximation.}

\ames neglect some higher-order correlations, but still contain information beyond individual pairs of nodes. For example, $i_{k,m}$ is the probability that a node of degree $k$ is infected and has $m$ infected neighbors -- such term contains information about all $k$ neighbors.

The pair approximation implies a further reduction, considering quantities that relate to nodes and pairs of nodes, but not those involving three or more nodes. In other words, the links emanating from a node are treated as independent. The probability that a particular node has $m$ infected neighbors is then binomial (which is not necessarily the case for solutions of the \ames above). The relevant variables are: (i) the probability $\rho(t)$ that a node is infected (as opposed to recovered); (ii) the probability $p(t)$ that a neighbor of an infected node is infected; and (iii) the probability $q(t)$ that a neighbor of a susceptible node is infected. 

There is a further distinction between so-called {\em heterogeneous} and {\em homogeneous} versions of the pair approximation. The heterogeneous pair approximation explicitly distinguishes between nodes of different degrees. That is to say, the probabilities above depend on the degree $k$, and thus we write $\rho_k(t)$, $p_k(t)$ and $q_k(t)$. Both $p_k$ and $q_k$ can be written in terms of sums (over $m$) of $m s_{k,m}$ and $m i_{k,m}$, respectively, relating the pair approximation to the \ames \cite{gleeson_high-accuracy_2011,gleeson_binary-state_2013}. Indeed, a system of coupled differential equations for  $\rho_k$, $p_k$, and $q_k$ can be derived by asserting that $s_{k, m}= (1 - \rho_k) \ell$ and $i_{k,m}=\rho_k \ell'$, where the integer random variables $\ell$ and $\ell'$ follow binomial distributions $B_{k, \ell}(p_k)$ and  $B_{k, \ell'}(q_k)$, respectively, with $B_{k,\ell}(p) = \binom{k}{\ell} p^\ell (1-p)^{k-\ell}$. In a network with maximal degree $k_{\rm max}$, the heterogeneous pair approximation is a closed system with $3k_m+1$ degrees of freedom (these are $\rho_k$ for $k=0,\dots,k_{\rm max}$, and $p_k$ and $q_k$ for $k=1,\dots,k_{\rm max}$).

The homogeneous pair approximation is applicable to random regular networks (all nodes with the same degree $k$). The three relevant variables $\rho$, $p$, and $q$ are not independent degrees of freedom, leading to a closed system of two coupled equations for the fraction of infected nodes $\rho$ and the density of active links (the probability that a randomly chosen link connects nodes in different states) \cite{vazquez_analytical_2008}. A similar reduction can also be used as an approximation for heterogeneous networks (if the degree distribution is sufficiently narrow) by neglecting the dependence of $\rho_k$, $p_k$ and $q_k$ on $k$.

\paragraph{Heterogeneous and homogeneous mean-field approximation.} 

As one main step, mean-field approximations neglect correlations between the states of elements of a system, and replace the effects upon a particle by a ``mean field''. 

In models of opinion dynamics, the term ``mean field theory'' is used loosely. Mean-field approximations usually involve quantities only relating to nodes, but not to groups of two or more nodes. Compared to pair approximations, this reduces the mathematical complexity of the description. Just like with the pair approximation, there are heterogeneous and homogeneous versions of mean-field theory. In heterogeneous mean field theory, the only relevant variables are the probabilities $\rho_k$ that a node of degree $k$ is infected. A closed set of equations for these variables can be found by replacing $p_k$ and $q_k$ with $\omega = z^{-1} \sum_k k P_k \rho_k$, where $z = \sum_k k P_k$ is the mean degree in the network. We obtain \cite{gleeson_high-accuracy_2011,gleeson_binary-state_2013}
\begin{equation}
\label{eq:MF}
\frac{d \rho_k}{dt} = -\rho_k \sum_m R_{k, m} B_{k, m}(\omega) + (1 - \rho_k) \sum_m F_{k, m} B_{k, m} (\omega),
\end{equation}
where $B_{k,m}(\omega)$ is again a binomial distribution. \eref{eq:MF} constitutes a
a closed system of $k_m + 1$ coupled differential equations known as the degree-based or heterogeneous mean-field approach in epidemics \cite{pastor-satorras_epidemic_2015}.

Homogeneous mean-field theory entails a further reduction to a single degree of freedom, the probability $\rho(t)$ that any randomly selected node is infected. This quantity is assumed to be independent of node degree (i.e. $\rho_k\equiv \rho$). To obtain a closed equation for $\rho$, we set $\omega=\rho$ and then average the right-hand side of \eref{eq:MF} over $k$ with the degree distribution $P_k$.

\paragraph{Extensions and generalizations.}

\ames (and the coarser pair and mean-field approximations) capture some of the effects of network structure on models of opinion dynamics. The approximations can be extended to more general networks, such as signed or weighted networks, or networks with multilayer interactions. To do this, we replace $(k, m)$ by vectors $(\mathbf{k}, \mathbf{m})$ containing the degrees and numbers of infected neighbors for different edge types, weight bins, or layers in the network \cite{unicomb_threshold_2018,unicomb_reentrant_2019}. In temporal networks, the time-dependent degree of a node can be defined as the number of interactions in a sliding time window, leading to bursty changes in node degrees throughout time \cite{unicomb_dynamics_2021}. For monotonic social contagion models (i.e., models with $R_{k, m} = 0$, see \tref{tab:AMEmodels}), \ames can be reduced to a system of two differential equations \cite{porter_dynamical_2016,ruan_kinetics_2015}. This accurately captures the role of weight heterogeneity and layer overlap in cascading contagion \cite{unicomb_threshold_2018,unicomb_reentrant_2019}. The AME framework has also been generalized to multistate dynamics \cite{fennell_multistate_2019} [e.g., models with more than two opinion states \cite{bolzern_multi-opinion_2023}].

We also highlight extensions of the pair approximation to finite networks, proposed by \textcite{peralta_stochastic_2018, peralta_binary-state_2020}. These combine the above approximation methods with Kramers--Moyal expansions, and result in stochastic rather than deterministic differential equations for the node-based and pair-based degrees of freedom.

\subsubsection{Moment-closure methods}

The description of stochastic dynamics frequently leads to a hierarchy of equations for different moments of the distribution generated by the process. \textcite{newman_networks_2018} gives a simple example as follows [also reported by \textcite{porter_dynamical_2016}]. Consider a spreading process on a network, where each node can be susceptible ($s_i=0$) or infected ($s_i=1$), and where infected nodes infect each of their neighbors with rate $\lambda$. If $A_{ij}$ denotes the elements of the adjaceny matrix, we have 
\begin{equation}
    \frac{d}{dt}\langle s_i\rangle = \lambda\sum_{j} A_{ij} \langle (1-s_i)s_j\rangle,
\end{equation}
where the angle brackets stands for an average over realizations. This is an equation for the first moment $\langle s_i\rangle$, but on the right-hand side we find the second-order moment $\langle s_i s_j\rangle$. If we were to write down a similar equation for $\frac{d}{dt}\langle s_i s_j\rangle$, we would find third-order moments and so on. The term ``moment-closure'' refers to mathematical approximations with which these hierarchies of equations are truncated. The key step is to approximate higher-order moments by expressions involving lower-order moments [similarly to the Kirkwood closure method in physics \cite{kirkwood_statistical_1935}]. For example, writing $\langle s_i s_j\rangle \approx \langle s_i\rangle \langle s_j \rangle$ leads to a closed description for first-order moments only. This neglects correlations between the states of neighboring nodes, amounting to a type of mean-field description. Similarly, retaining first and second order moments is akin to making a pair approximation. Thus, moment-closure methods are not an alternative to approximate master equations, pair-approximation methods or mean-field theories, but there is overlap between the approaches. We point the reader to \textcite{scholl_moment_2016} for a mathematical review.

In the study of individual-based models on static or adaptive networks, moment-closure methods are used to truncate equations for so-called ``network moments'' [see \cite{demirel_moment-closure_2014} for an insightful pedagogical account; further resources can be found in \cite{zschaler_adaptive-network_2012}]. The network moments are counts of certain subgraphs with nodes in specific states. These could for example be nodes in state $A$, pairs $AB$ of nearest-neighbor nodes in states $A$ and $B$, or triplets $ABA$ (i.e. a triplet with a node in state $B$ in the center, connected to two nodes in state $A$). In models of opinion dynamics, these network moments describe quantities such as the mean number of nodes in a particular opinion state,  the density of connected pairs of nodes who are in the same opinion state or who disagree respectively, triplets of nodes who are all in a particular opinion state, etc.

Closed equations for the time evolution of these abundances can then be formulated by truncating the hierarchy at a given level, for example by using approximations such as $[ABB]\approx \kappa  2[BB][AB]/[B]$, where $\kappa$ is a time-dependent factor (to be determined for the model at hand) capturing details of the network structure \cite{demirel_moment-closure_2014}. We have used square brackets to indicate the density of a particular motif. Similarly to mean-field theories and pair approximations, there are homogeneous and heterogeneous variants of moment-closure approaches.

Moment-closure methods have been used in the modeling of epidemic spreading \cite{miller_epidemic_2014,porter_dynamical_2016,kiss_mathematics_2017}, and can be formalized via algebraic theory \cite{house_algebraic_2015} or entropy maximization arguments \cite{rogers_maximum-entropy_2011}. Applications to models of opinion dynamics on static and adaptive networks include work by \textcite{demirel_moment-closure_2014,silk_exploring_2014, peng_multilayer_2021,zschaler_adaptive-network_2012}.  Moment-closure methods are particularly adept at characterizing critical behavior of coevolving opinion models (see \sref{subsec:coevolution}), and to study aspects such as the role of heterogeneity and correlations in the fragmentation transition of the \avm. We point to \textcite{wuyts_mean-field_2022} for an automated procedure to carry out moment closure.

\subsection{Numerical simulations}
\label{sec:simulations}

\subsubsection{Synchronous and asynchronous updating}
\label{sec:updating}

There are two main updating procedures for simulations of agent-based models of opinion dynamics. These are referred to as ``asynchronous'' or ``synchronous'', respectively. The former is also known as ``sequential'' updating, and the latter as ``parallel'' updating.  

In \textit{asynchronous or sequential updating}, nodes (or edges) are updated in sequence. At each step of the dynamics, only one node is considered for a state change (or one link for rewiring). Nodes and links can be considered in a fixed sequence (e.g., starting with node $1$, then node $2$ and so on, and then repeating the sequence). Alternatively, at each step one node or link is chosen at random for a potential update (``random sequential updating''). The time step associated with an opinion update is usually set to $\Delta t = 1/N$. This means that, on average, each node is considered for update once per unit time. A collection of $N$ attempted updates is referred to as a ``generation'' or a ``Monte Carlo sweep''. The choice of $\Delta t=1/N$ for each microscopic event allows for comparisons between systems of different size and continuous-time descriptions when $N\to\infty$. For large $N$ it is thus natural to think of (random) sequential updating as a dynamics in (quasi-) continuous time.

In \textit{synchronous or parallel updating}, all nodes (or links) are considered simultaneously for update. This means that any number of nodes (links) can change at the same time in one step (typically, a finite fraction of all degrees of freedom). A time step here corresponds to one generation, i.e. $\Delta t=1$. Parallel updating thus describes discrete-time dynamics.
 
A smooth interpolation between these two schemes can be achieved by updating a randomly selected fraction $f$ of nodes for update at time $t$, with $f = 1/N$ and $f=1$ corresponding to sequential and parallel updating, respectively \cite{porter_dynamical_2016}. Parallel and sequential updating procedures in a given model can lead to different outcomes, as discussed by \textcite{fennell_limitations_2016}.  

\subsubsection{Event-driven simulations}
\label{sec:gillespie}

Asynchronous updating can be computationally inefficient. For example, in the voter model one would first select an agent at random, and then choose a random interaction partner. To do this the algorithm has to generate two random numbers. If the two agents are in the same state no actual event occurs (as the voter model dynamics consists of copying the state of the neighbor), and the state of the system is exactly the same. Thus, two random numbers have been ``wasted'' and the algorithm proceeds.

Event-driven simulation algorithms ensure that a change of state occurs at each step of the simulation. In turn, the length of each time step is a random quantity, generated from the transition rates. The key quantities are thus survival times, i.e. times to the next opinion update, also referred to as inter-event times in the temporal networks literature \cite{holme_temporal_2013}.

\paragraph{Gillespie algorithm.}

The best-known method is the stochastic simulation algorithm by
\textcite{gillespie_general_1976,gillespie_exact_1977}, initially explored for  coupled chemical reactions \cite{anderson_modified_2007}, and related to what is now known as kinetic Monte Carlo algorithms \cite{sickafus_introduction_2007}. We describe the Gillespie algorithm for a model with binary node states and a fixed network, but generalizations are easy to construct. 

At each step of the simulation, a rate of change $\lambda_i$ is determined for each node $i=1,\dots,N$. These rates are based on the present state of the system. In the voter model for example, $\lambda_i$ would be the proportion of neighbors of node $i$ that have opposite opinion. The Gillespie algorithm then generates a sequence of events such that precisely one node changes state in each event. The time to the next event, $\tau$, at a given step is drawn from an exponential distribution with parameter given by the {\em total} rate of change in the system, $\lambda=\sum_{i=1}^N\lambda_i$. Time is incremented by $\tau$. Subsequently the algorithm decides what node changes state, with node $i$ chosen with probability $\lambda_i/\lambda$, and the state of this node flips. All reaction rates are updated, and the process repeats. A description in pseudo-code is given by \textcite{anderson_modified_2007}.

In systems with all-to-all interactions (and assuming binary states), there are only two possible events, $n\to n+1$ and $n\to n-1$ (with $n$ the number of individuals in state $1$). Denoting the corresponding rates as $\lambda^\pm(n)$, the time to the next event is exponentially distributed with parameter $\lambda(n)=\lambda^+(n)+\lambda^-(n)$ if the current state is $n$. The next event is of the type $n\to n+1$ with probability $\lambda^+(n)/\lambda(n)$, and of type $n\to n-1$ with probability $\lambda^-(n)/\lambda(n)$. 

The algorithm can be used to generate stochastic trajectories of the system. This provides a statistically faithful ensemble of samples of the stochastic process defined by the reaction rates. 

\paragraph{\texorpdfstring{$\tau$}{t}-leaping.}

Although more efficient than direct asynchronous simulation, the Gillespie algorithm can become costly in terms of computing time when there are many reaction channels, and/or when the system is large.  On networks, for example, the state of the system cannot be described by simple occupation numbers. Reaction rates have to be formulated for all nodes, so there are ${\cal O}(N)$ events to choose from at each time step, increasing the computational cost for the selection of a node for update. 

The so-called $\tau$-leaping algorithm is often an appropriate alternative \cite{gillespie_approximate_2001}. This algorithm operates in discrete time [with time step $\Delta t={\cal O}(N^0)$], and is no longer exact (i.e., it does not produce a statistically faithful sample of trajectories of the underlying continuous-time model). $\tau$-leaping algorithms are based on the approximation that all rates $\lambda_i$ are constant during each time step. Node $i$ changes state in the next step with probability $\lambda_i \Delta t$, and opinion updates occur independently from those of other agents so that multiple nodes can change state in the same time step. For systems with all-to-all interactions and binary states, this translates into binomially distributed numbers of individuals changing from state $0$ to state $1$ (and vice versa) in each time step.

\paragraph{Non-exponential inter-event distributions, time-dependent rates and Lewis' thinning algorithm.}
\label{para:thinning}

The Gillespie algorithm is remarkably flexible in simulating an array of reaction-like, non-equilibrium processes with or without absorbing states. It can be used not only in opinion dynamics, but also to simulate epidemic spreading and information diffusion. The algorithm can be adapted to heavy-tailed inter-event time distributions, e.g., those initially observed in empirical communication dynamics by \textcite{barabasi_origin_2005}. Event-driven simulations can also capture memory effects in non-Markovian stochastic processes \cite{anderson_modified_2007, boguna_simulating_2014,masuda_gillespie_2018,baron_analytical_2022}, as well as epidemic and social contagion dynamics on bursty temporal networks \cite{vestergaard_temporal_2015,holme_fast_2021,unicomb_dynamics_2021}.

A very clever exact method to deal with time-dependent rate coefficients is Lewis' ``thinning algorithm'' \cite{lewis_simulation_1979}. This addresses models in which the reaction rates $\lambda_i$ change between discrete update events (i.e., they have their own explicit time dependence). In opinion dynamics, this could model situations in which the agents' rate to imitate behavior has a seasonal time-dependence, or in which external factors favoring one opinion over the other vary in time \cite{mobilia_polarization_2023, caligiuri_noisy_2023}. The issue also arises in models with aging \cite{baron_analytical_2022}. The central idea of the thinning algorithm is to add a ``null reaction'' to the system. This null reaction comes with a suitably chosen time-dependent rate that guarantees that the total rate of all possible events (including the null reaction) is constant in time. The standard Gillespie algorithm can then be used for the system including the null reaction. Whenever a null reaction is selected for implementation, time is incremented as in the standard Gillespie method, but no state change occurs.

\subsection{Optimization methods}
\label{subsec:algo-appr}

The convergence of opinion dynamics within a population towards a collective macroscopic state, such as consensus or polarization, is sometimes treated as an optimization problem in the field of computer science.
This approach leads to the development of various algorithmic solutions we describe below. 
Many of these works focus on the Friedkin--Johnsen (FJ) \cite{friedkin_social_1990} or DeGroot \cite{degroot_reaching_1974} models as a workhorse, probably due to their traceability.

\subsubsection{Inference of model parameters}

Recent research has increasingly focused on learning the parameters of opinion dynamics models from data \cite{peel_statistical_2022}.
This task is equivalent to estimating the model parameters under the assumption that the opinion dynamics model has generated the observed data.
In a linear influence model, learning the pairwise influences between agents, given the opinions, is framed as a linear optimization problem.
\textcite{de_learning_2014} derive the exact analytical form of the likelihood function for linear models, like DeGroot.

In more complex, nonlinear, discrete models, the derivation of the likelihood typically relies on Probabilistic Generative Models (PGMs) and differentiable approximations of the discrete rules of models under scrutiny.
In this context, the bounded confidence interval of a bounded confidence model is estimated by maximizing the likelihood function with gradient-descent methods~\cite{lenti_likelihood-based_2024}.
Likewise, in a similar model including a backfire effect, \textcite{monti_learning_2020} estimated the backfire thresholds by means of expectation maximization algorithms.
By formulating the opinion dynamics models as PGMs, such an estimation task can be effectively approached by variational inference.
In this way, one can estimate the bounded confidence interval, the backfire threshold, and other discrete attributes of the agents in different extensions of the bounded confidence model~\cite{lenti_variational_2025}.

In contrast, in continuous-time opinion dynamics models, parameter estimation draws inspiration from stochastic differential equations.
In this realm, the algorithms SLANT~\cite{de_learning_2016} and SLANT+~\cite{kulkarni_slant_2017} use Poisson and Hawkes processes for models with private opinions and public actions.
Given the tractability of these stochastic processes, the parameters are estimated by maximizing the likelihood function.

Instead of learning the model parameters, Sociologically Informed Neural Networks (SINNs) learn to predict the trajectories of opinions~\cite {okawa_predicting_2022}.
SINNs are a sociological adaptation of the wider class of Physics Informed Neural Networks (PINNs).
SINNs consider opinion dynamics models that can be described by an ordinary differential equation, such as DeGroot and Friedkin-Johnsen.
SINNs are then trained by minimizing a loss function that comprises: (i) a data loss, which measures the discrepancy between the output of the neural network and the data; and (ii) a differential equation loss, which forces the constraints imposed by the equation.
As a result, the trained SINNs produce trajectories that are both data-driven and consistent with the theoretical opinion dynamics model.

\subsubsection{Opinion maximization}

One early contribution in this area is by \textcite{gionis_opinion_2013}. 
This study addresses an opinion optimization problem on a social network where opinions evolve according to the FJ model.
Similar to the well-known influence maximization problem \cite{chen_efficient_2009}, the objective is to select a set of $k$ initial nodes to influence, thereby maximizing the overall average opinion in the network. 
A practical application can be a company hiring selected influencers to maximize positive opinions about a product in the overall population. 
The authors propose a greedy algorithm for this task, drawing on the connection between opinion maximization and absorbing random walks.

Within this research line, \textcite{zhou_sublinear_2023} address the opinion maximization problem for the leader-follower DeGroot model on a directed network. 
They propose a deterministic greedy algorithm and a faster sampling algorithm based on absorbing random walks. 
In contrast to standard opinion maximization,  \textcite{he_dynamic_2023} examine a scenario where the opinions of initially selected nodes are not static and can evolve over time due to negative influence from their neighbors. 
To address this dynamic opinion maximization framework, the authors introduce an adaptive cooperation model using reinforcement learning, proving its convergence properties. 
Furthermore, they develop an algorithm for the dynamic generation of the initial set of seed nodes targeted for influence.

\textcite{abebe_opinion_2018} investigate an alternative perspective on the opinion maximization problem, focusing on interventions that modify individuals' susceptibility to persuasion rather than altering the social network structure. 
Within the context of the FJ model, this means altering the susceptibility of a selected set of nodes, as opposed to adjusting their internal opinions. 
When the number of nodes whose susceptibility can be modified is unconstrained, the authors provide a polynomial-time algorithm to determine the optimal target set. 
Conversely, they prove that the problem of identifying a limited-size optimal target set is NP-hard, and the authors introduce a heuristic solution to address this complexity.

\subsubsection{Reducing opinion polarization}

The reduction of opinion polarization has often been framed as an optimization problem as well. 
One of the foundational works in this area is by \textcite{musco_minimizing_2018}. 
In this paper, the authors pose the question of which social network structure simultaneously minimizes disagreement and avoids the emergence of echo-chambers (called polarization, or controversy in the paper), given a certain opinion distribution. 
The problem is non-trivial due to the trade-off between a low disagreement, quantified by a small number of connections between individuals holding different opinions, and a high segregation between communities of like-minded nodes. 
Utilizing the FJ model to simulate opinion dynamics, the authors developed an exact algorithm to identify network structures that achieve this minimization, revealing that sparsely connected networks are optimal for reducing both disagreement and polarization.

The research question of minimizing polarization and disagreement has also been explored by~\textcite{zhu_minimizing_2021}.
This work takes a different approach from~\textcite{musco_minimizing_2018}, by focusing on link recommendation rather than network design from scratch. 
The authors formulate an optimization problem aimed at selecting $k$ new links to add to an existing social network, intending to minimize the sum of polarization and disagreement. 
To solve this, they present a simple greedy algorithm that provides a constant-factor approximation guarantee and has a cubic time complexity.
\textcite{cinus_rebalancing_2023} also consider the same optimization problem, unfolding on directed networks. 
Different from the previous work, the authors consider a directed input graph and focus on re-weighting edges to minimize the level of polarization and disagreement in the network.

The problem of reducing polarization has been addressed by \textcite{haddadan_repbublik_2021} on hyperlink graphs among web pages, where readers can be trapped in polarized bubbles.
The authors tackle the problem of reducing this bias through edge insertions, focusing on finding the best $k$ edges to add to the hyperlink graph to maximize reduction. 
The proposed algorithm, RePBubLik, uses a variant of random walk closeness centrality to select edges, achieving a constant-factor approximation and outperforming existing methods in reducing structural bias.

Unlike works focusing on direct disagreement,~\textcite{chen_quantifying_2018} study the risk of conflict. 
Employing the FJ model to describe opinion dynamics, the authors differentiate between internal conflict (the discrepancy between an individual's internal belief and their expressed opinion) and external conflict (disagreement with others). 
Departing from approaches that rely on specific initial opinions, this work investigates conflict risk  depending purely on the topology
of the network.
They quantify conflict under extreme and average internal opinion distributions, showing that both cases can be minimized through local network editing.
They also show that the sum of internal conflict, twice the external conflict, and controversy remains constant throughout the opinion formation process.

Along the same line,~ \textcite{wang_relationship_2023} study how popular link recommendation algorithms impact conflict risk (quantified as the sum of polarization and disagreement) on a social network.
The authors study the amount of change in opinion conflict caused by link additions. 
By deriving closed-form analytical expressions for this, they show that purely adding new links cannot increase opinion conflict.
Then they show that some recommendation algorithms have a better ability to reduce conflict more effectively. 

\subsubsection{Adversarial approaches}

In contrast to minimizing disagreement, \textcite{gaitonde_adversarial_2020}
investigate the adversarial manipulation of opinion dynamics to maximize discord within social networks. 
These attacks aim to destabilize online communities and can originate from external malicious entities, for example, networks of bots disseminating fake news.
Using spectral graph theory, the authors demonstrate that various parts of the network's spectrum, not just extreme values, determine the adversary's ability to create discord. By linking spectral properties to network structure, they identify vulnerable and resilient network types and propose a convex programming approach for network defenders to efficiently mitigate attacks through heterogeneous node insulation.

The adversarial approach has also been addressed by~\textcite{tu_adversaries_2023}.
In contrast with previous works, which assumed the adversary possessed full knowledge of network topology and user opinions, the authors consider the more realistic scenario where only the network topology is known, affirmatively demonstrating that an attacker can still sow discord. 
They present approximation algorithms for the FJ model that identify influential users, showing their effectiveness in increasing disagreement, even when initial network polarization is low.

\section{Outlook}
\label{sec:outlook}

In the following we highlight some directions in which we envision the field of opinion dynamics to advance.

\subsection{More data to model opinion dynamics}

Opinions are key to how we perceive the world and those around us, shaping our behaviors and relationships with others. 
They are an essential part of political and cultural discourse worldwide, leading to a myriad of collective social phenomena from cooperation and consensus to polarization and conflict. 
Together with other types of self-organized dynamics such as epidemic spreading \cite{pastor-satorras_epidemic_2015}, synchronization \cite{arenas_synchronization_2008}, collective intelligence/adaptation \cite{centola_network_2022,galesic_beyond_2023}, games \cite{szabo_evolutionary_2007}, and social influence \cite{flache_models_2017,centola_damon_how_2018}, opinion formation is an archetypal example of how the interconnected decisions of individuals lead to emergent collective behavior at the level of entire populations. 

Unlike biological phenomena, where models are informed by established theories and extensive empirical data, modelling opinion dynamics is complicated by the interplay of psychological, social, and cultural factors.
This makes the definition, measurement, and prediction of opinions complex.
Empirical data on opinions, though varied in source and collection method—--ranging from detailed individual-level experiments and surveys to aggregated population statistics from polls and online platforms—--remains sparse compared to the abundance of opinion dynamics models. 
These models span from simple, analytically tractable equations inspired by psychosocial experiments or physics analogies, to complex network-based and agent-based models requiring computational exploration.

For more than a decade, reviews on opinion dynamics have noted the asymmetry between experimental and theoretical efforts \citep{castellano_statistical_2009,sobkowicz_modelling_2009,loreto_participatory_2017,peralta_opinion_2022}. 
One reason for this can perhaps be found in the contributions of the statistical physics community in the field. Physicists are accustomed to a wide array of experimental results and fundamental laws around which to construct models and theories. But there is also an inherent value in the idealized, mechanistic modeling of opinion dynamics \citep{holme_mechanistic_2015}, as these models allow us to ask \textit{what-if} questions, probe scenarios, test descriptive theories for consistency, and explore possible mechanisms of emergent phenomena like consensus, fragmentation, and polarization. 
There is an undisputed need for as much data as can reasonably be obtained. And yet, despite some opposing views \cite{anderson_end_2008}, in an area as interdisciplinary as opinion dynamics, theory still matters to build credible narratives of what the data actually reveals \cite{gonzalezbailon_social_2013, epstein_why_2008}.
Such a theoretical effort will allow us to both better understand humans and use computational social science \cite{conte_manifesto_2012} as a tool to guide policy making \cite{lazer_computational_2020} and ethical algorithm development \cite{wagner_measuring_2021}.

As data and models of opinion dynamics get ever closer to one another, we expect an increase in attempts to quantitatively validate progressively less idealized mechanisms of social interaction that lead to individual opinion changes and their corresponding population-level distributions. 
These efforts might include more in-the-lab experiments with designs similar to those used to test game-theoretical notions of cooperation \cite{gracia-lazaro_heterogeneous_2012}, as well as in-vivo, \textit{citizen-science} experiments \cite{sagarra_citizen_2016} and data donation projects \cite{kmetty_public_2025} where researchers collaborate with the general population to collect data on their opinions and interactions. 
Regarding observation data, more empirical studies are needed from diverse platforms. 
Research has concentrated on Twitter-like platforms (or, more recently, Reddit), yet dynamics may differ significantly on algorithm-heavy platforms such as TikTok. 
This is crucial as public discourse shifts, and findings may not generalize \cite{pera_measuring_2023}.

Detecting causal relations between opinion change and various confounding factors is also crucial. For instance, causal pathways to opinion change could be estimated by combining signals from social interaction and news media influence \cite{lenti_causal_2025}.
Likewise, causality can help selecting between competing mechanisms of interaction. 
Examples are the current controversy as to whether exposure to opposing political views on social media increases \citep{bail_exposure_2018} or decreases \citep{balietti_reducing_2021} polarization, or the challenge of distinguishing between simple and complex contagion in information diffusion \cite{cencetti_distinguishing_2023,st-onge_nonlinear_2024,andres_distinguishing_2025}. 
Mixing data sources in opinion dynamics will also go hand in hand with the development of tools to automatically infer microscopic rules of behavior \cite{chen_flexible_2020} and opinion states \cite{masuda_detecting_2019} from noisy and incomplete data. 
The statistical inference of model rules from empirical opinion dynamics might benefit from tools such as approximate Bayesian computation \cite{gutmann_bayesian_2016} and network-aware nonparametric inference \cite{peel_statistical_2022}.

Overall, a better mutual knowledge between people in different
fields (physics, mathematics, political science, computer science) working
on the same problems will be key for the development of the field of opinion dynamics in the years to come.

\subsection{Artificial Intelligence and opinion dynamics}
\label{sec:AI_op_dyn}

Recent advancements in Artificial Intelligence (AI), particularly Large Language Models (LLMs), are driving new developments in social simulation. 
One avenue involves using LLMs to simulate individual human behavior, such as opinion formation and expression (\sref{sec:LLMs}). 
Concurrently, LLMs are being deployed in autonomous AI agents that interact without necessarily representing human behavior (e.g., as purchasing assistants or in coordination games). 
These AI agents can also hold opinions, leading to an emerging application of opinion dynamics focused on understanding the collective opinions of AI agent societies, rather than human ones (\sref{sec:AI_agents}).

\subsubsection{Large language models in opinion dynamics models}
\label{sec:LLMs}

\paragraph{Agent dynamics and environments.}

Social simulations can be built where LLM agents freely describe their actions, with a simulated environment using another LLM to execute the changes described in the text produced by agents  \cite{park_generative_2023}. 
This has motivated the current development of generative agent frameworks in which the dynamics of agents are specified by an LLM rather than by explicit equations as in a traditional agent-based model \cite{vezhnevets_generative_2023}. 
An application of LLMs to social simulation bears the promise of simplifying assumptions about microscopic dynamics in agent-based modelling \cite{larooij_large_2025}. 
But the use of LLMs in social simulation, in particular opinion dynamics models, is contingent on the validity of LLMs to reproduce individual and social human behavior \cite{lu_llms_2024}.
An early example of this used the GPT-2 language model (a predecessor of ChatGPT) to allow agents in an opinion dynamics model to verbalize their opinions during interaction \cite{betz_natural-language_2022}. 
In that design of an agent, LLMs are used in two steps of the simulation. 
First, the LLM interprets the opinions expressed through text by other agents, which is then used to update the opinions of the agent based on various principles, including bounded confidence (see \ref{subsubsec:bounded_confidence}).  
Agents express their opinions through text, which is part of the prompt of other agents when they perceive the opinions of others.

LLMs can function beyond mere communication components in agents, determining actions and opinion updates. 
\textcite{piao_emergence_2025} used LLMs to model discussion dynamics, reproducing human-like polarization. 
While replacing explicit behavioral assumptions with LLM prompts is promising, it faces challenges like the "self-consistency problem" \cite{piao_emergence_2025}, where agents exhibit paradoxical behavior. 
Future work must integrate consistency checks, perhaps by prompting agents to rectify detected inconsistencies.

\paragraph{Social media simulation.}

LLMs offer potential for modeling online communities, especially text-based ones, enabling new social simulation approaches.
They can mimic user behavior and interactions on a social platform, processing textual inputs (like posts or comments) and generating textual outputs, effectively simulating content creation and communication flows. 
The OASIS framework \cite{yang_oasis_2025} exemplifies this, simulating platforms like Reddit and Twitter to analyze the effects of interventions and platform design on collective behavior. 
Similarly, \textcite{park_social_2022} used LLMs to simulate Reddit users, demonstrating how moderation rules influence agent behavior. 
This paradigm allows agents to operate as text-to-text functions, eliminating explicit opinion representations and enabling macroscopic state measurement via textual output, like toxicity or network structures.

Using LLMs in social media simulation opens the possibility of not having to use an explicit representation of the opinion of the agent.
Instead, agents operate as functions that map an input, such as the text they see on social media, to an output, the text they produce and can be seen by other agents. 
Macroscopic states can then be measured from agent-produced text using methods akin to observational social media analysis (see Section \ref{subsec:opinion_data_sm}), such as assessing text toxicity or network structures. 
This enables testing different feed algorithm designs \cite{tornberg_simulating_2023} to evaluate their impact on phenomena like echo chambers or cross-party toxicity.
LLMs in social simulation also enable nuanced agent diversity through textual "personas" \cite{argyle_out_2023, park_generative_2024}. 
These personas, derived from individual-level survey data (including demographics and attitudes) \cite{tornberg_simulating_2023}, can be integrated as prompt prefixes to simulate specific agent properties.

Replacing agent dynamics with LLMs is a promising avenue, but their lack of interpretability needs to be outweighed by their realism in reproducing the behavior of individuals and their interactions. 
LLMs can represent average inter-group opinion differences \cite{argyle_out_2023} but may reduce intra-group variance, leading to stereotyping \cite{bisbee_synthetic_2024}. 
 Imbalances in training data also cause differential group representation \cite{santurkar_whose_2023}. 
Despite initial excitement, recent studies highlight challenges in accurately representing diverse individuals across groups due to biases \cite{sen_missing_2025-1}, lack of heterogeneity, and issues like 
temporal sensitivity and language transferability \cite{kozlowski_simulating_2024}. 
Nevertheless, LLMs can reproduce individual experimental dynamics, with strong correlations to human behavior \cite{hewitt_predicting_2024}, though they often show amplified effect sizes. 
A promising approach to overcoming this is to go beyond personas and to fine-tune models for different groups of people.
For example, using Reddit data by partisanship \cite{haller_opiniongpt_2024}, or by steering models with explainability methods that show how the LLM state can be configured towards different ideologies \cite{kim_linear_2025}.

\subsubsection{Artificially intelligent agents}
\label{sec:AI_agents}

In contrast to modeling human behavior, one can study AI agents as entities of interest, whose opinions and behaviors have societal consequences. 
For example, LLMs can complement opinion polling, representing the political preferences of individuals, thus
augmenting opinion polling \cite{gudino_large_2024}. 
If these AI agents interact to represent members of parliament, as in an AI-augmented parliament, their opinion dynamics and collective behaviors present both opportunities and risks.

The analysis of opinion dynamics among AI agents in interacting scenarios functions as a testbed where autonomous LLMs directly engage in controlled settings, rather than simulating human behavior. 
For example, \textcite{marzo_ai_2025} found that even in minimal, fully connected communities with no incentives or external information, LLMs tend to reach consensus. 
This consensus is limited by group size and LLM linguistic abilities, with larger communities hindering consensus unless language capabilities are high. 
This phenomenon is analogous to Dunbar's scaling \cite{dunbar_neocortex_1992}, where more advanced social brains support coordination of larger groups. 
Notably, recent LLMs enable AI agents to surpass this scaling, coordinating in groups significantly larger than typical human informal groups (150-250 individuals).
Within this research line, \textcite{flint_ashery_emergent_2025} analyzes how conventions emerge among AI agents in the naming game,
where agents collectively establish a shared vocabulary for an object through local interactions.
The authors show how minorities can influence a collective outcome among AI agents.

Decades of opinion dynamics research offer valuable tools for analyzing AI agent interactions and their societal impact. 
LLMs can be trained to promote human consensus \cite{bakker_fine-tuning_2022}, generate persuasive arguments \cite{breum_persuasive_2024}, and even counter conspiracy theories \cite{costello_durably_2024}, highlighting the need to assess risks from AI agent interactions. 
For instance, consensus in LLM agents can be modeled by a voter model, showing agreement between analytic solutions and LLM benchmarks \cite{marzo_ai_2025}. 
These research lines show promise in applying opinion dynamics knowledge to predict macro-level outcomes of interacting AI agent systems. 
This is crucial for mitigating risks, such as preventing malicious AI swarms from creating artificial consensus in online media \cite{schroeder_how_2026}, or averting undesirable spontaneous synchronization among AI agents (e.g., flash crashes in stock trading).

Using tools for the study of opinion dynamics as testbeds for these scenarios, researchers can predict when these macro-level behaviors can appear and propose policies or interventions to prevent them, such as circuit breakers for trading. 
While consensus among AI agents can be desirable (e.g., in political decision-making), research on opinion dynamics can also clarify its conditions and side effects, like human understandability or enforceability. 
As AI agents become more prevalent, opinion dynamics research gains relevance. %
However, future approaches must account for these agents being language models, not full behavioral representations, trained from potentially biased or incomplete data, and run on hardware with high energy consumption.

\begin{acknowledgments}
We thank Luca Aiello, Andrea Baronchelli, Claudio Castellano, Bikas Chakrabarti, Guillaume Deffuant, Gianmarco De Franscisci Morales, Andreas Flache, Serge Galam, Mirta Galesic, James Gleeson, Rainer Hegselmann, Iacopo Iacopini, Hang-Hyun Jo, J\'anos Kert\'esz, Jacopo Lenti, Mauro Mobilia, Henrik Olsson, Romualdo Pastor-Satorras, Agnieszka Rusinowska, Maxi San Miguel, Ra\'ul Toral, \& Andr\'e Vilela for helpful comments on earlier drafts of the manuscript.

M.S. acknowledges support from Grants No.
RYC2022-037932-I and CNS2023-144156 funded by
MCIN/AEI/10.13039/501100011033 and the European
Union NextGenerationEU/PRTR. 
F.B. acknowledges support from a University of Pennsylvania MindCORE fellowship.
T.G. acknowledges financial support by the the Agencia Estatal de Investigaci\'on and Fondo Europeo de Desarrollo Regional (FEDER, UE) under projects APASOS (PID2021-122256NB-C22) and COSASTI (PID2024-157493NB-C22), and the Mar\'ia de Maeztu programme for Units of Excellence, CEX2021-001164-M funded by  MCIN/AEI/10.13039/501100011033. 
M.K. acknowledges funding from the National Laboratory for Health Security (RRF-2.3.1-21-2022-00006); the MOMA WWTF; and COLINE DUT-FFG projects. J.L. acknowledges funding from the European Commission under project EuropeanCity2 Grant ID 101178170. K.S-W acknowledges funding from the National Science Centre, Poland under the OPUS call in the Weave programme, project no. 2023/51/I/HS6/02269.

\end{acknowledgments}

\end{document}